\documentclass[twocolumn,natbib]{svjour3}
\usepackage{ragged2e}
\usepackage{amsmath}
\usepackage{caption}
\usepackage{subcaption}
\usepackage[utf8]{inputenc}
\usepackage{graphicx}
\usepackage{physics}
\usepackage{natbib}
\usepackage{color}

\definecolor{grey}{RGB}{2,72,115}
\usepackage[colorlinks,citecolor=blue,urlcolor=blue,bookmarks=false,hypertexnames=true]{hyperref} 
\begin{document}

\title{Towards Optimal Heterogeneity in Lattice Structures}
\author{Yash Agrawal \and G. K. Ananthasuresh}
\institute{M2D2 Laboratory, Mechanical Engineering, IISc Bengaluru \\\email{\{suresh,yashagrawal\}@iisc.ac.in}
\\}
\titlerunning{Towards Optimal Heterogeneity in Lattice Structures}
\authorrunning{Agrawal et al.}

\maketitle

\begin{abstract}

We present a multi-phase design parameterization to obtain optimized heterogeneous lattice structures. The 3D domain is discretized into a cubical grid wherein each cube has eight distinct unit cell types or \textit{phases}. When all phases are present, the  domain resembles a densely connected ground structure. The cross-section area of beam segments in lattice units, modelled using Timoshenko beam theory, are the design variables. All beam segments in a particular lattice phase have the same area of cross-section to keep the number of design variables low. The optimization problem is formulated for stiff structures and is solved using the optimality criteria algorithm. We present a case study to show the superiority of topology-optimized heterogeneous structures over uniform lattices of a single phase. In order to interpret the phase composition, we perform four basic load tests on single phases, namely, tension or compression, shear, bending, and torsion. The phases are ranked based on the stiffness corresponding to individual loading conditions. The results show that in the optimized structure, the local internal load configuration drives the selection of phases. We also note that micropolar elasticity captures the bulk behaviour of heterogeneous lattice structures, and helps not only to interpret the optimality of phases but also to improve the computational efficiency of the proposed optimization technique.\\
\end{abstract}

\begin{section}{Introduction}
Lattice structures fall in the category of heterogeneous cellular materials composed of slender elements such as beams and plates. They are useful for the light-weight design of components and are manufactured using additive manufacturing. There are many desirable properties of lattice structures, which include resistance to buckling \citep{refSigmund}, improved thermal performance, enhanced strength-to-weight ratio and good energy absorption characteristics \citep{refAshby}. The seminal work by A. G. M. Michel (\citeyear{refMichell}) leads to an important learning that optimal structures have a grillage pattern composed of slender components. There is a detail present at different levels of scales. To design such structures, we are required to specify the structure at both macro- and microscales. This premise is evident in various materials present in nature (e.g., bones and biological matter), as well as human-made structures (e.g., Eiffel Tower, Maria Pia bridge) \citep{refLakesHier,refSundaram}. This is the basis for our investigation in designing multiscale materials, using topology optimization tools. We restrict our scope to lattice structures comprising circular cross-section beams, even though cross-section shape provides additional freedom in optimizing beams \citep{refAshby2}.

To define a general cellular structure, we need to define the tessellation of the domain, the topology of the unit cell, and the cross-section area i.e. size of the individual elements \citep{refDhruv}. The microstructure is the part of the structure inside a small subdivision of the domain, usually taken to be a cube (3D) or a square (2D). We define the first level of heterogeneity as follows: A single microstructure repeated over all of the spatial directions to get a porous material with a periodic structure. The bulk behaviour can be modelled using homogenization theory when the size of the unit cell size is much smaller than the size of the structure. When the unit cell topology is kept uniform, but the cross-section sizes of the constituting structural elements are allowed to vary, we get to the second level of heterogeneity. The size variation enriches the design space as the unit cell density can now spatially vary. In the third level of heterogeneity, the unit cell topology is allowed to vary spatially, and the cross-section size of the frame elements can also be varied. The exploration of this level is the contribution of this work.

\begin{figure*}[t]
\centering
    \begin{subfigure}[b]{0.11\textwidth}
    \centering
    \includegraphics[width=\textwidth]{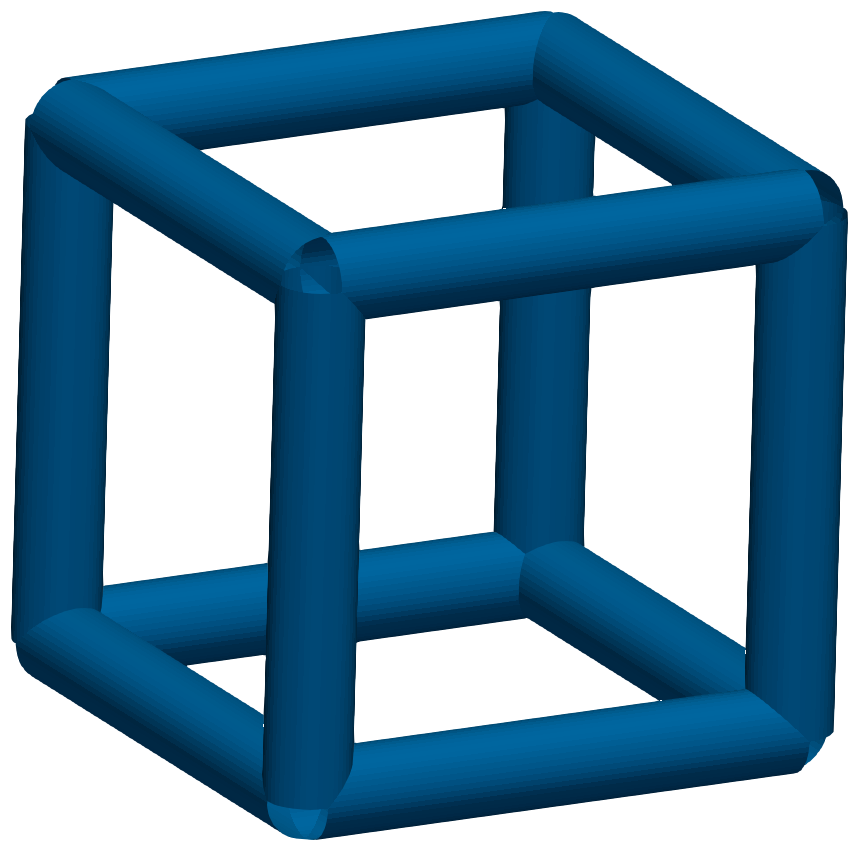}
    \caption{Phase 1}
    \label{Ph1}
    \end{subfigure}
    \begin{subfigure}[b]{0.11\textwidth}
    \centering
    \includegraphics[width=\textwidth]{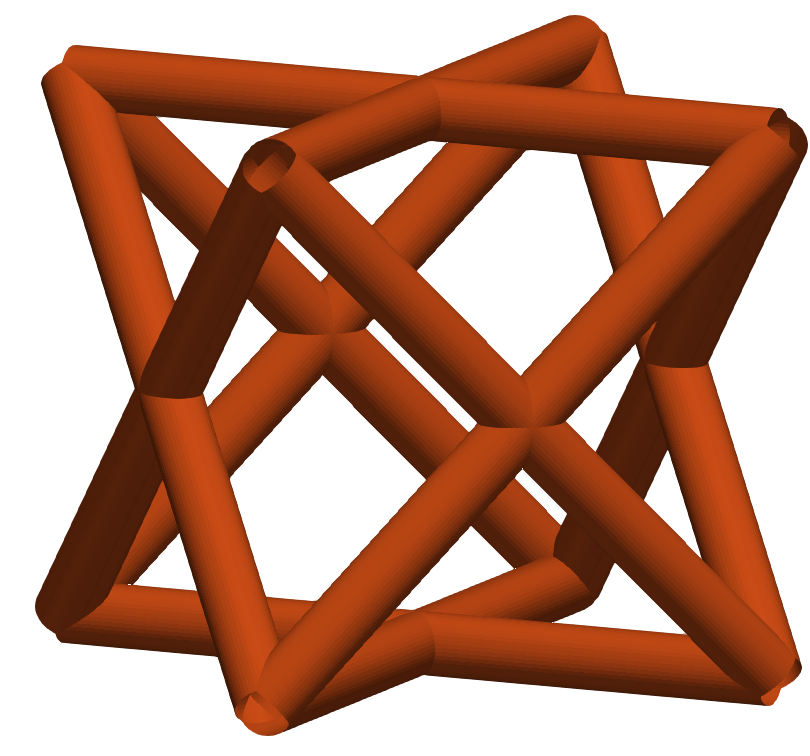}
    \caption{Phase 2}
    \label{Ph2}
    \end{subfigure}
    \begin{subfigure}[b]{0.11\textwidth}
    \centering
    \includegraphics[width=\textwidth]{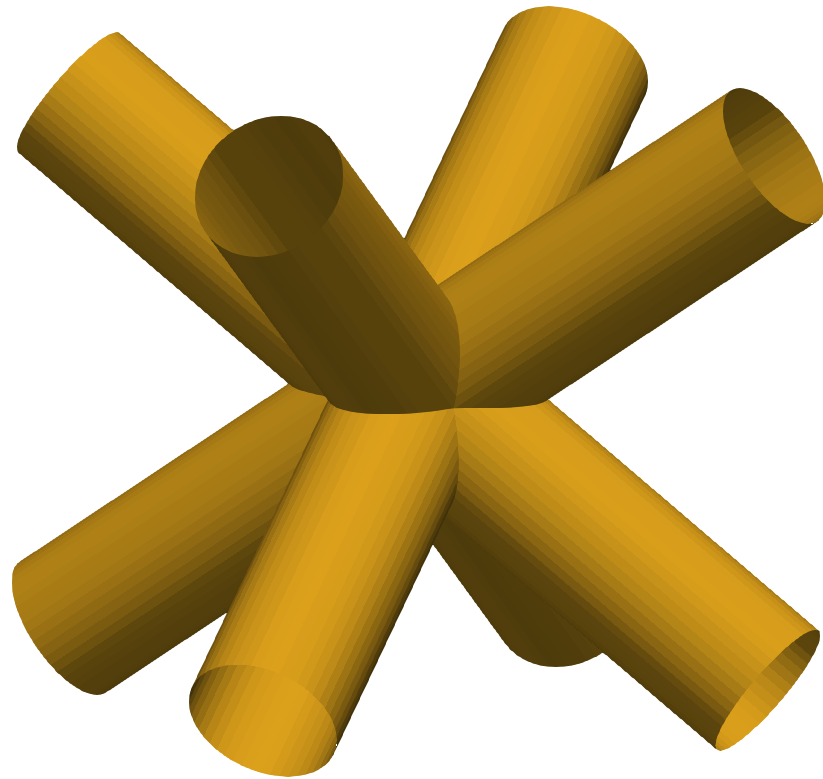}
    \caption{Phase 3}
    \label{Ph3}
    \end{subfigure}
    \begin{subfigure}[b]{0.11\textwidth}
    \centering
    \includegraphics[width=\textwidth]{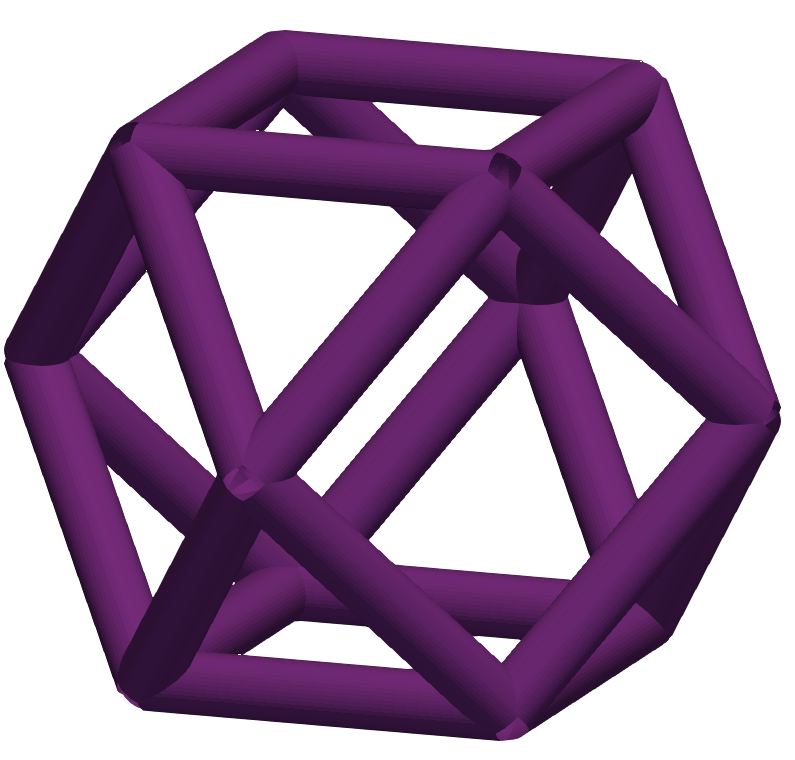}
    \caption{Phase 4}
    \label{Ph4}
    \end{subfigure}
    \begin{subfigure}[b]{0.11\textwidth}
    \centering
    \includegraphics[width=\textwidth]{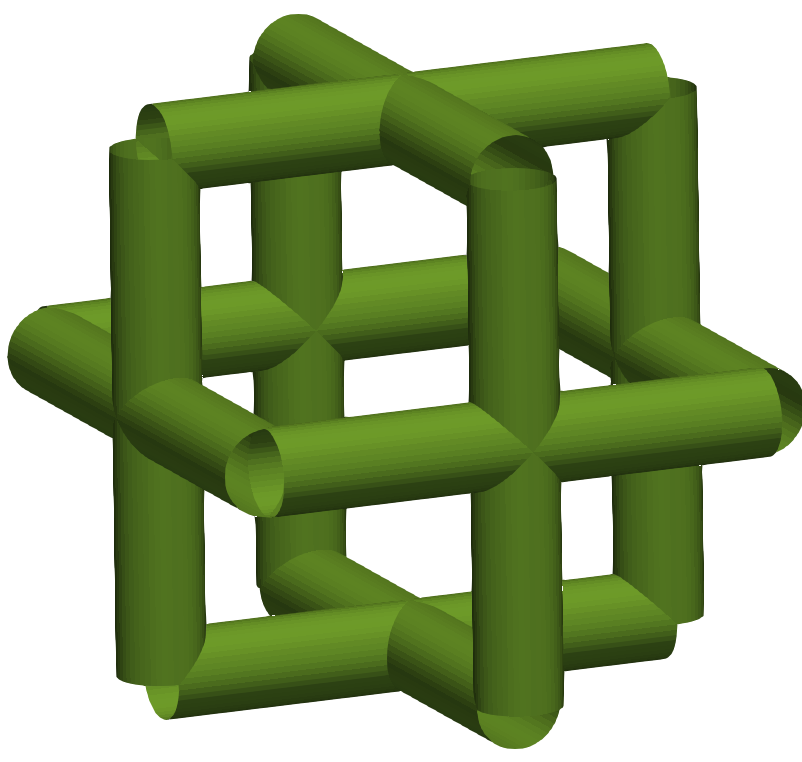}
    \caption{Phase 5}
    \label{Ph5}
    \end{subfigure}
    \begin{subfigure}[b]{0.11\textwidth}
    \centering
    \includegraphics[width=\textwidth]{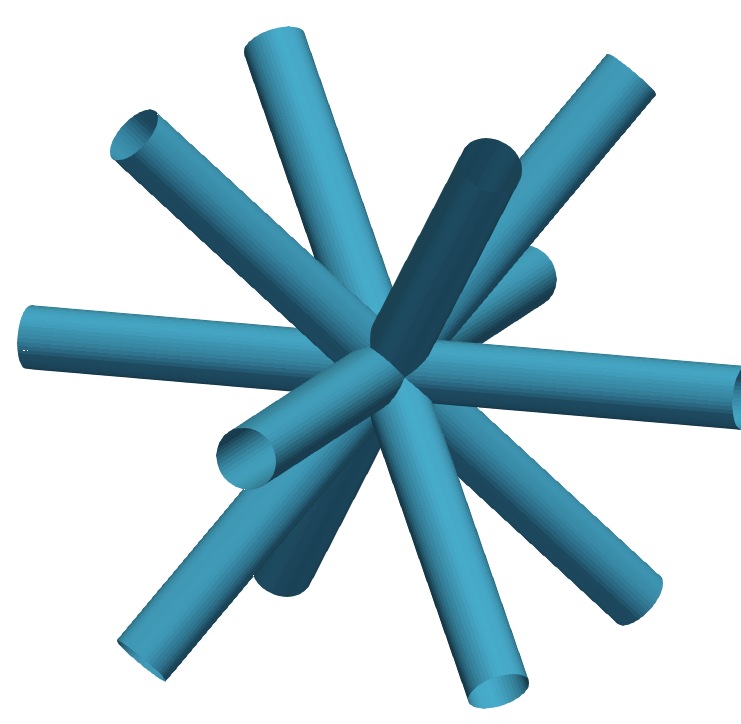}
    \caption{Phase 6}
    \label{Ph6}
    \end{subfigure}
    \begin{subfigure}[b]{0.11\textwidth}
    \centering
    \includegraphics[width=\textwidth]{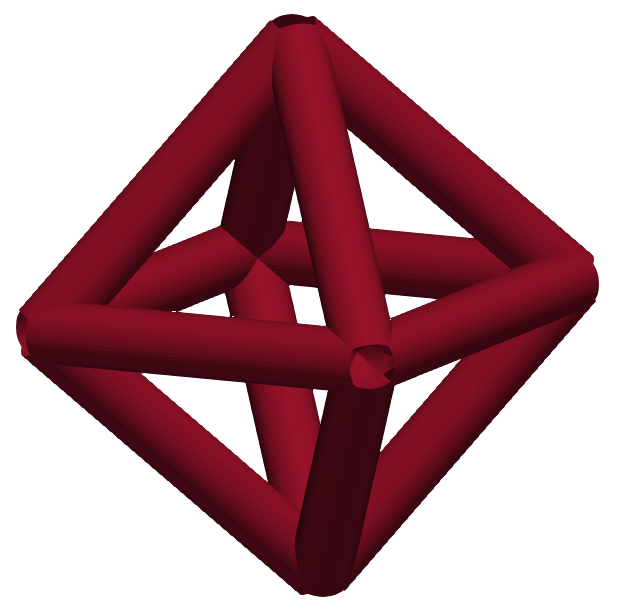}
    \caption{Phase 7}
    \label{Ph7}
    \end{subfigure}
    \begin{subfigure}[b]{0.11\textwidth}
    \centering
    \includegraphics[width=\textwidth]{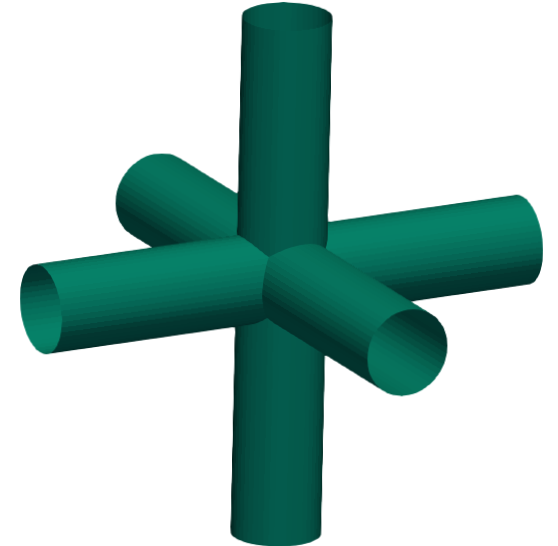}
    \caption{Phase 8}
    \label{Ph8}
    \end{subfigure}
\caption{The eight identified phases i.e. unit cells, each shown in a different colour.}
\label{IndPhases}
\end{figure*}

\begin{table*}
\centering
\caption{Geometric information of the fundamental phases, i.e. unit cells}
\label{gmi}       
\begin{tabular}{|c|c|c|c|}
\hline 
Phase & Number of beam segments & Beam segment length & Redundancy index $n_k$  \\[5pt]
\hline
1 & 12 & 1 & 4 \\[5pt]
2 & 24 & $\frac{1}{\sqrt[]{2}}$ & 2 \\[5pt]
3 & 8 & $\frac{\sqrt[]{3}}{2}$ & 1 \\[5pt]
4 & 24 & $\frac{1}{\sqrt[]{2}}$ & 2 \\[5pt]
5 & 24 & $\frac{1}{2}$ & 2 \\[5pt]
6 & 12 & $\frac{1}{\sqrt[]{2}}$ & 1 \\[5pt]
7 & 12 & $\frac{1}{\sqrt[]{2}}$ & 1 \\[5pt]
8 & 6 & $\frac{1}{2}$ & 1 \\[5pt]
\hline
\end{tabular}
\end{table*}

We discuss recent attempts in design parameterization schemes to obtain structures at different levels of heterogeneity. Various unit cell geometries have been suggested by researchers, for e.g. \cite{refSelect}. Starting with the first level of heterogeneity, \cite{refSandia,refSigmund} use the conventional topology optimization tools to get the optimized structures, using the homogenized material properties for a single unit cell. For the second level of heterogeneity, the homogenization is performed for various densities of a selected microstructure topology and material interpolation is done in terms of a physically meaningful density variable, or a discrete design variables. Then, using a modified sensitivity analysis, the optimization is performed and the structure is designed, called graded structure \citep{refATo1,refATo2,refShi}. One reference puts forth a novel and interesting clustering approach, where density clusters are formed and the unit cell is assigned based on the densities initiated from SIMP method \citep{refKSuresh}. Another approach improves the design space by introducing a possibility of rotating the structure in each subdivision, based on one of the two things: principle stress/strain direction \citep{refStr1}, or conforming to the outer geometry \citep{refChair}. 

Coming to the schemes that address the third level of heterogeneity, we first discuss a few attempts that use concurrent optimization technique. The work initiated with \cite{refRodrigues}, where both the micro- and the macrostructure are designed such that each subdivision has a completely different and optimized geometry, using a density-based approach. In \cite{refSiva2016} a segregated domain is considered, with a distinct unit cell geometry designed for each partition, using a level-set approach. \cite{refDeng} explore a robust design of cellular material. Each subdivision is assigned one of a finite number of microstructures that are designed using discrete material optimization, and a combined SIMP and PAMP interpolation technique. \cite{refMatKang} presented a similar material-distribution type problem, parameterizing microgeometry by density-type variables, and the macrogeometry using velocity field level set. A clustering approach is used in \citep{refJia2020}, where the elemental strain energies serve as a basis to form finite amount of clusters at macroscale, each cluster being assigned a distinct unit cell, designed concurrently. The design space generated by these techniques is similar in the sense that a finite number of unit cells with distinctly different geometries are treated as materials, and distributed in the design domain. So far, concurrent techniques do not consider spatial density variation in the structure. 

The non-concurrent approaches use predefined unit cell types and sometimes allow the density variation in the subdivisions that further enriches the design space. In \cite{refWang2018}, two variables are employed, one being a density variable in the classical-SIMP sense, and the second a discrete variable with distinct predefined unit cell geometries assigned. The elastic properties are interpolated using the discrete variable. A preconceived parameterised level set defines the microstructure in  \cite{refZhang2019}, and the level is adjusted at different points in the design domain to get the density variation aspect in the design. \cite{refChikwesiri} take an exhaustive approach by establishing a elastic property workspace with respect to the radii of various beam segments in a 3D lattice, and then treating those radii as design variables. Similarly, \cite{refWang2020} use a 5\textsuperscript{th} order polynomial for fitting the elastic properties in two variables, first a physically-meaningful density, and a second aspect ratio variable defining the unit cell topology. Moving away from homogenization, \cite{refLiu2020} model the microstructure by static condensation method, with a physically-meaningful density variable and another discrete variable for microstructure definition. The common thread in all of the techniques, except \cite{refRodrigues}, is the connectivity treatment, and the design space. The information about connectivity is lost during homogenization, which leads to the contiguous subdivisions having non-connecting microstructures. The microstructure connectivity is ensured either by a connectivity constraint, or by non-participating design measures.  Futhermore, the methods explored thus far do not permit more than one unit cell type to co-exist in any and every cube of the discretized domain. 

Our design parameterization is a non-concurrent approach, where the design space resembles \cite{refWang2020} and \cite{refLiu2020}, with the extension that many unit cell types are permitted to occupy a location, depending on the optimality of the solution. Thus, our approach is capable of giving optimal heterogeneity in lattice structures. We begin by identifying eight different unit cell types, or phases for cubes in 3D (Figure \ref{IndPhases}). They are centrosymmetric and all of the beam segments in a phase have the same length (Table \ref{gmi}). Junctions exists at corners, midpoints of edges and faces, as well as the centre of the cube. Furthermore, no two phases share a beam segment, although they can intersect. We assign a single density parameter to each of these phases such that each beam in a single phase has the same cross-section area, proportional the density parameter. All eight phases are permitted to exist in the cube and hence there are eight design parameters per cube, one for each phase. When a density parameter becomes small, we consider that the corresponding phase is absent in that cube. In this way, effectively $2^8$ combinations of unit cell topologies are possible, including a void. The parameterization retains the sense of density variation. When all of the phases are present, the microstructure resembles a densely connected ground structure. The cubical grid is populated with eight density parameters belonging to each cube, similar to \cite{refProtein}.

As there is a lot of detail in the structure, modelling with continuum elements is computationally expensive. Homogenization theory is not applicable when the unit cell size (a cube here) is not small enough as compared to the size of the structure. The bulk behaviour of lattice structures is better captured by micropolar elasticity \citep{refMathMult}. Scale effects have been experimentally observed in lattice-structure components that are not explained by classical elasticity; stochastic polymer foams \citep{refOCPFLakes} and lattice structures \citep{refCossLakes}. Discrete asymptotic homogenization \citep{refTollenaere} is a technique that yields the material properties for unit cells with discrete elements such as beams. Since we need to simplify the computation and preserve the physics of the system at the same time, we prefer to model the structure using Timoshenko 3D frame elements \citep{refSpaceAna}. We note that  connectivity is guaranteed by virtue of frame finite element analysis (FEA) in our procedure. We do away with the extensive computations for the homogenization and the material interpolation step. Also, the sensitivity analysis is a straightforward application of the chain rule, is analytically exact, and there is no requirement of sensitivity filters. In this model, there is a source of error due to the stiffening at the junctions i.e. the structure is actually stiffer than predicted. So we err on the conservative side. One way to design the junctions is by using a convex hull approach as in \cite{refKsuresh2}. 

We consider the problem of designing stiff structures by minimising the mean compliance with linear constraints. The optimality criteria algorithm \citep{refOCRozvany,refYin} is found to be sufficient for this problem. In mean compliance minimization, the volume constraint acts as the source of conflict. We also add a minimum porosity limit in the form of a second constraint to get practically meaningful structures. The algorithm does not face convergence issues, and the scaling of density variables is convenient. 

We solve five test cases using our approach and present one case study. The optimized structures have admissible heterogeneity. There is a form of continuity, expected in size optimization, where some regions have thick beams and some have very thin beams or voids. Some aspects of the macrogeometry are understandable by intuition. In the case study, we compare the designs where the unit cell topology is restricted (second level of heterogeneity, uniform or homogeneous topology) and the designs with all of the phases allowed (heterogeneous topology). Here the optimality of heterogeneous designs is highlighted. We finally interpret the optimal structures using unit cell tests based on micropolar elasticity. 

The paper is organised as follows: Section 2 outlines the identification method of the phases and the design parameterization. Section 3 is devoted to the optimization problem formulation, followed by the sensitivity analysis and the algorithm used to numerically solve the problem. Section 4 is for the discussion of the results for the various test cases and the case-study. In Section 5, we try to interpret the designs by putting forth a hypothesis. The final section is the closure where we summarize the contribution of the paper and discuss extensions where the method is applicable. The appendix summarises the relations necessary for the frame FEA, derived from the first principles.  
\end{section}

\begin{section}{Design Parameterization}
In each of the cubic subdivision of the domain, we allow the microstructure topology to be constructed from a basic set of phase topologies. There are three core ideas in our design parametrization.
\begin{enumerate}
\item	Decomposing the densely connected ground structure topology into a finite number of groups.
\item	Assigning single density parameter to all segments of a group.
\item	Connecting the slenderness ratio of a beam segment to each density parameter.
\end{enumerate}
We define a connectivity scheme that yields a connectivity graph from a set of nodes as the input. In this scheme, at each node, we mark its connection with all of the nodes closest to it. Then all of the redundant connections are removed. We consider four types of nodes in a cube, the eight vertices (V), the six face centres (F), the twelve edge centres (E) and one body centre (B). Using the described scheme and feeding in the coordinates in the set of two node types, we can form a basis set of phases. There are eleven distinct combinations, namely, VV, EE, FF, VE, VF, VB, EF, EB, FB, BB and void. BB yields only a single point, each of the remaining are considered individually.

\paragraph{Vertex-Vertex and Vertex-Edge}: When we input only the vertex node coordinates, we get the simple cubic topology. The same topology emerges when both vertices and edge-centres are input. We merge these into a single phase, phase 1 (Figure \ref{Ph1}).

\paragraph{Face-Vertex and Face-Face}: With face centre and vertex coordinates as the input, the figure is a combination of the stellated tetrahedron and the octahedron. When the face centres are the sole input, the octahedron unit cell comes out. Thus, we form two distinct unit cells, stellated tetrahedron and the octahedron , respectively called phase 2 (Figure \ref{Ph2}) and phase 7 (Figure \ref{Ph7}). 

\paragraph{Body-Vertex and Body-Edge and Body-Face}: There are three distinct topologies formed here, connecting body centre to each of the node types. These are numbered: vertex, phase 3 (Figure \ref{Ph3}), edge centre, phase 6 (Figure \ref{Ph6}), and face centre, phase 8 (Figure \ref{Ph8}).

\paragraph{Edge-Edge}: With only edge centre coordinates as the input, the cuboctahedron unit cell is formed. We call this phase 4 (Figure \ref{Ph4}). 

\paragraph{Face-Edge}: With face centre and edge centre coordinates as the input, the segments are arranged in a plus (+) sign on all of the six faces of the cube. We call this phase 5 (Figure \ref{Ph5}).

Next, we assign a density parameter to each of the phases such that all of the segments grouped in a single phase have equal cross-section area, proportional to $\rho$. To each of the fundamental phases we attach $l$, the length of the segment in that group. For each cubic subdivision there are eight (in our case) phases and density parameters. When all of the eight phases are present, the exhaustive connectivity emerges (Figure \ref{PhAll}). With this design parameterization, we have: 

\begin{equation}
    \rho =  \begin{bmatrix}
    \rho_{11} & \rho_{12} & \dots & \dots & \rho_{1n} \\
    \rho_{21} & \ddots & & & \rho_{2n} \\
    \vdots & & \ddots & & \\
    \vdots & & & \ddots & \\
    \rho_{m1} & \rho_{m2} & \dots & \dots & \rho_{mn} \\
    \end{bmatrix}
\end{equation}
  
\begin{equation} 
    l =  \begin{bmatrix}
    l_{11} & l_{12} & \dots & \dots & l_{1n} \\
    l_{21} & \ddots & & & l_{2n} \\
    \vdots & & \ddots & & \\
    \vdots & & & \ddots & \\
    l_{m1} & l_{m2} & \dots & \dots & l_{mn} \\
    \end{bmatrix}
\end{equation}
Here, $m$ is the number of phases, which is eight in our case. $n$ is the number of cubic subdivisions in the design domain. 

In each phase, the types of the two nodes at the end of each segment is unique. Consequently, no beam segment is shared among different phases of a single cube. They intersect and we add nodes at those places.  The basic set of phases chosen by us is a representative set used to demonstrate the approach. There can be another set that is optimal for certain boundary conditions and objective functions. The phases can also be asymmetric, directional, be designed by a concurrent approach, or come from manufacturing considerations. 

Next, we define the proportionality constant for connecting the density parameters to the area of cross-section of individual beam segments. The optimization constraints need to be inspired from the model’s limitations on the slenderness ratio. Appropriate scaling of the density parameter will provide us a direct handle on the ratio, which simplifies the constraints. When $\rho$ is unity, we have chosen for the slenderness ratio to be 3. The area of circular cross-section of a beam segment is given by, 

\begin{equation}
    \vb*{A_{k}} = \frac{\pi l_{ij}^2}{36} \rho_{ij}
\end{equation}

When we consider all of the cubic subdivisions to be populated using the same phases, the adjacent cubic subdivisions might share some beam segments. To address this, we define a redundancy index $n_{k}$. The shared segments are either present on the cube’s edges or faces.  The beams on the edges will have $n_{k}=4$ and the ones on faces will have $n_{k}=2$. The segments completely inside the cube will have $n_{k}=1$. On the boundary of the design domain, these indices are again set to 1. To get the area of cross-section of a beam segment, we follow the following equation.

\begin{figure}[t]
    \centering
    \includegraphics[width=0.4\textwidth]{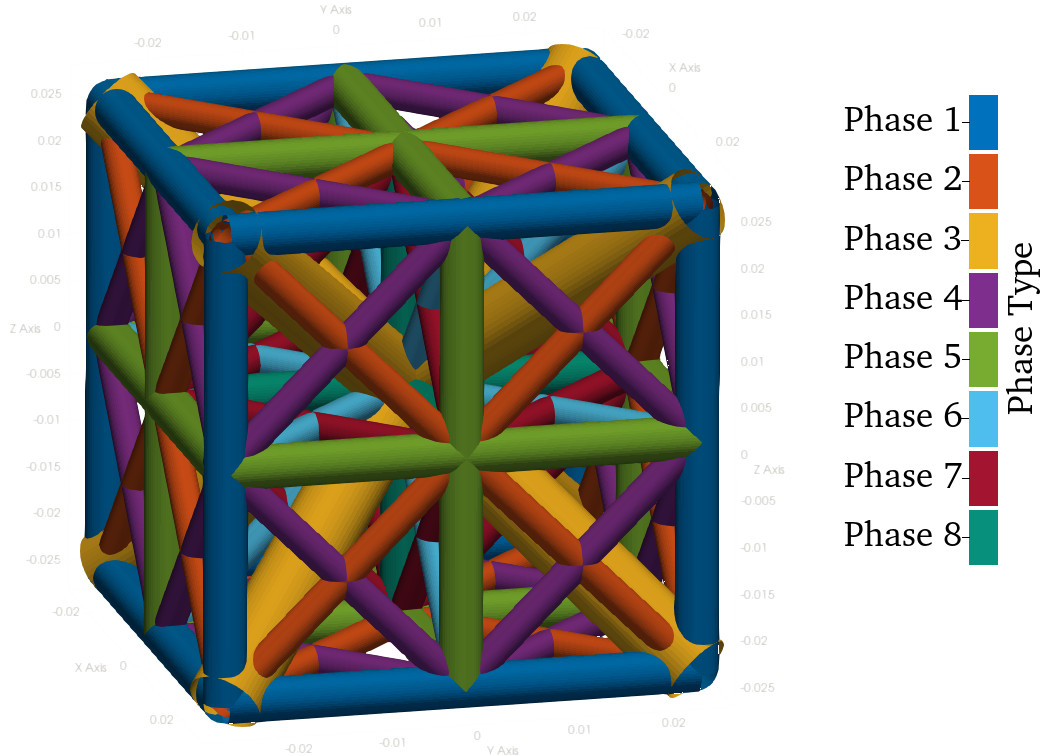}
    \caption{A combination of all of the phases}
    \label{PhAll}
\end{figure}

\begin{equation}
\vb*{A_{k}} = \sum_{j = \ n_{k} \ values}^{} \frac{\pi l_{ij}^2}{36 n_{k}} \rho_{ij}
\end{equation}

To summarize, we define fundamental phases as topologies and then let the microstructure topology be a combination of them, also letting the size to vary spatially. This parametrization yields a design space within the third stage of heterogeneity including spatial variation of phase topologies and sizes.
\begin{figure*}[t]
\centering
\begin{subfigure}[b]{0.45\textwidth}
\centering
\includegraphics[width=\textwidth]{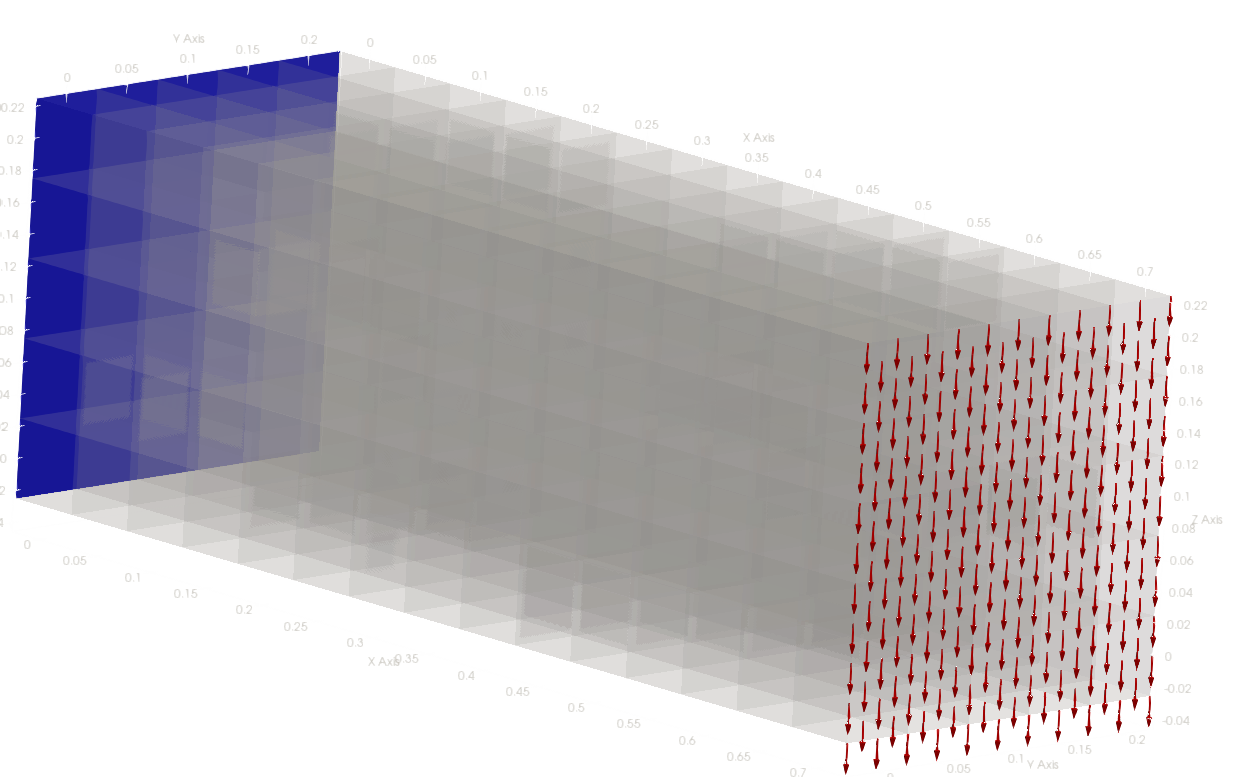}
\caption{Boundary condition}
\label{CantBC}
\end{subfigure}
\begin{subfigure}[b]{0.45\textwidth}
\centering
\includegraphics[width=\textwidth]{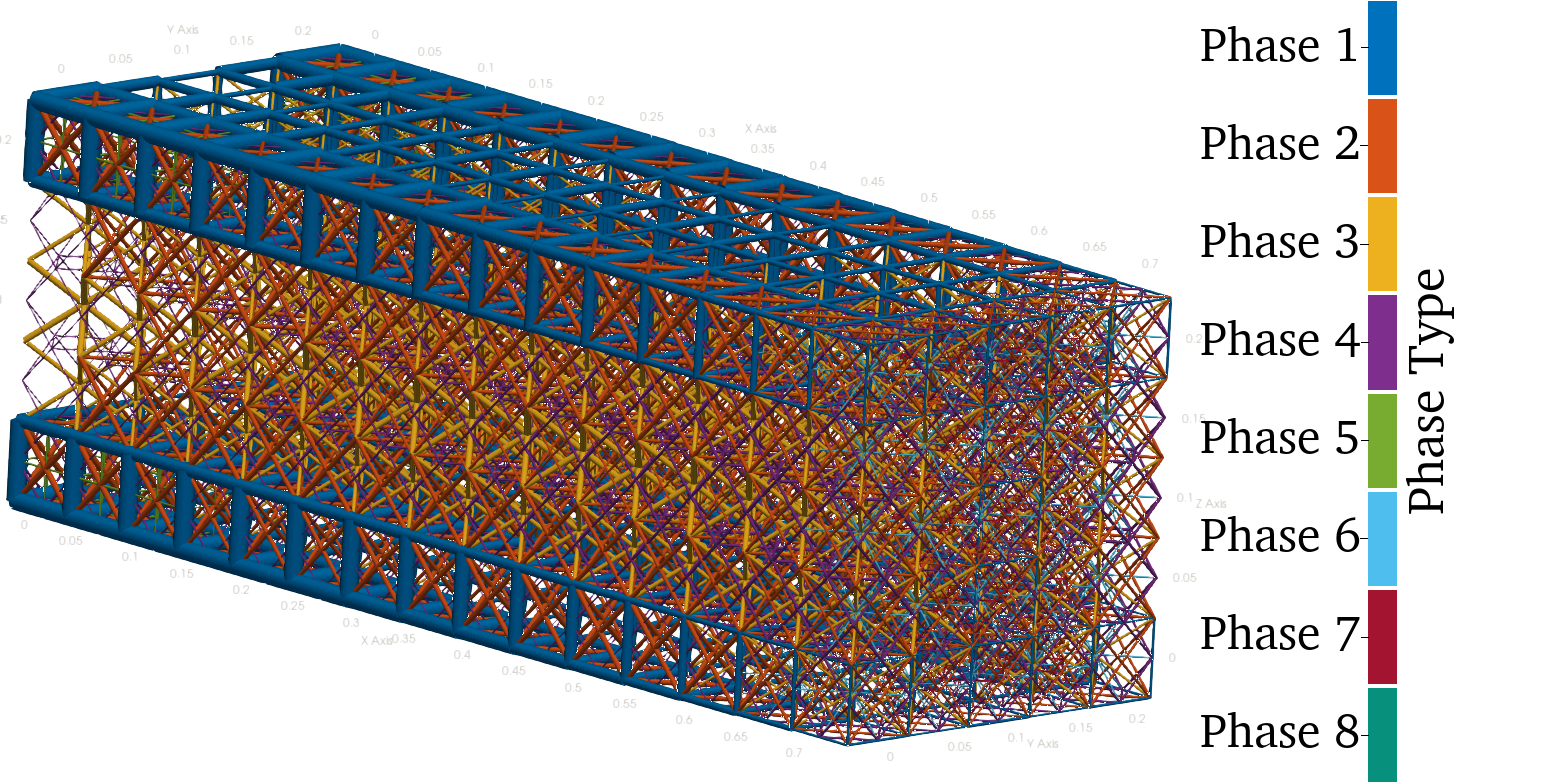}
\caption{Optimized design}
\label{CantSol}
\end{subfigure}
\caption{Cantilever case boundary condition and result}
\end{figure*}

\begin{figure*}
\centering
\begin{subfigure}[b]{0.45\textwidth}
\centering
\includegraphics[width=\textwidth]{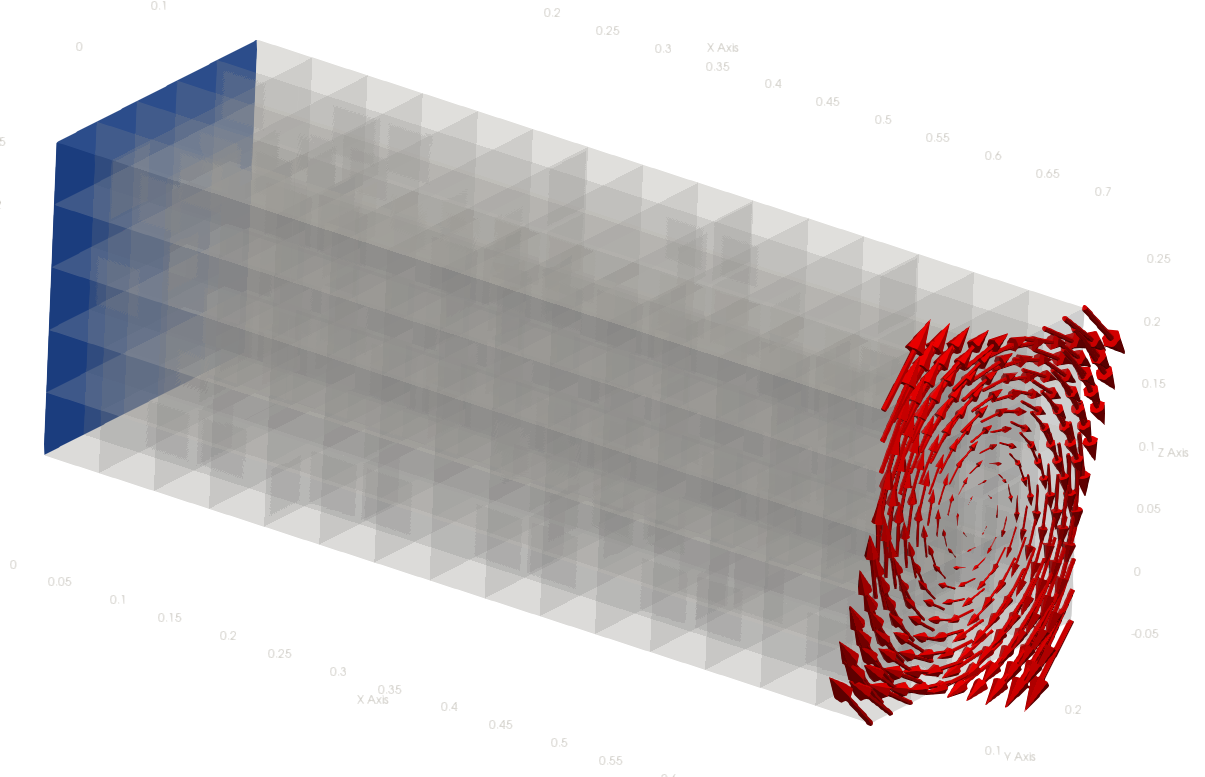}
\caption{Boundary condition}
\label{TorBC}
\end{subfigure}
\begin{subfigure}[b]{0.45\textwidth}
\centering
\includegraphics[width=\textwidth]{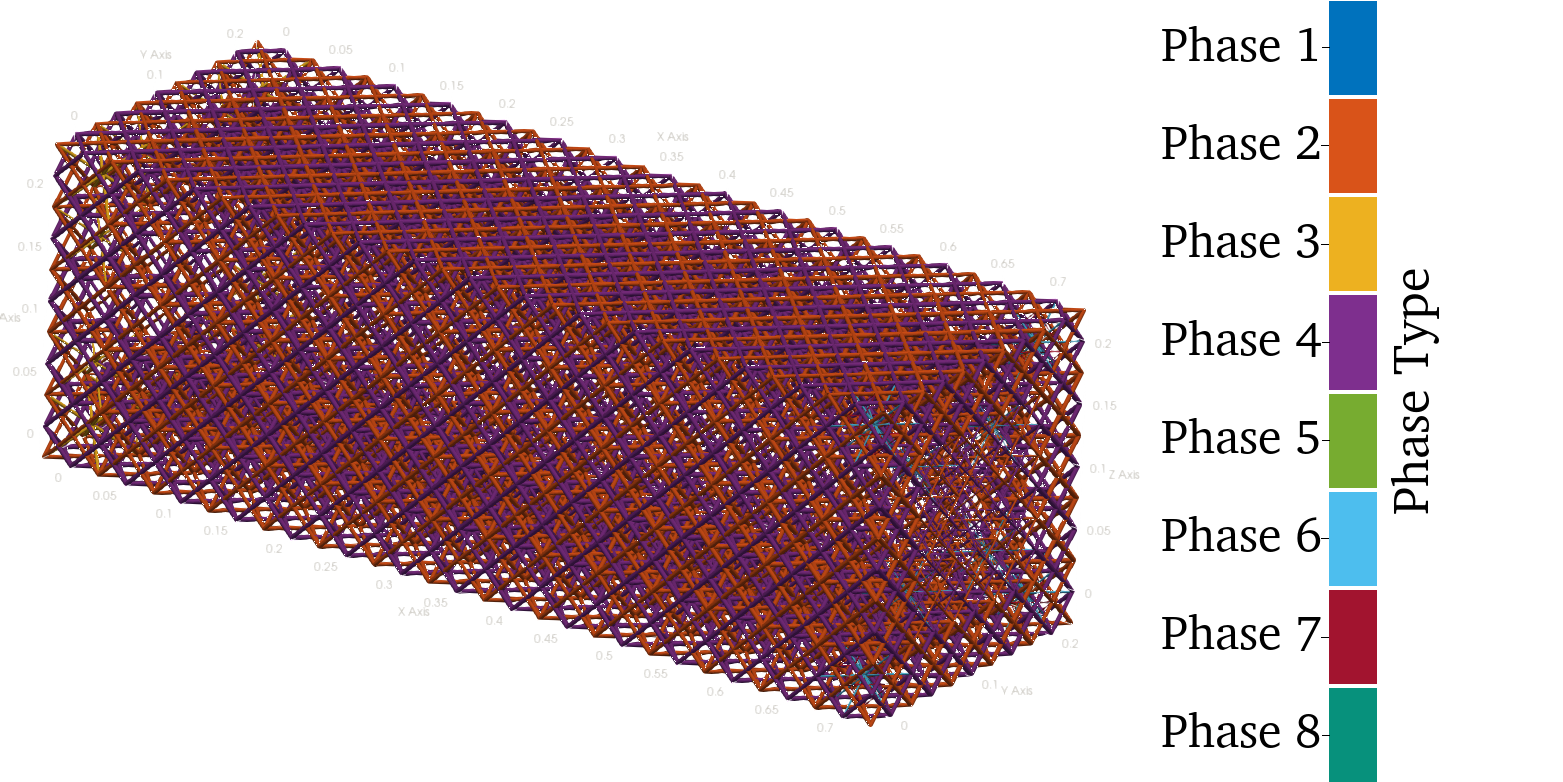}
\caption{Optimized design}
\label{TorSol}
\end{subfigure}
\caption{Torsion case boundary condition and result}
\end{figure*}
\end{section}

\begin{section}{Problem Formulation, Sensitivity Analysis, and Numerical Algorithm}
\begin{subsection}{Problem Formulation}
The optimization problem is framed with the minimization of the mean compliance as the objective. 

\begin{equation}
\underset{\genfrac{}{}{0pt}{2}{\rho_{ij}}{\vb*{u}}}{\mathrm{min}} \ \ \vb*{f}^{T}\vb*{u} \nonumber
\end{equation}
\begin{equation}
\mathrm{Subject \ To:} \nonumber
\end{equation}
\begin{equation} \label{EqGov}
\lambda \ : \ \vb{K}(\vb*{A}(\rho_{ij}))\vb*{u}-\vb*{f}=0
\end{equation}
\begin{equation} \label{EqVol}
\Lambda \ : \ \sum_{k \ = \ 1}^{P} \vb*{A_k}(\rho_{ij})\vb*{L_k}-\mu V^{*}=0
\end{equation}
\begin{equation} \label{EqPor}
\phi_j \ : \ \sum_{i \ = \ 1}^{m} \frac{\sum_{k}^{}\frac{\pi \vb*{L_k}}{36 n_k} l_{ij}^2 \rho_{ij}}{V_j} - \kappa \le 0    
\end{equation}
\begin{equation}
\ \ \ \ \varepsilon \le \rho_{ij} \le \rho_{\mathrm{max}} \nonumber
\end{equation}
\begin{equation}
\mathrm{Data:} \nonumber
\end{equation}
\begin{equation}
\vb*{f}, \ V^{*}, \ l_{ij}, \ \vb*{L_{k}}, \ n_{k}, \ V_{j}, \ \vb{K}, \ \mu , \ \varepsilon, \ \rho_{\mathrm{max}}, \kappa \nonumber    
\end{equation}

The constraint corresponding to the Lagrange multiplier $\lambda$ is for the governing equation, $\vb{K}$ being the global stiffness matrix and $\vb*{u}$ the displacement. The constraint corresponding to the Lagrange multiplier $\Lambda$ is the global volume constraint, $\vb*{L_{k}}$ the length of the beam segment k, $V^{*}$ is the volume of the design domain, and $\mu$ being the volume ratio. The constraint corresponding to the Lagrange multiplier $\phi$ is for maintaining a minimum porosity in the structure, defined by $\kappa$ that is same for all cubic subdivisions. $V_{j}$ is the volume of cubic subdivision j. $\varepsilon$ and $\rho_{\mathrm{max}}$ are the upper and lower bounds on the density parameter. $\varepsilon$ is uniformly taken as $0.0001$ while optimization and $0.01$ while visualization.
\end{subsection}
\begin{subsection}{Sensitivity Analysis}

\begin{figure*}
\centering
\begin{subfigure}[b]{0.3\textwidth}
\centering
\includegraphics[width=\textwidth]{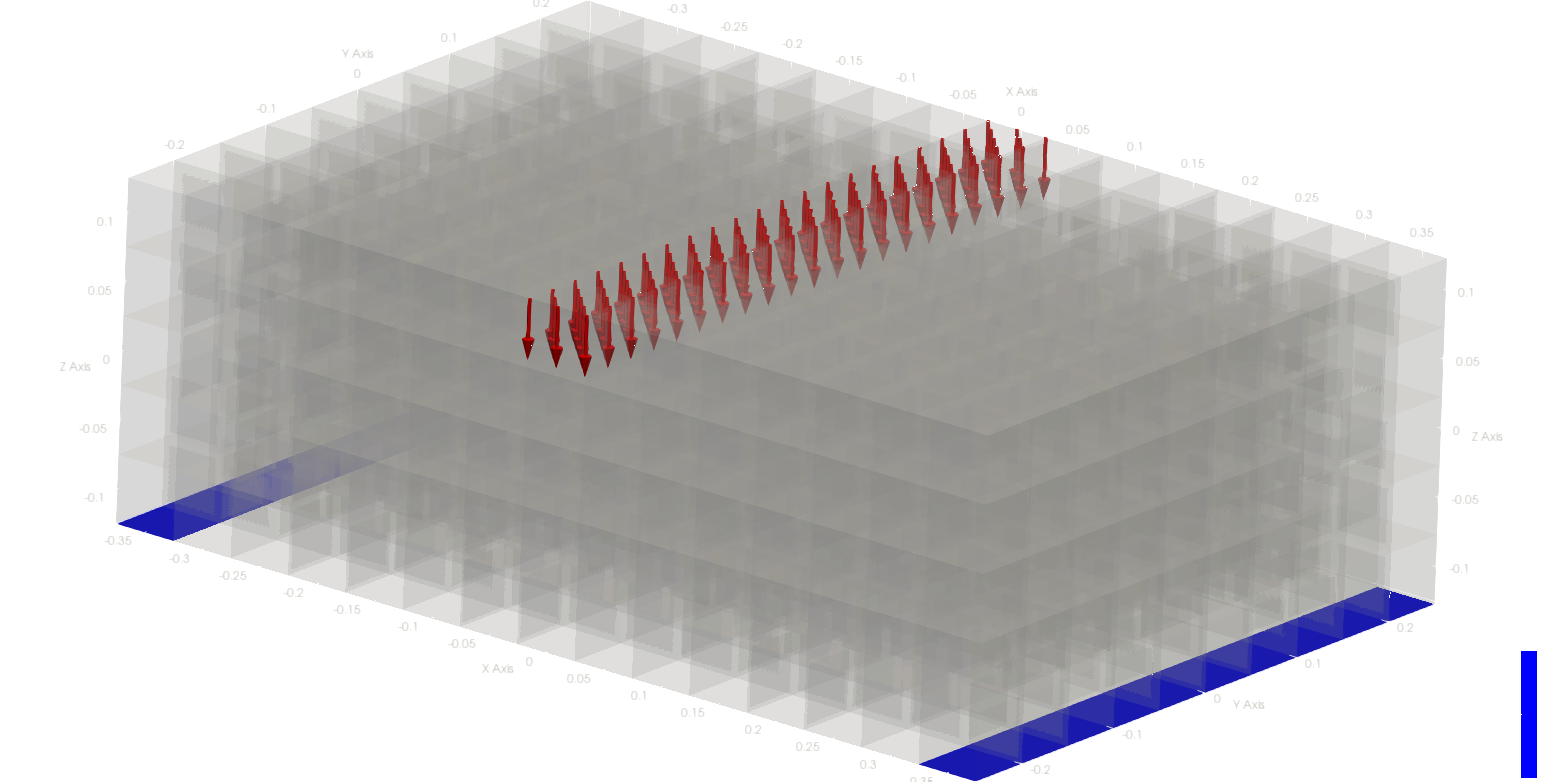}
\caption{Boundary condition: MBB}
\label{MBBBC}
\end{subfigure}
\begin{subfigure}[b]{0.3\textwidth}
\centering
\includegraphics[width=\textwidth]{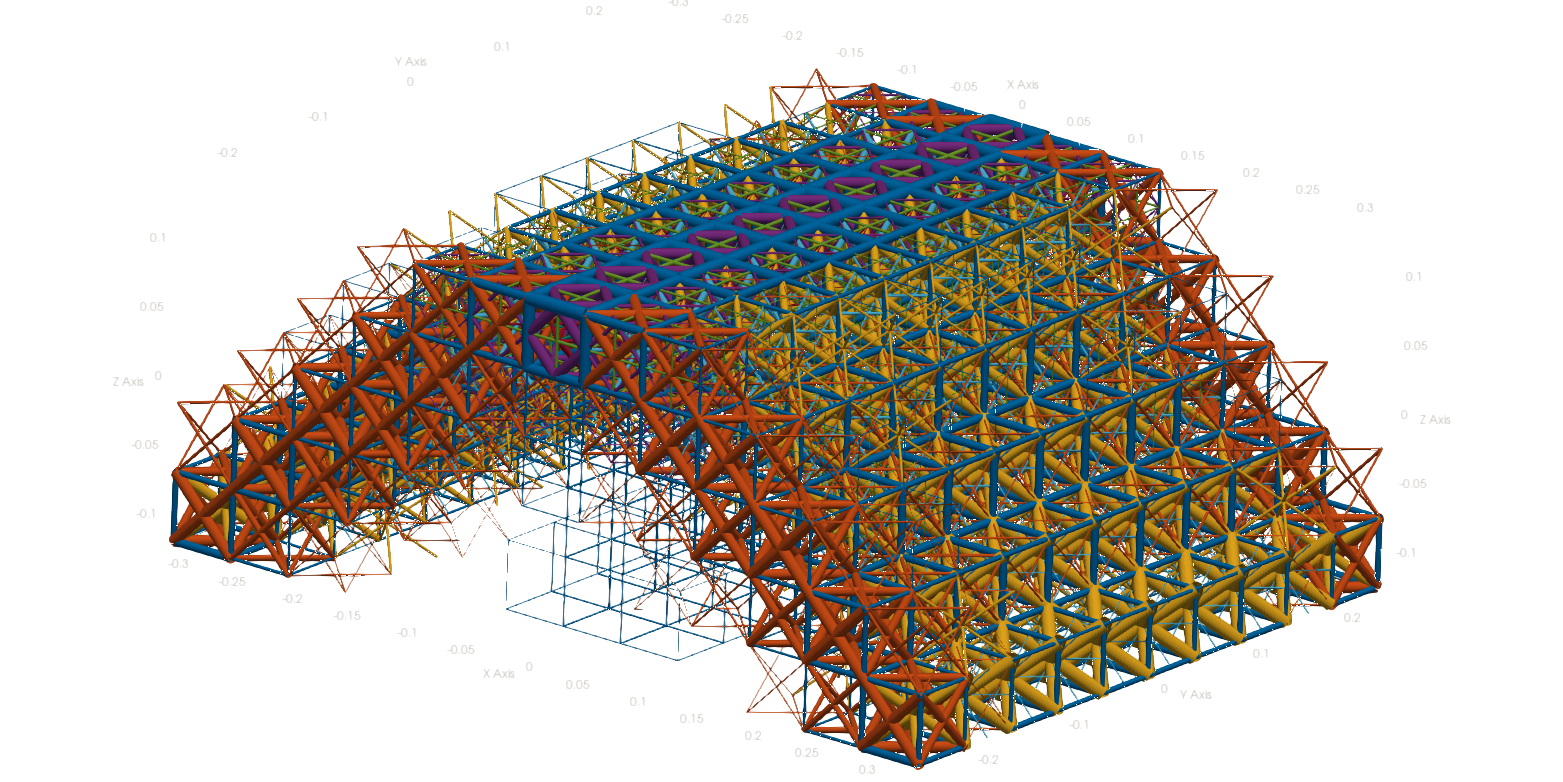}
\caption{Optimized design: MBB}
\label{MBBSol1}
\end{subfigure}
\begin{subfigure}[b]{0.3\textwidth}
\centering
\includegraphics[width=\textwidth]{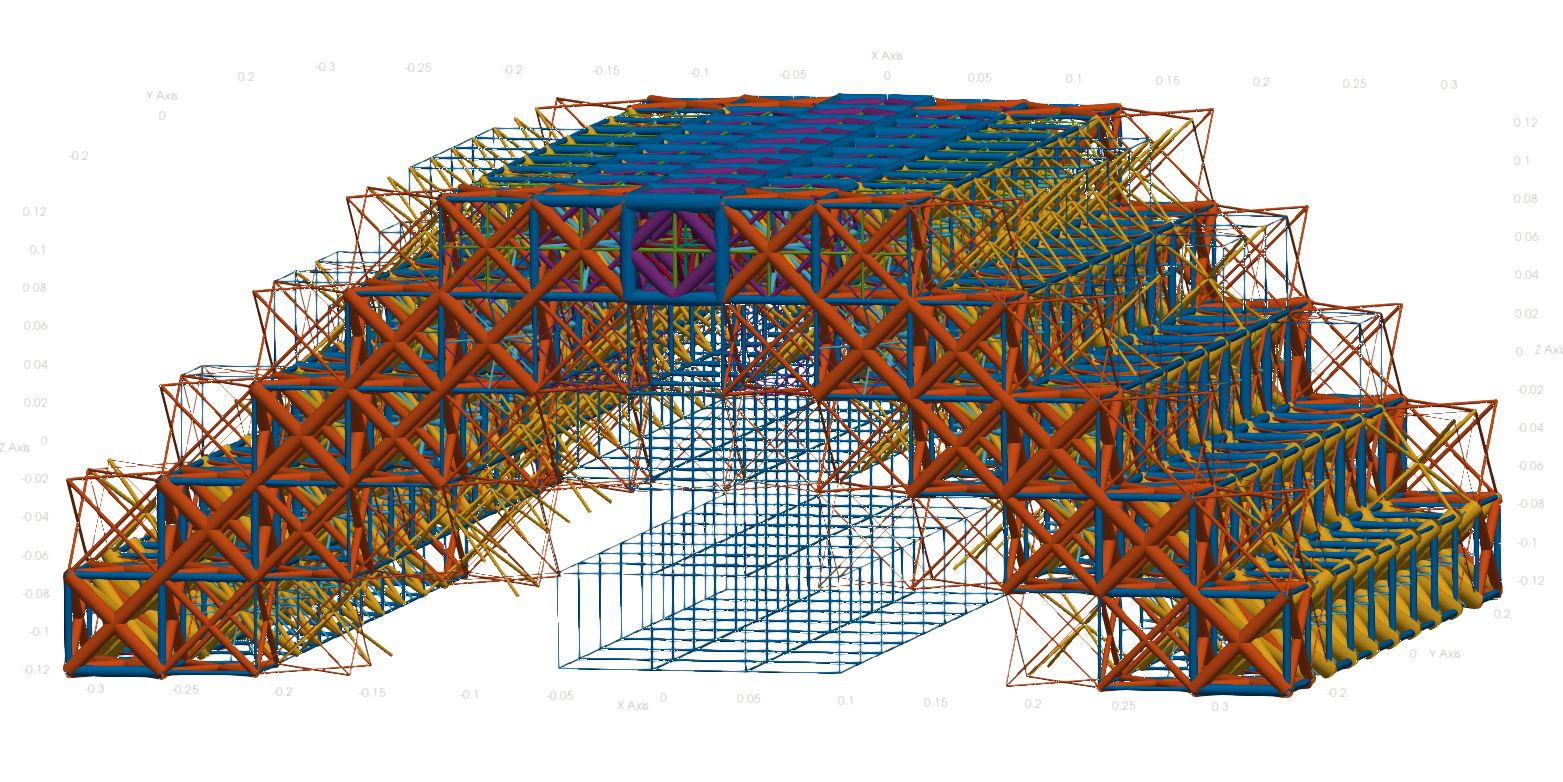}
\caption{Optimized design: MBB}
\label{MBBSol2}
\end{subfigure}
\begin{subfigure}[b]{0.3\textwidth}
\centering
\includegraphics[width=\textwidth]{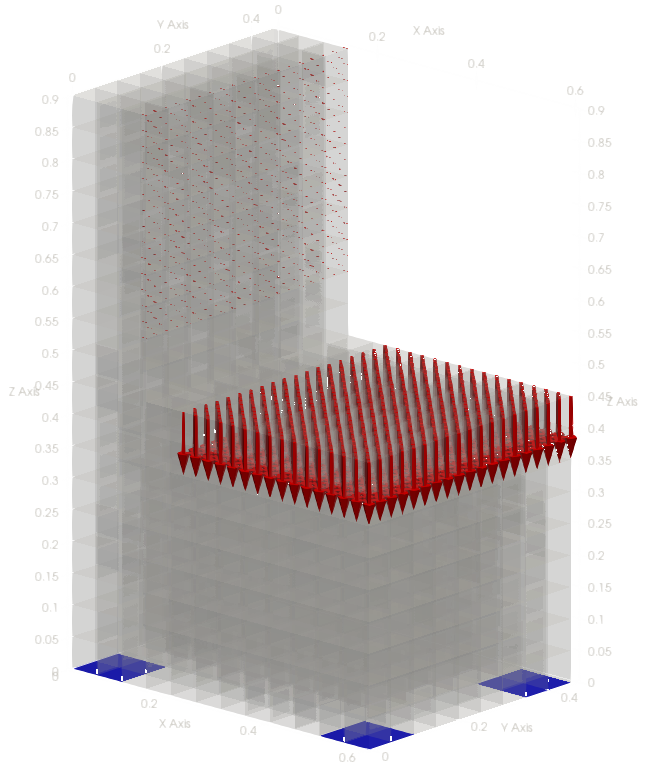}
\caption{Boundary condition: Chair}
\label{ChairBC}
\end{subfigure}
\begin{subfigure}[b]{0.3\textwidth}
\centering
\includegraphics[width=\textwidth]{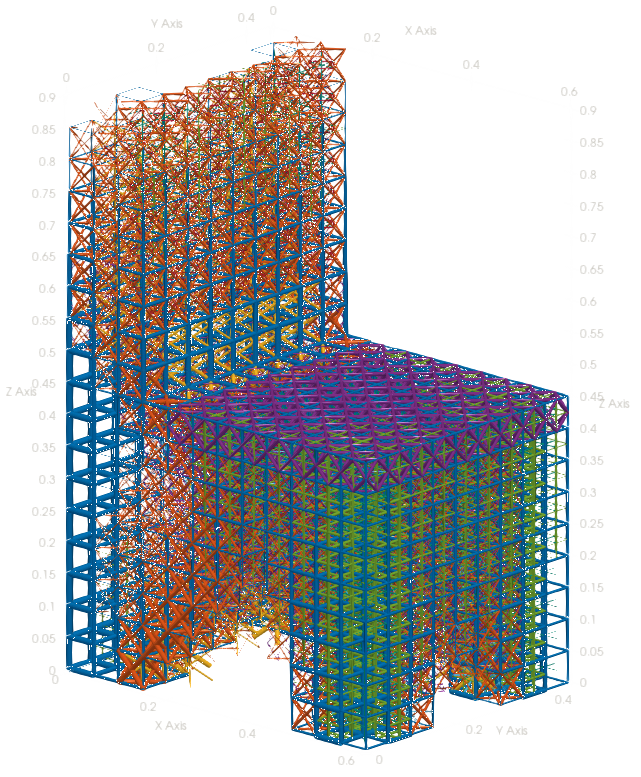}
\caption{Optimized design: Chair}
\label{ChairSol1}
\end{subfigure}
\begin{subfigure}[b]{0.3\textwidth}
\centering
\includegraphics[width=\textwidth]{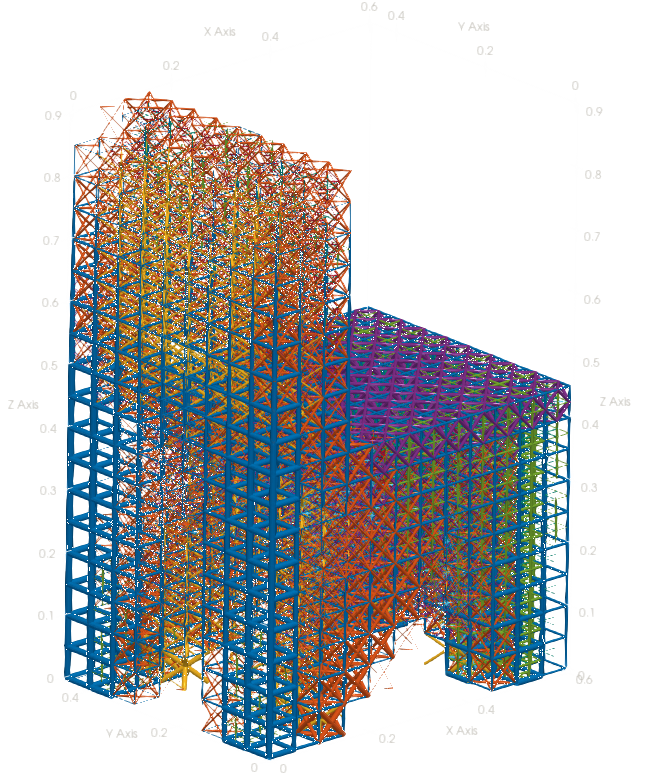}
\caption{Optimized design: Chair}
\label{ChairSol2}
\end{subfigure}
\caption{Results for MBB and Chair cases}
\end{figure*}

The Karush-Kuhn-Tucker (KKT) conditions are derived and the sensitivities are calculated using the adjoint method. We first find the sensitivity of the governing equation with respect to the density parameters. For that, we differentiate the potential energy w.r.t. the area of cross-section to get  

\begin{equation}
\begin{split}
\frac{\partial \Pi}{\partial \vb*{A_k}} & =\int_{0}^{L_e} \bigg[\frac{E}{2} \bigg({{u_{x}}^\prime} ^2 + \frac{A}{2\pi} {{\theta_z}^\prime} ^2 + \frac{A}{2\pi} {{\theta_y}^ \prime} ^2\bigg) \\ & + \frac{K_e G}{2}\bigg({{u_{y}}^\prime} ^2+{\theta_z}^2-2\theta_z {u_y}^\prime\bigg) \\ & + \frac{K_e G}{2}\bigg({{u_{z}}^\prime} ^2+{\theta_y}^2+2\theta_y {u_z}^\prime\bigg) + \frac{GA}{2\pi}{{\theta_x}^\prime}^2\bigg]dl
\end{split}
\end{equation}

Now we apply the Ritz method to find the elemental sensitivity to area of cross-section in the form of a matrix $\frac{\partial \vb{K_{k}}}{\partial \vb*{A_k}}$. The expression is present in the appendix with the matrix values in the terms of the beam parameters. The elemental sensitivity matrix is to be appropriately rotated. Using the chain rule, we perform a summation of the sensitivity of all beam segments present in a phase to finally find the sensitivity.

\begin{equation}
\frac{\partial (R^T \vb{K} R)}{\partial \rho_{ij}} = \sum_{k \ = \ \mathrm{all \ beams \ in \ phase}}^{} R^T \frac{\partial \vb{K_{k}}}{\partial \vb*{A_k}} R \frac{\partial \vb*{A_k}}{\partial \rho_{ij}}     
\end{equation}
\begin{equation}
\frac{\partial \vb*{A_k}}{\partial \rho_{ij}}=\frac{\pi l_{ij}^2}{36n_k}    
\end{equation}
\end{subsection}

The rest of the sensitivities are also calculated following a similar procedure.

\begin{subsection}{Numerical Algorithm}
We use the optimality criteria (OC) algorithm that is tailored for linear density-based structural optimization problems \citep{refoncompl}.

\begin{figure*}
\centering
\begin{subfigure}[b]{0.23\textwidth}
\centering
\includegraphics[width=\textwidth]{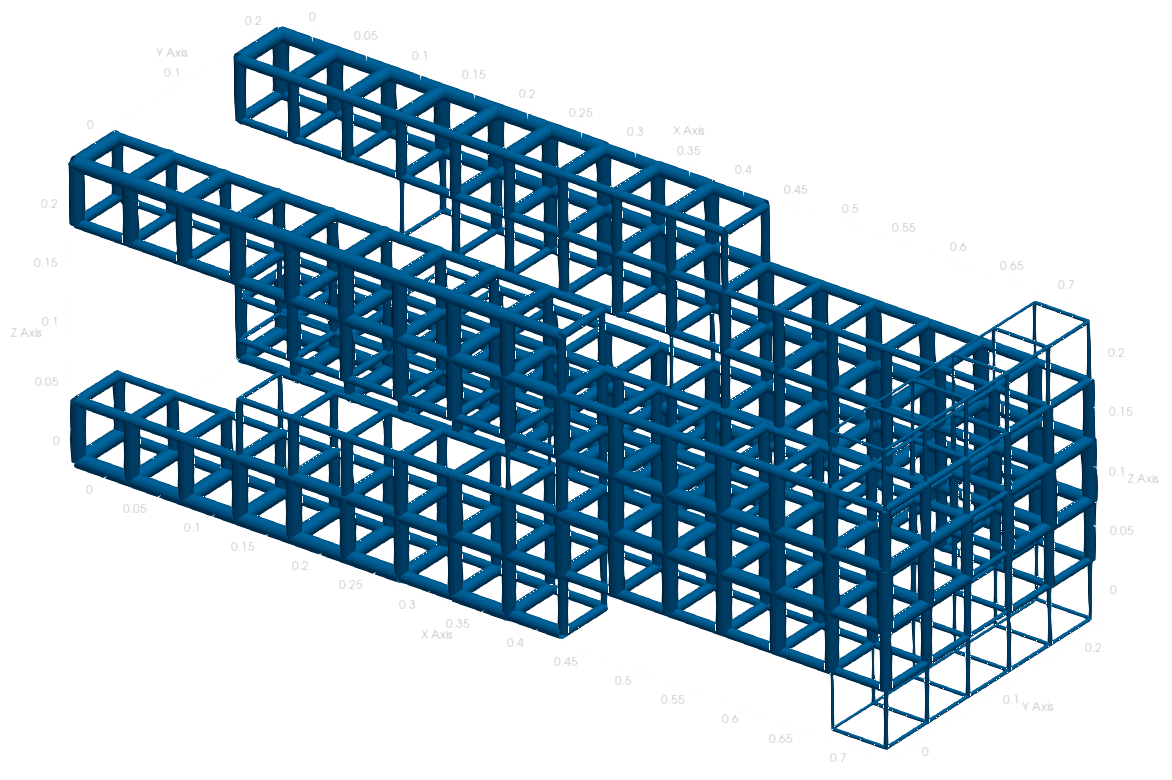}
\caption{Phase 1}
\label{CantPh1}
\end{subfigure}
\begin{subfigure}[b]{0.23\textwidth}
\centering
\includegraphics[width=\textwidth]{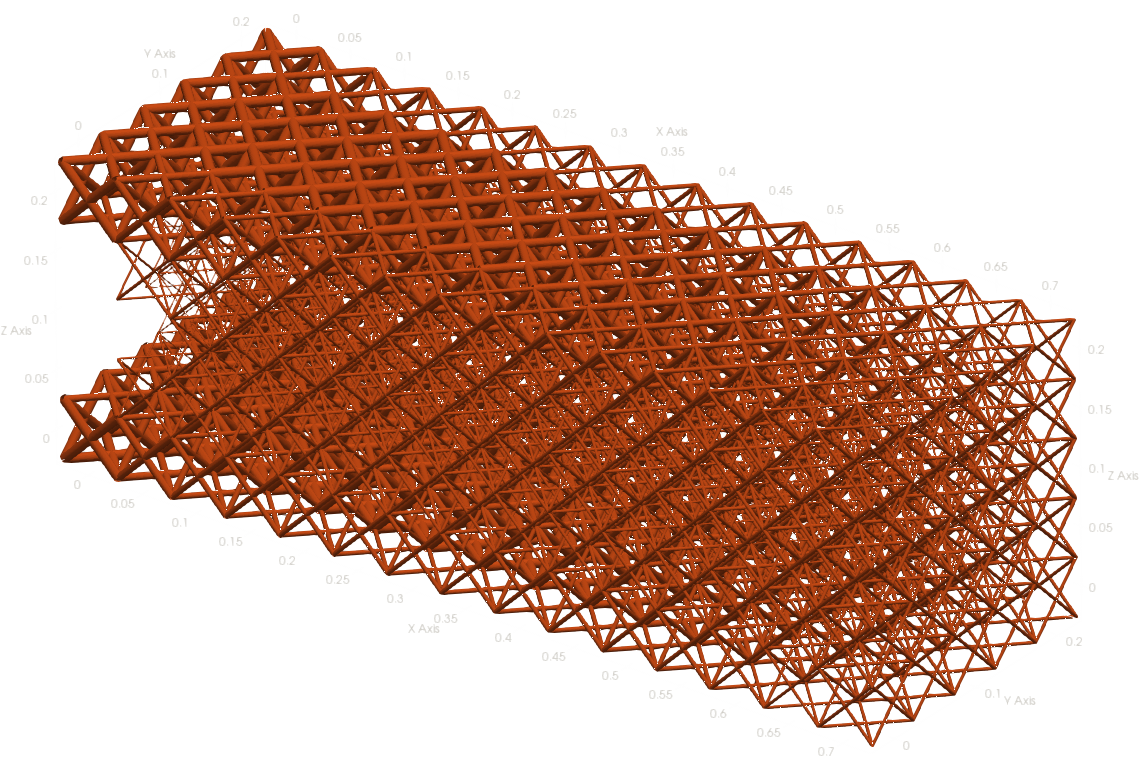}
\caption{Phase 2}
\label{CantPh2}
\end{subfigure}
\begin{subfigure}[b]{0.23\textwidth}
\centering
\includegraphics[width=\textwidth]{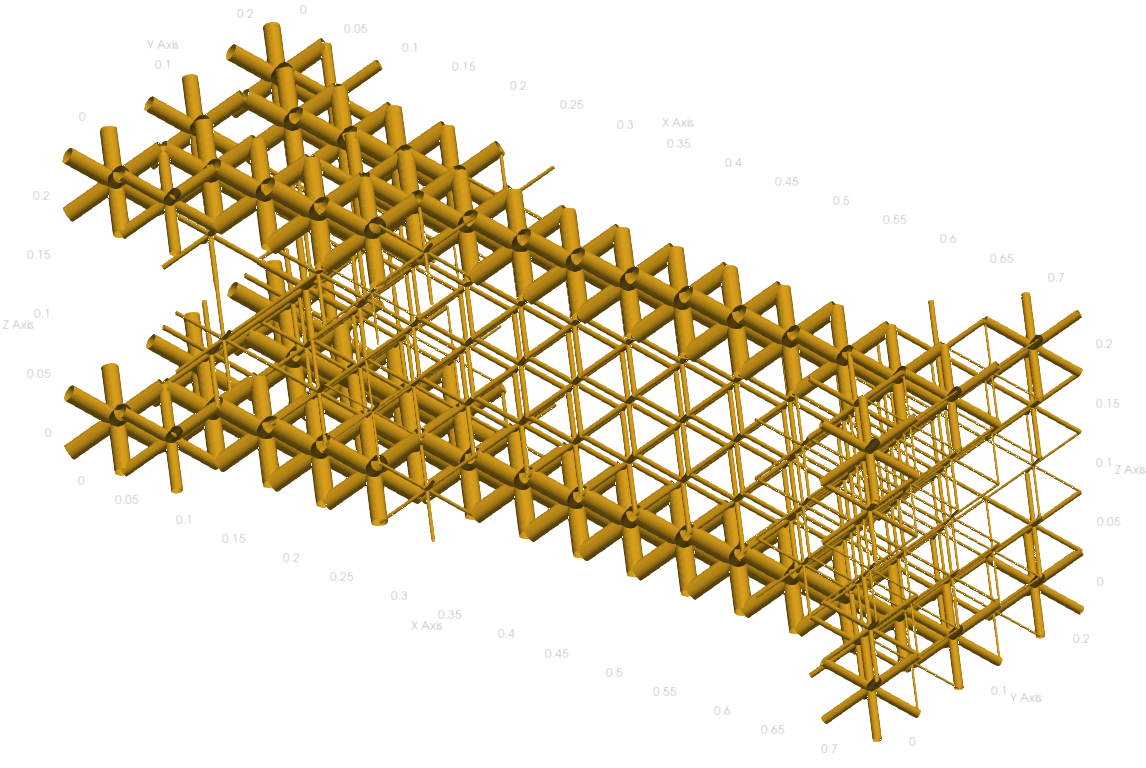}
\caption{Phase 3}
\label{CantPh3}
\end{subfigure}
\begin{subfigure}[b]{0.23\textwidth}
\centering
\includegraphics[width=\textwidth]{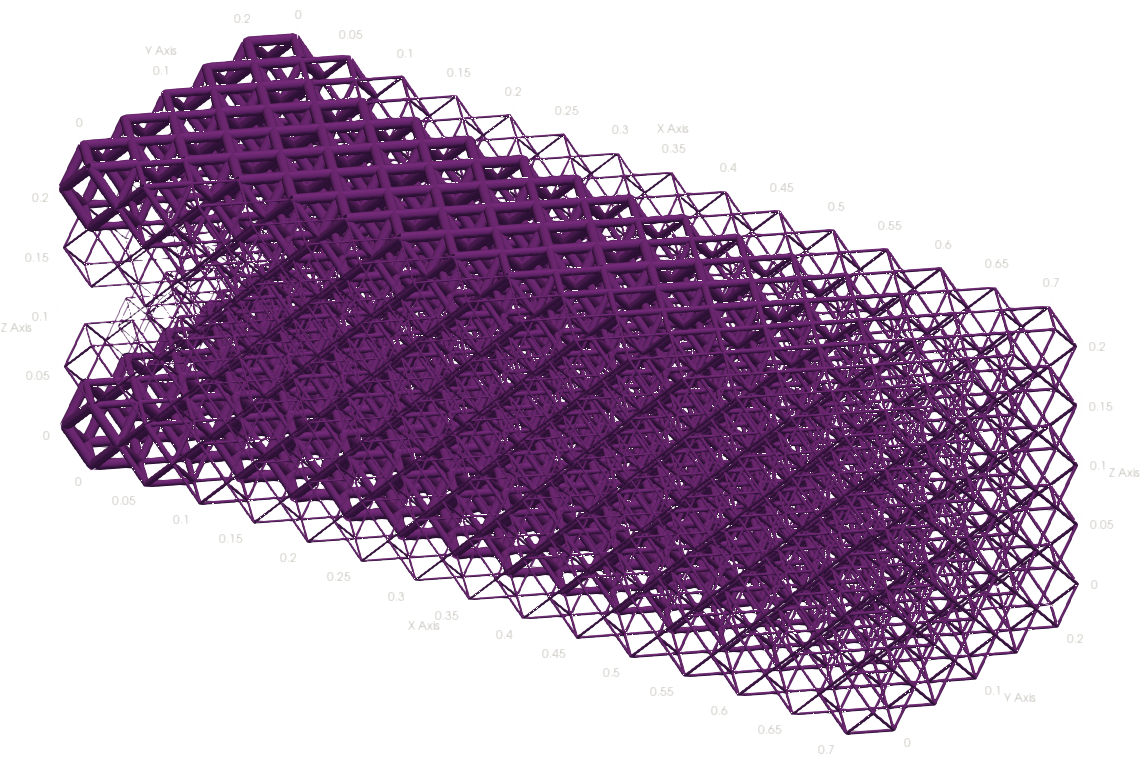}
\caption{Phase 4}
\label{CantPh4}
\end{subfigure}
\begin{subfigure}[b]{0.23\textwidth}
\centering
\includegraphics[width=\textwidth]{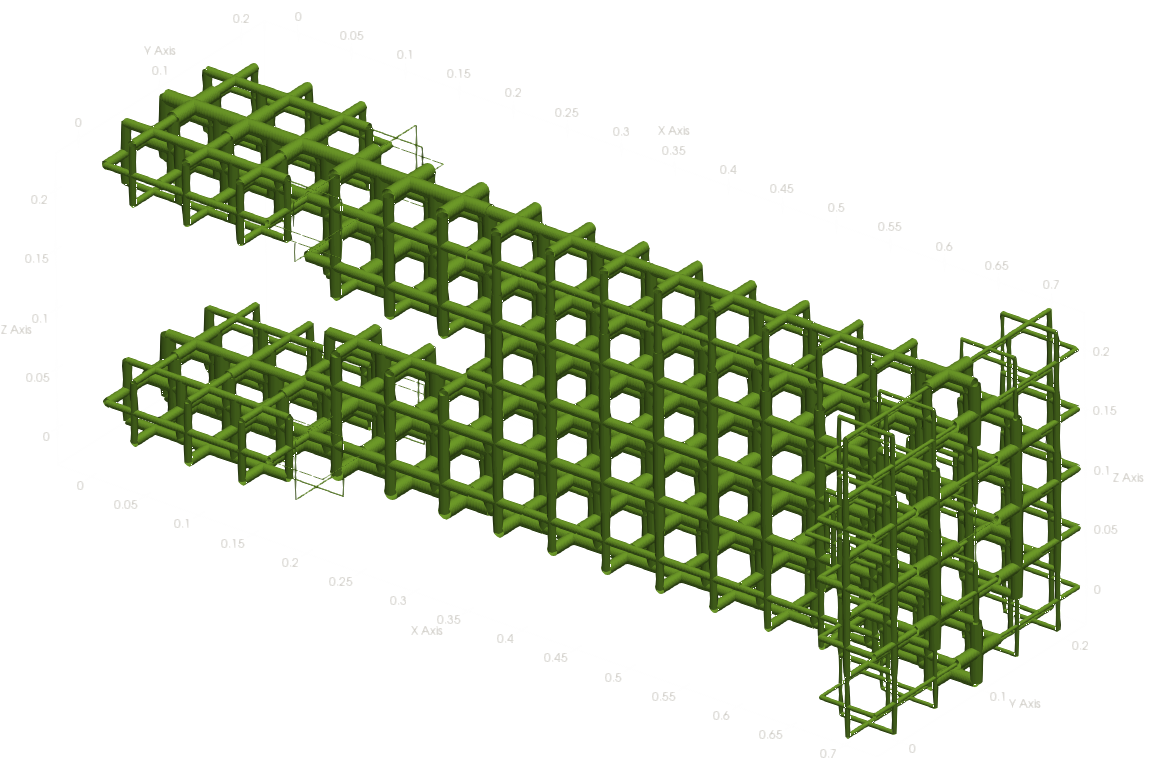}
\caption{Phase 5}
\label{CantPh5}
\end{subfigure}
\begin{subfigure}[b]{0.23\textwidth}
\centering
\includegraphics[width=\textwidth]{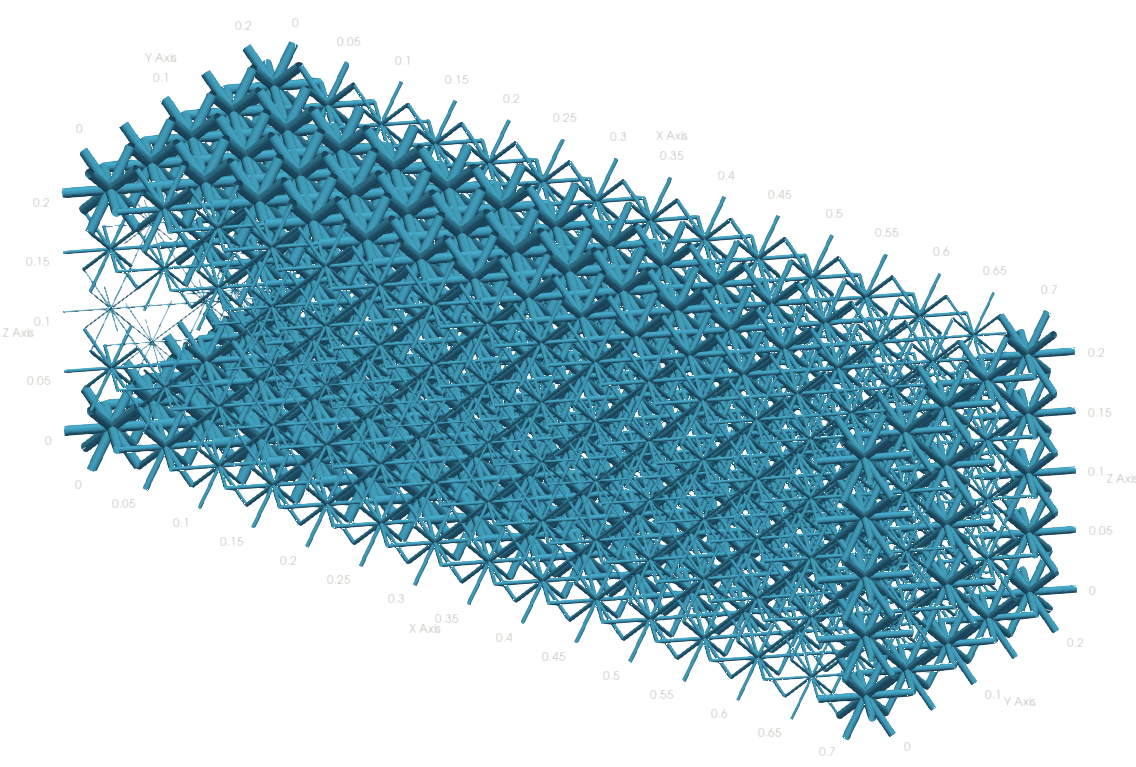}
\caption{Phase 6}
\label{CantPh6}
\end{subfigure}
\begin{subfigure}[b]{0.23\textwidth}
\centering
\includegraphics[width=\textwidth]{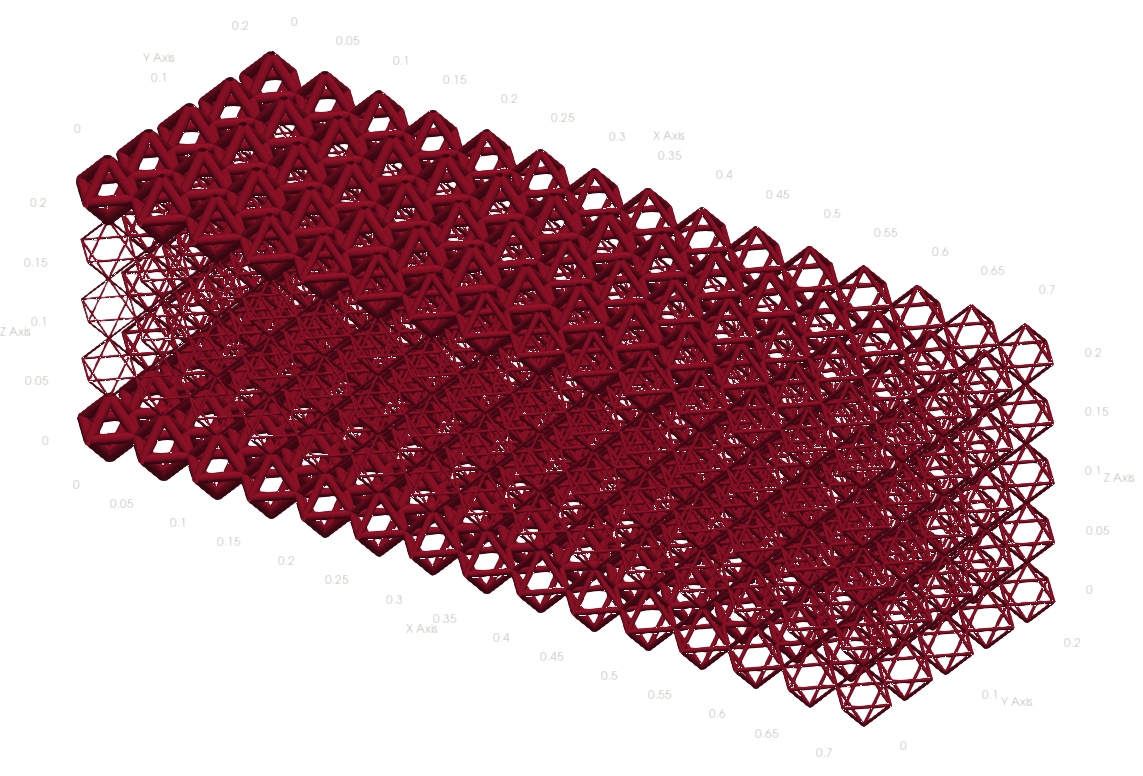}
\caption{Phase 7}
\label{CantPh7}
\end{subfigure}
\begin{subfigure}[b]{0.23\textwidth}
\centering
\includegraphics[width=\textwidth]{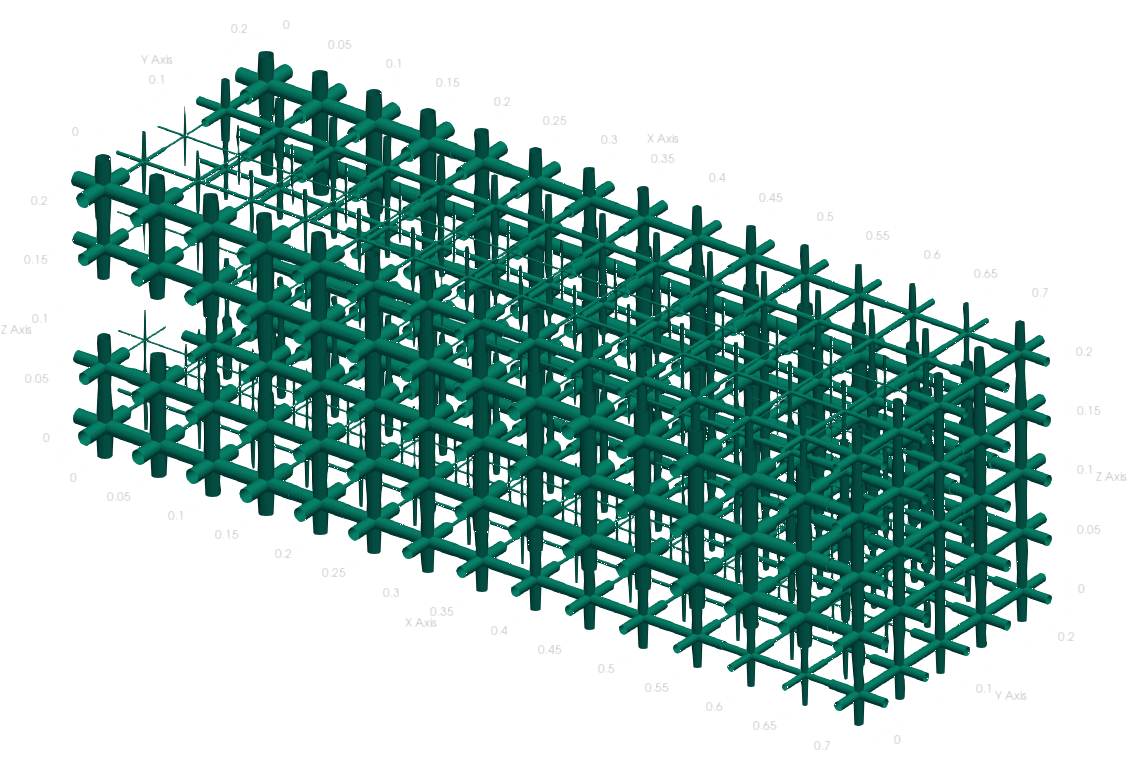}
\caption{Phase 8}
\label{CantPh8}
\end{subfigure}
\begin{subfigure}[b]{0.6\textwidth}
        \centering
        \includegraphics[width=\textwidth]{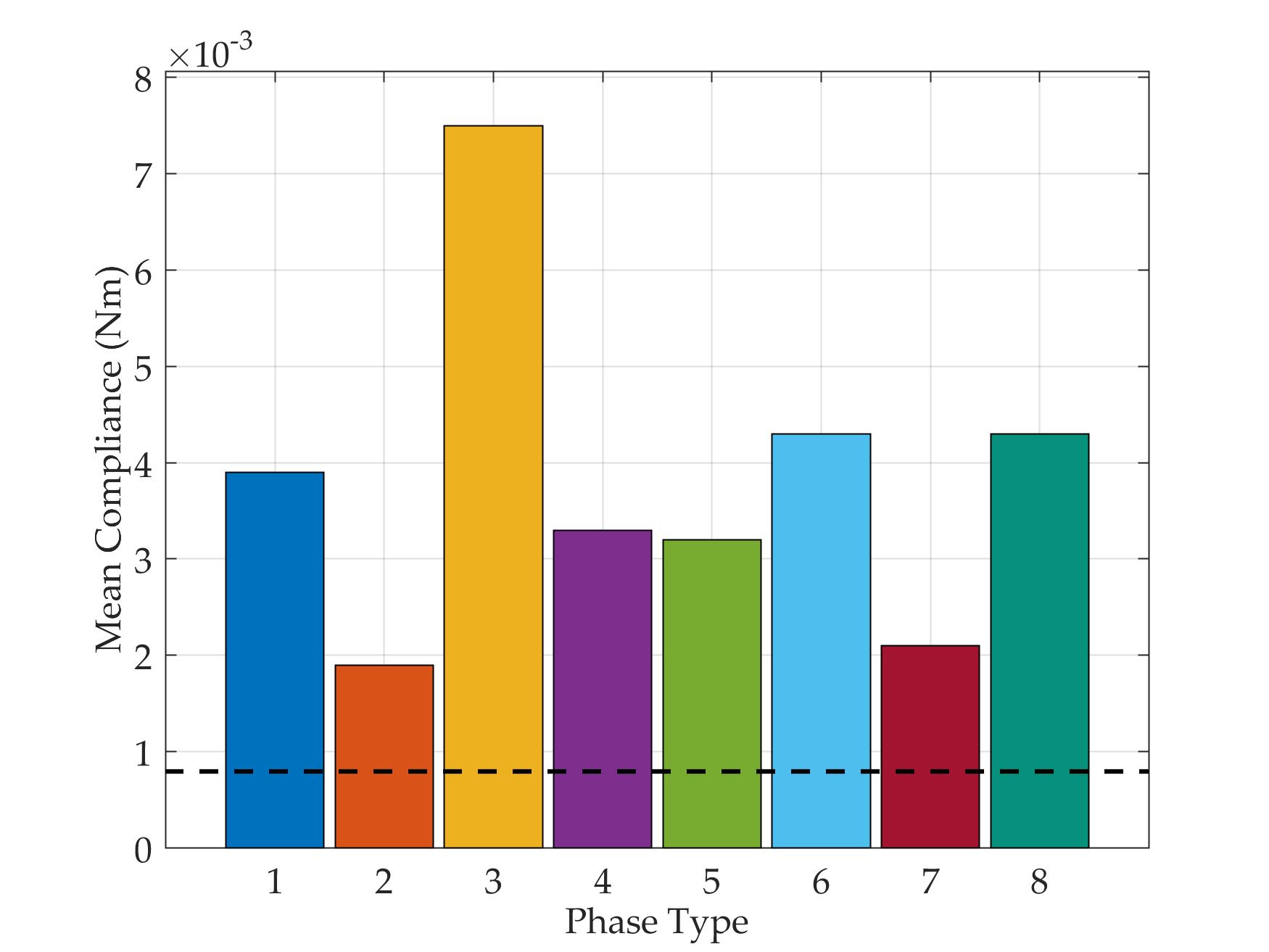}
        \caption{Comparison of stiffness in homogeneous designs (bars in color) and heterogeneous design (black dashed line, Figure \ref{CantSol})}
        \label{CanGraph}
    \end{subfigure}
\caption{Case-Study: Designs for Cantilever case}
\label{CantCS}
\end{figure*}

\begin{figure*}
\centering
\begin{subfigure}[b]{0.23\textwidth}
\centering
\includegraphics[width=\textwidth]{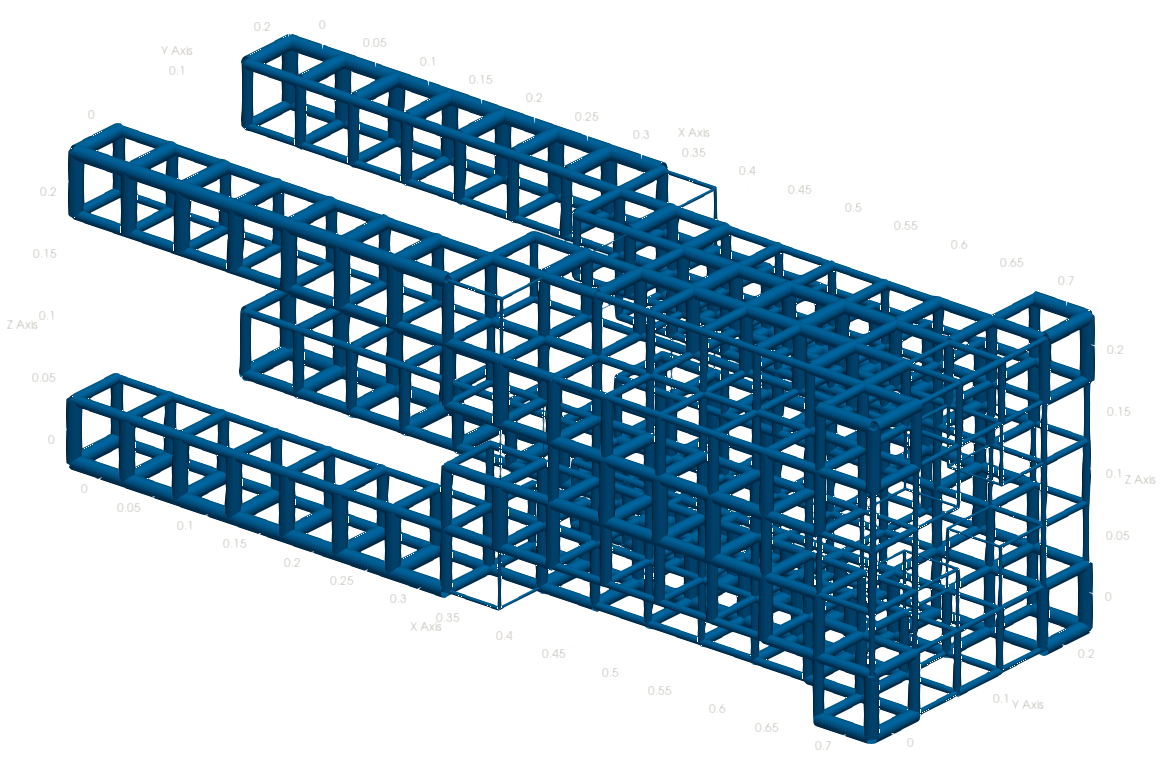}
\caption{Phase 1}
\label{TorPh1}
\end{subfigure}
\begin{subfigure}[b]{0.23\textwidth}
\centering
\includegraphics[width=\textwidth]{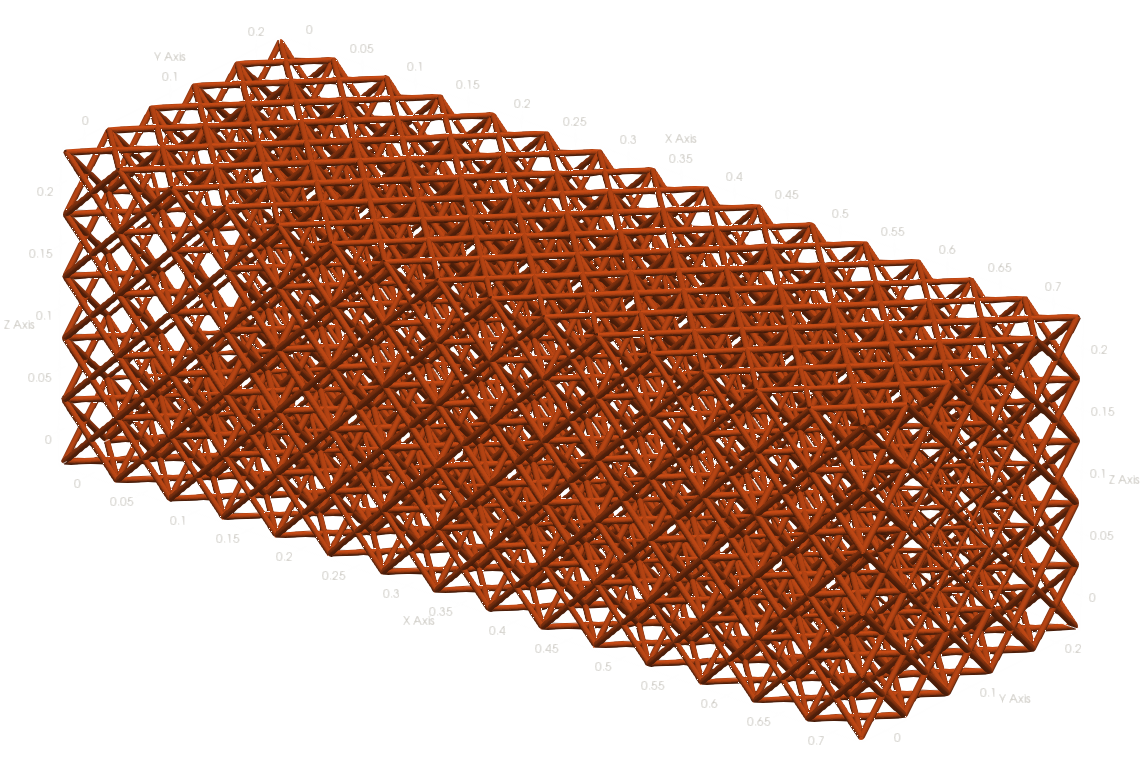}
\caption{Phase 2}
\label{TorPh2}
\end{subfigure}
\begin{subfigure}[b]{0.23\textwidth}
\centering
\includegraphics[width=\textwidth]{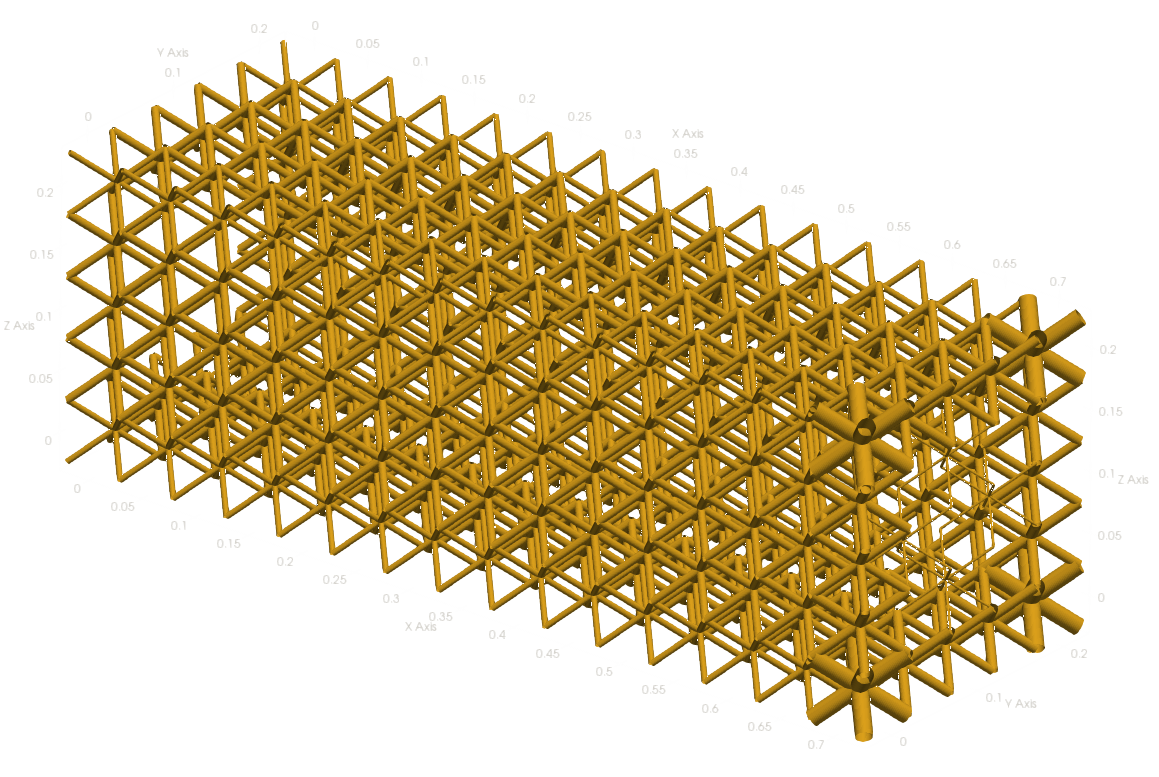}
\caption{Phase 3}
\label{TorPh3}
\end{subfigure}
\begin{subfigure}[b]{0.23\textwidth}
\centering
\includegraphics[width=\textwidth]{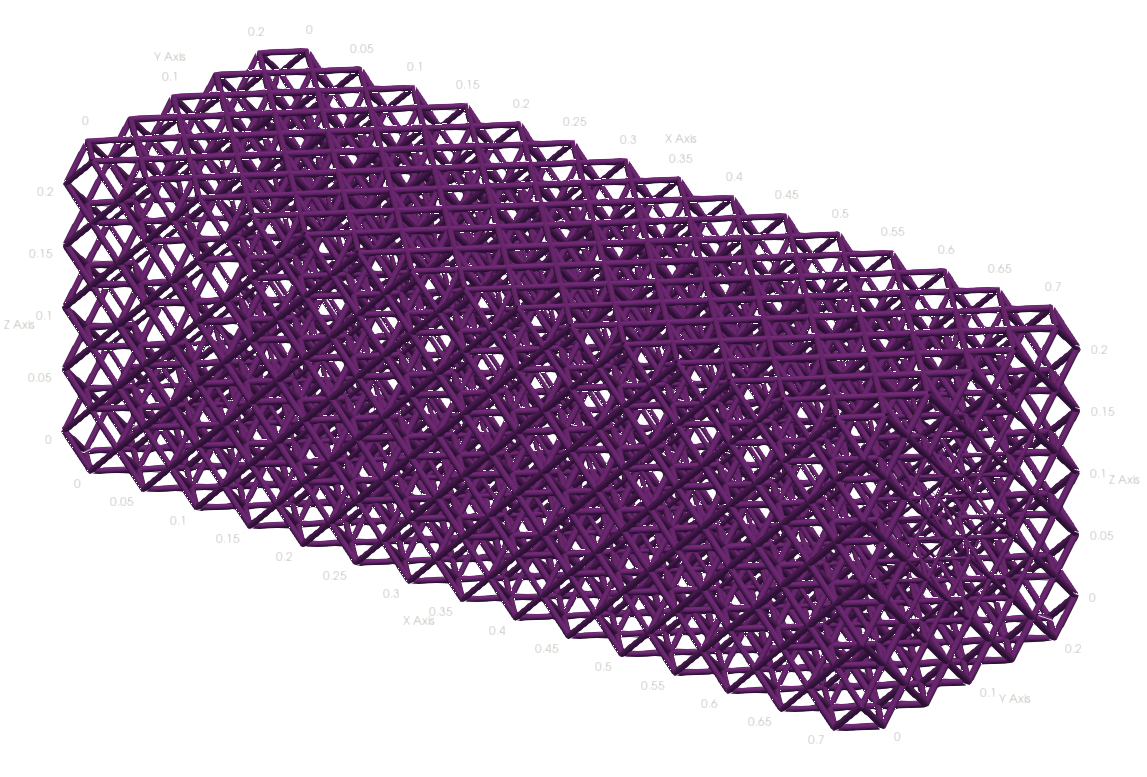}
\caption{Phase 4}
\label{TorPh4}
\end{subfigure}
\begin{subfigure}[b]{0.23\textwidth}
\centering
\includegraphics[width=\textwidth]{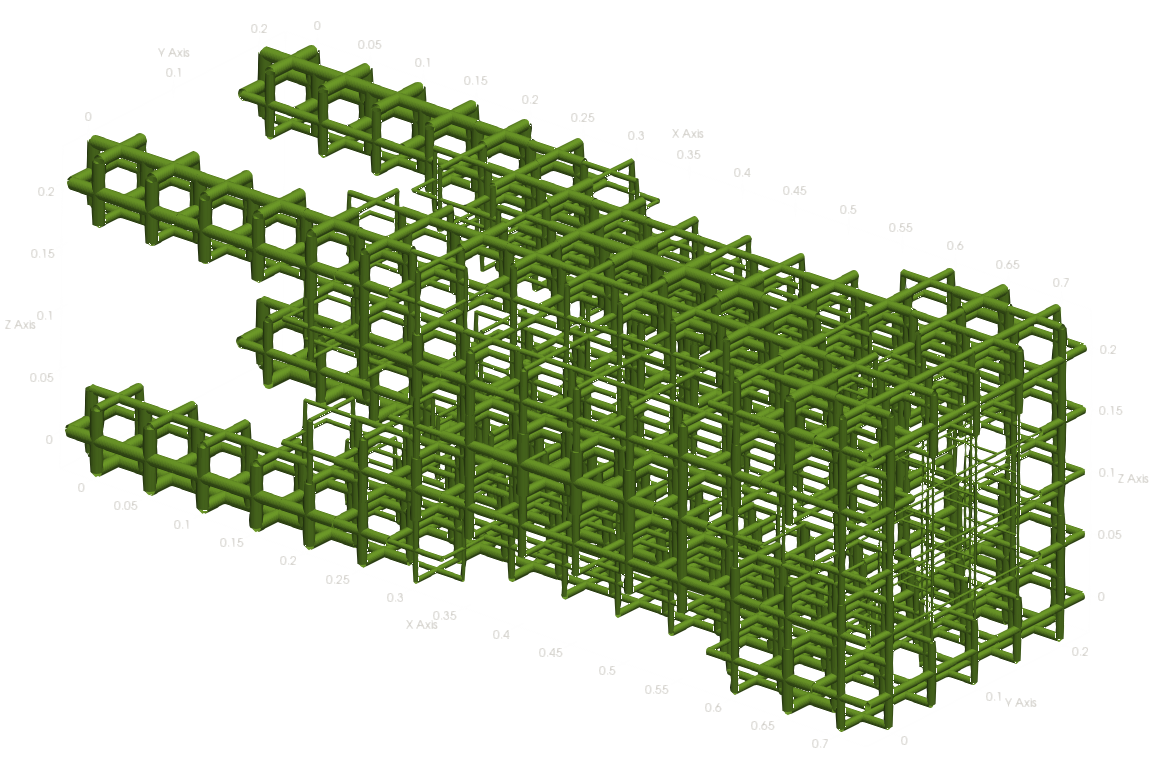}
\caption{Phase 5}
\label{TorPh5}
\end{subfigure}
\begin{subfigure}[b]{0.23\textwidth}
\centering
\includegraphics[width=\textwidth]{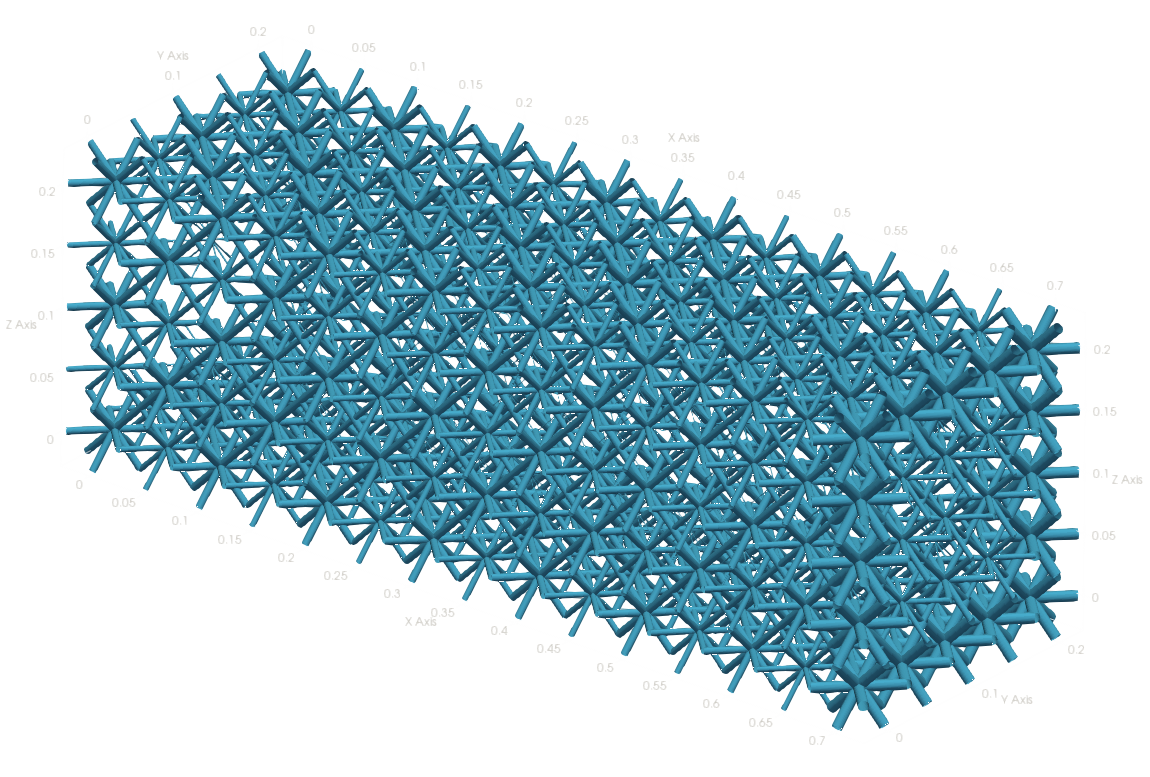}
\caption{Phase 6}
\label{TorPh6}
\end{subfigure}
\begin{subfigure}[b]{0.23\textwidth}
\centering
\includegraphics[width=\textwidth]{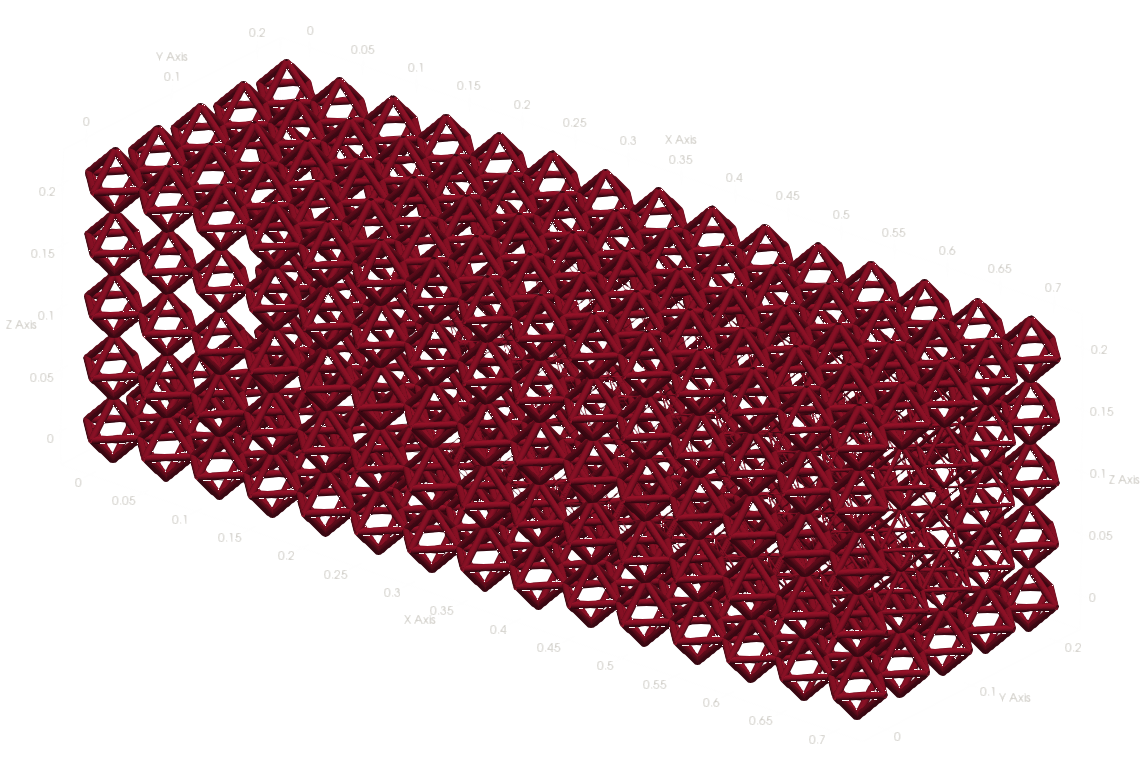}
\caption{Phase 7}
\label{TorPh7}
\end{subfigure}
\begin{subfigure}[b]{0.23\textwidth}
\centering
\includegraphics[width=\textwidth]{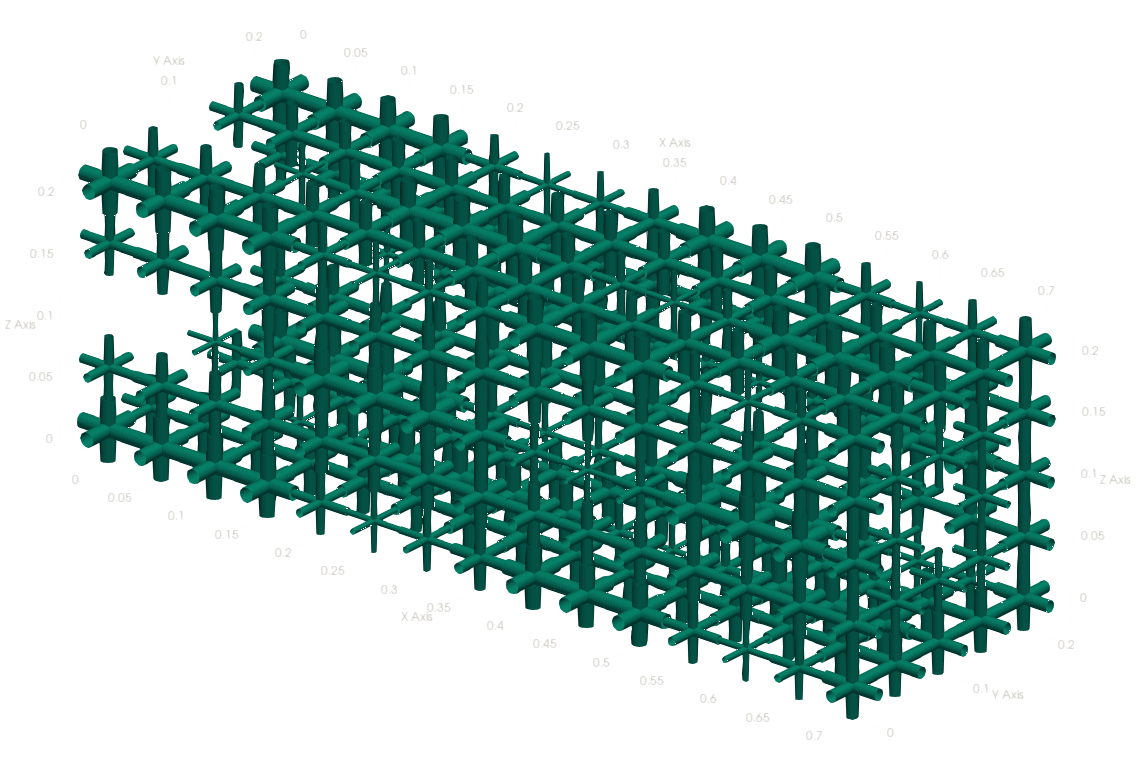}
\caption{Phase 8}
\label{TorPh8}
\end{subfigure}
\begin{subfigure}[b]{0.6\textwidth}
        \centering
        \includegraphics[width=\textwidth]{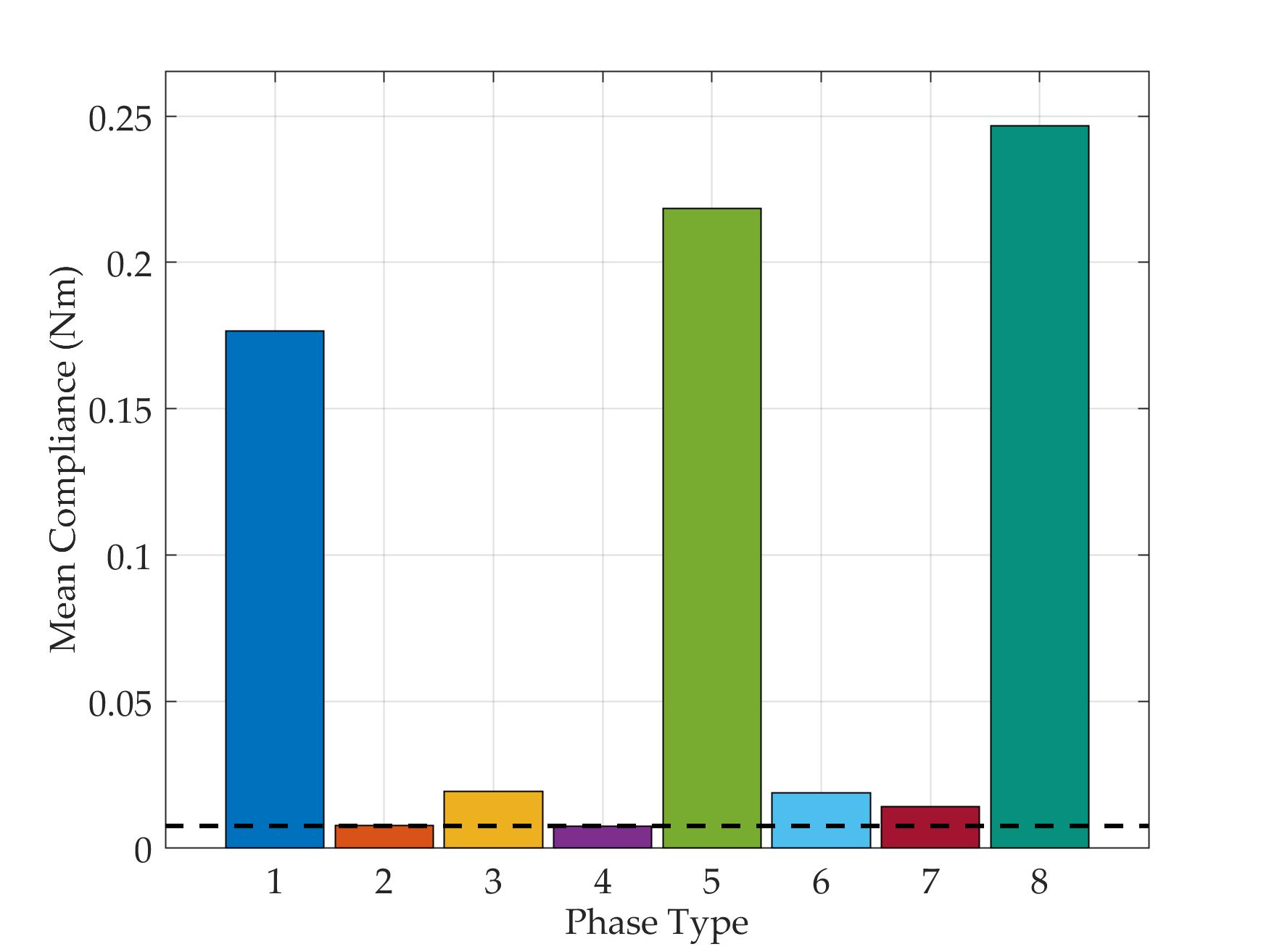}
        \caption{Comparison of stiffness in homogeneous designs (bars in color) and heterogeneous design (black dashed line, Figure \ref{TorSol})}
        \label{TorGraph}
    \end{subfigure}
\caption{Case-Study: Designs for Torsion case }
\label{TorCS}
\end{figure*}

\begin{equation}
\renewcommand{\arraystretch}{1.5}
\rho _{ij}^{(p+1)}=\rho _{ij}^{(p)} . \Bigg( \dfrac{u^T \dfrac{\partial (R^T \vb{K} R)}{\partial \rho_{ij}} u }{\bigg( \Lambda+\dfrac{\phi_k}{V_k} \bigg)\sum_{a}^{}\dfrac{\pi \vb*{L_a}}{36 n_a} l_{ij}^2} \Bigg)^\beta
\end{equation}
The initial guess is taken to be uniform densities with all of the eight phases present equally and the volume constraint satisfied. The number of cubic elements depends on the problem. The boundary conditions are applied appropriately. The update scheme is as follows.

\begin{enumerate}
    \item   Using the design variables perform function evaluation and sensitivity analysis.
    \item   Calculate the OC multipliers by assuming the value of $\bigg(\Lambda+\frac{\phi_k}{V_k}\bigg)$ as unity. Multiply all of the $\rho_{ij}$ with their respective multipliers.
    \item   Enforce the global volume constraint by dividing all of the $\rho_{ij}$ by an appropriate scalar.
    \item	Mark the design variables that have breached the upper or lower limits and enforce the limits. The marked variables will not be changed any further in the ongoing iteration. Once again satisfy the global volume constraint by dividing the unmarked variables by an appropriate scalar.
    \item   Find the active constraints in $\phi_j$ and mark the relevant design variables. Satisfy the constraint $\phi_j$ by dividing each of the unmarked design variable in the active constraints by respective scalars for each constraint.
    \item   Satisfy the volume constraint by dividing the unmarked design variables by a single scalar. Remove and reset all of the marks placed on variables. Go back to step (1) until convergence.  
\end{enumerate}
The convergence is reached when the difference in the objective function value for successive iterations is lower than a tolerance, which is taken in accordance with the problem. This algorithm leads us to a local minimum. Here, $\beta$ is a positive numerical handle less than unity. It is constant for an optimization run, but appropriately choosen to solve for different values of the volume ratio $\mu$. We have developed a vectorized script in MATLAB implementing the FEA and the algorithm.
\end{subsection}
\end{section}
\begin{section}{Numerical Results}
We solve several problems to illustrate the approach described. The mean compliance in all of the problems converges to 4 significant digits comfortably within 100 iterations. 

For the optimized results in this section, we have taken the minimum porosity limit $\kappa$ to be 40\% and the target volume ratio $\mu$ to be 5\%. The plots for the optimized design are rendered in ParaView \citep{refPV1,refPV2}. The dimensions are realistic and the area of cross section of the beam segments are up to scale. Each phase has a different color assigned to it for the purpose of visualization.
\begin{subsection}{Case 1: Cantilever Beam}
The first problem is the cantilever case (boundary conditions: Figure \ref{CantBC}). We encastre one end of the domain by completely fixing the nodes that fall in the blue region. A distributed transverse force is applied to all of the nodes on the other end, red arrows indicate the direction. The optimized structure is the adjoining Figure \ref{CantSol}.

There is a qualitative match with the intuitive expectations on the stiffest structure. The designed structure has a major amount of material concentrated at the top and bottom stacks. Furthermore, the density is more towards the support end, i.e. the beam segments are thicker towards the support end. The final value of mean compliance is 7.919E-4 Nm.
\end{subsection}
\begin{subsection}{Case 2: Torsion}
In this case, we take the same geometry as case 1 and apply a distributed torsion load on the load end (Figure \ref{TorBC}). The support end is encastre. The optimized design is found to be a hollow tube with a closed lid at the loaded end (Figure \ref{TorSol}). The final mean compliance is 7.403E-3 Nm.
\end{subsection}
\begin{subsection}{Case 3 and 4: MBB and Chair Problem}
\begin{figure*}
\centering
\begin{subfigure}[b]{0.23\textwidth}
\centering
\includegraphics[width=\textwidth]{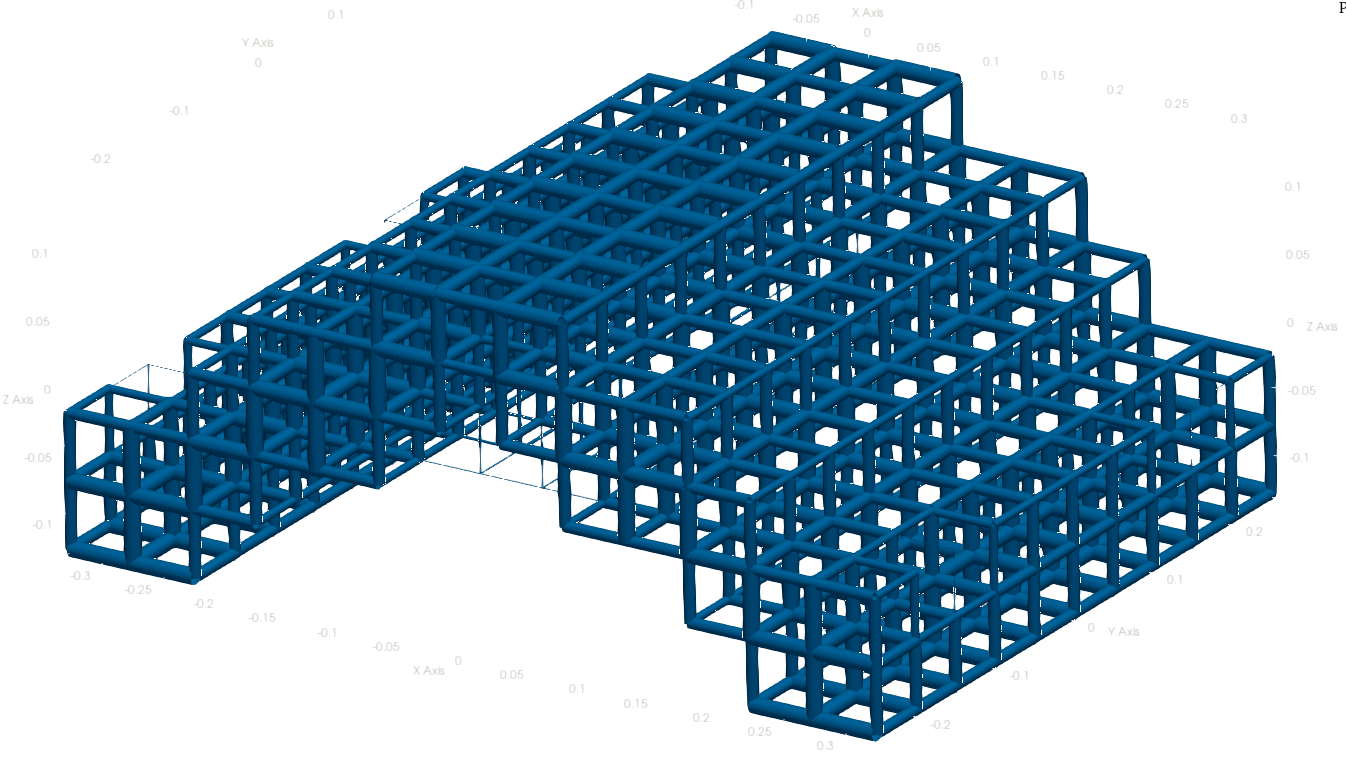}
\caption{Phase 1}
\label{MBBPh1}
\end{subfigure}
\begin{subfigure}[b]{0.23\textwidth}
\centering
\includegraphics[width=\textwidth]{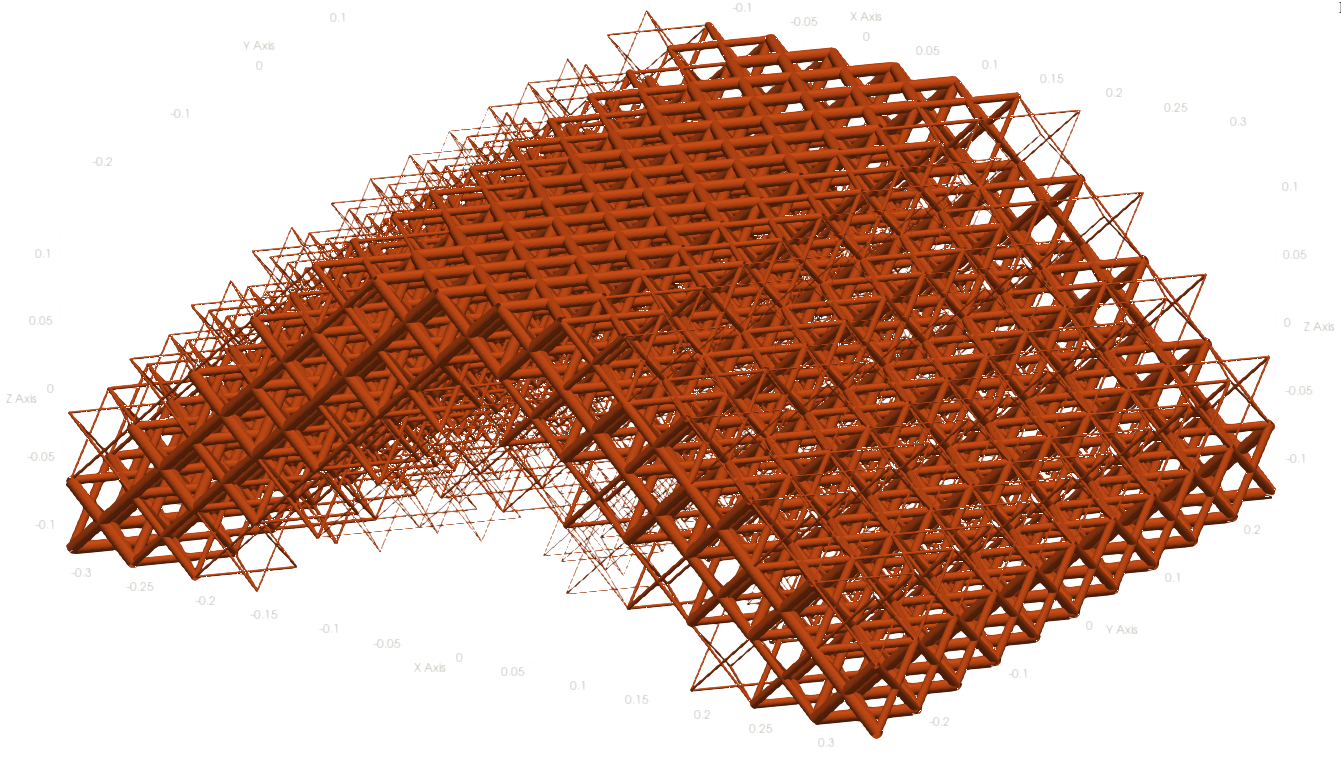}
\caption{Phase 2}
\label{MBBPh2}
\end{subfigure}
\begin{subfigure}[b]{0.23\textwidth}
\centering
\includegraphics[width=\textwidth]{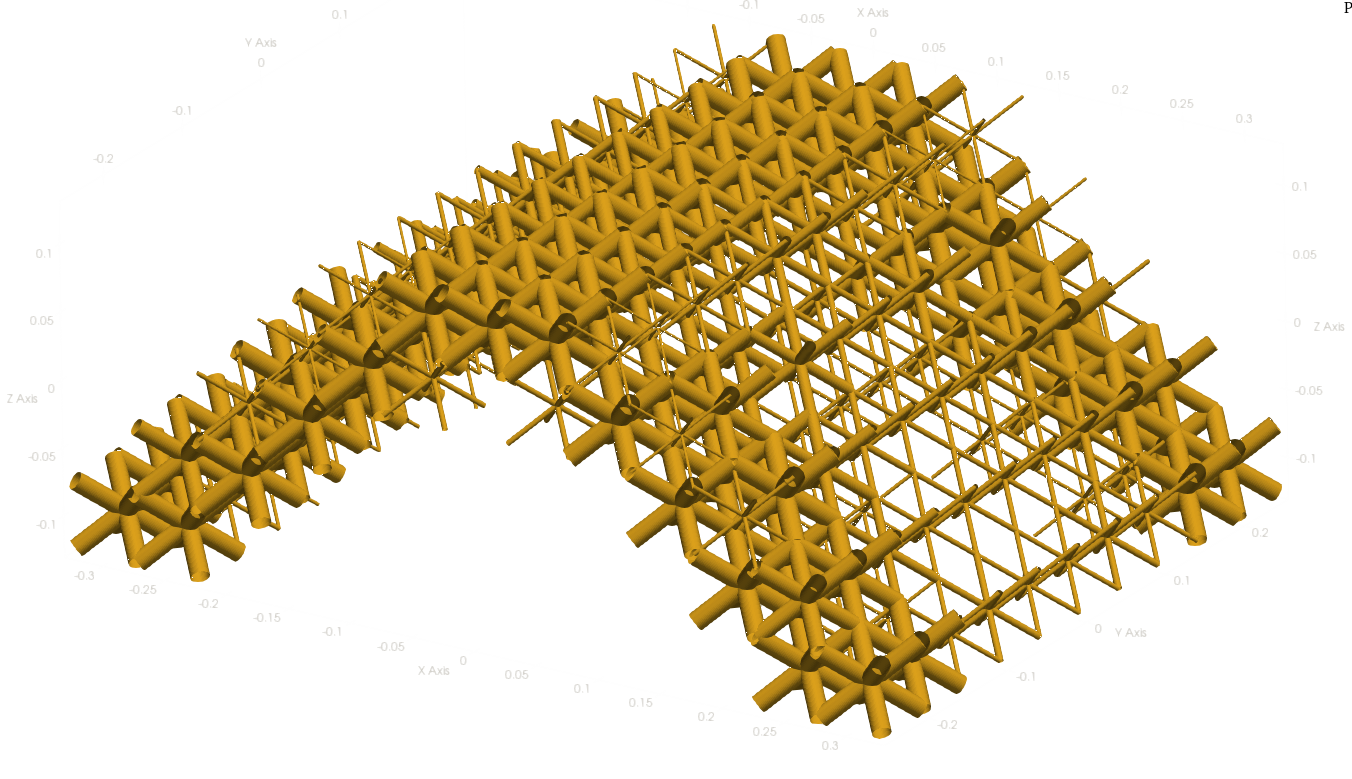}
\caption{Phase 3}
\label{MBBPh3}
\end{subfigure}
\begin{subfigure}[b]{0.23\textwidth}
\centering
\includegraphics[width=\textwidth]{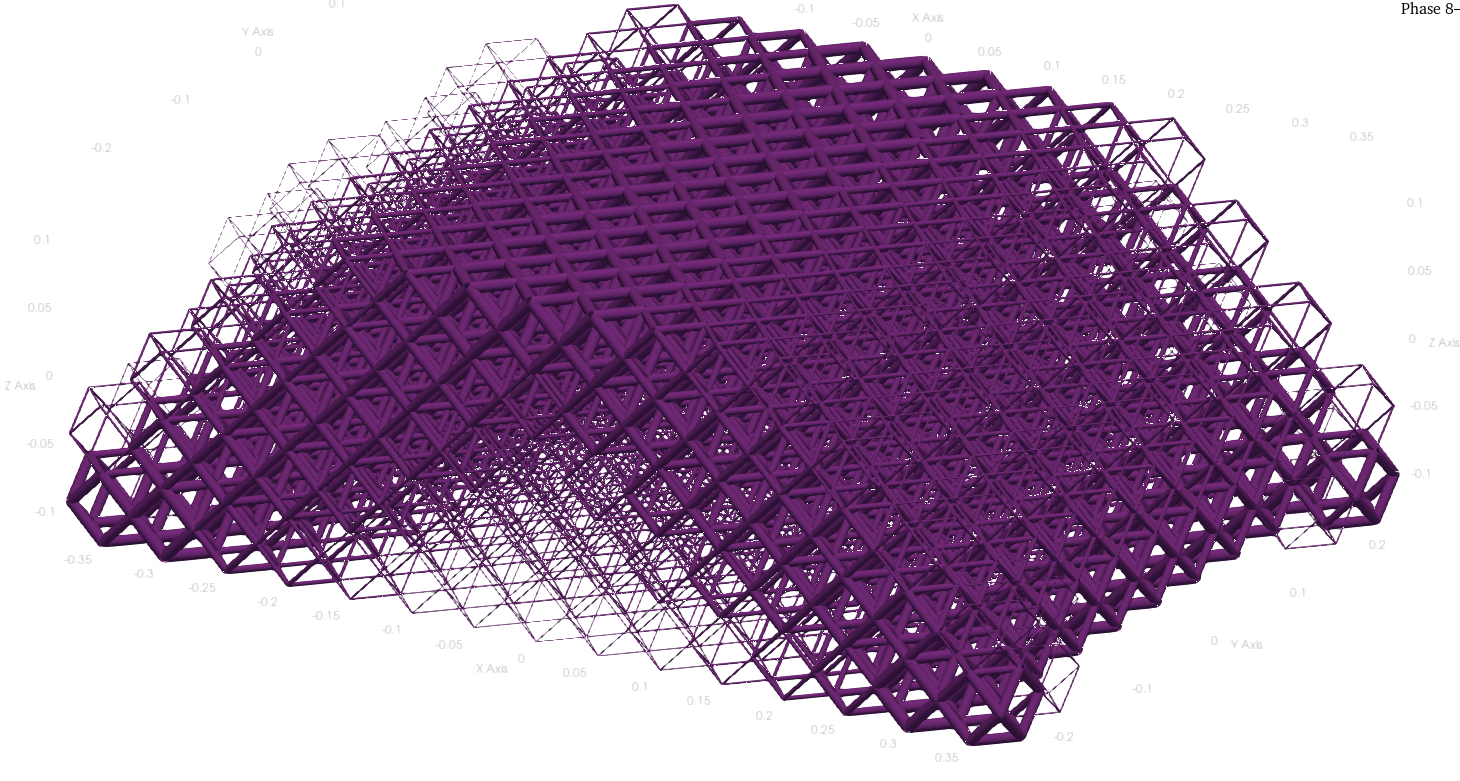}
\caption{Phase 4}
\label{MBBPh4}
\end{subfigure}
\begin{subfigure}[b]{0.23\textwidth}
\centering
\includegraphics[width=\textwidth]{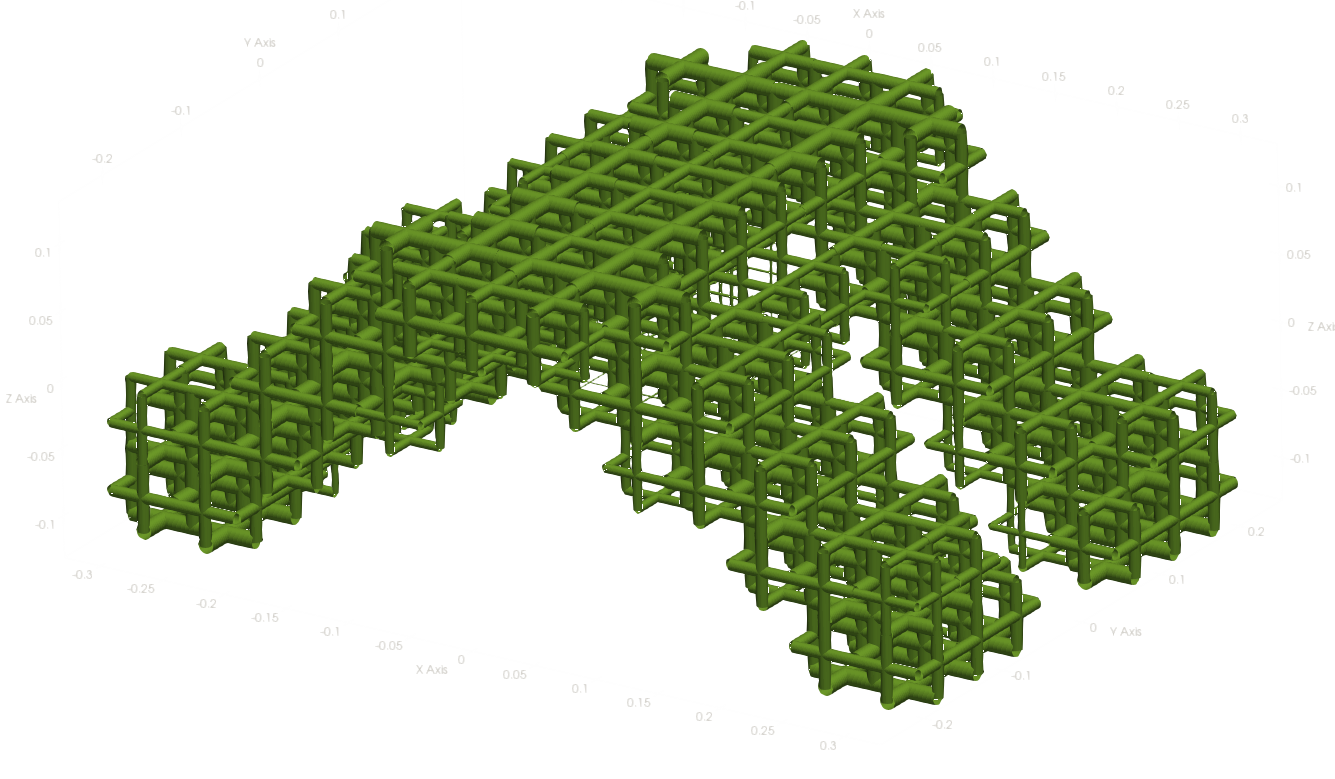}
\caption{Phase 5}
\label{MBBPh5}
\end{subfigure}
\begin{subfigure}[b]{0.23\textwidth}
\centering
\includegraphics[width=\textwidth]{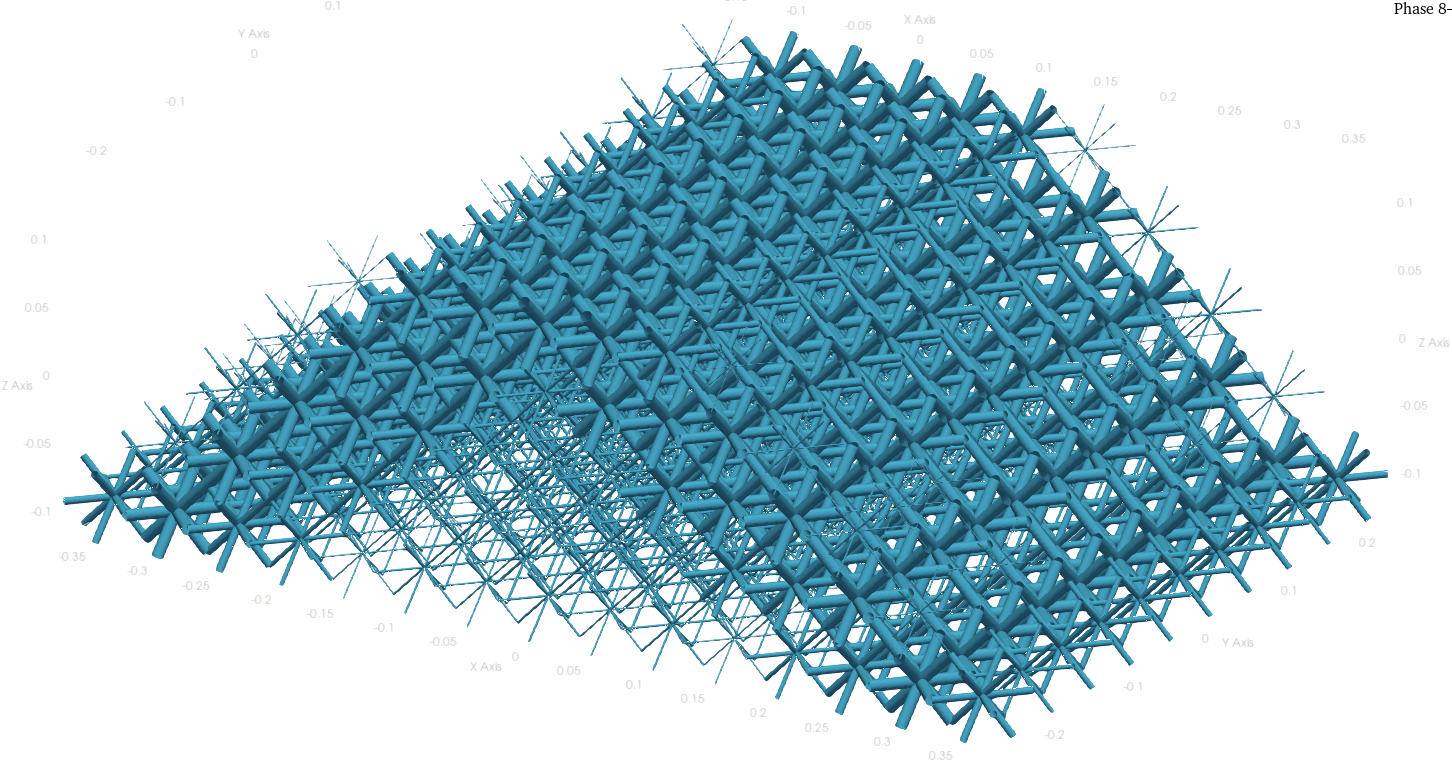}
\caption{Phase 6}
\label{MBBPh6}
\end{subfigure}
\begin{subfigure}[b]{0.23\textwidth}
\centering
\includegraphics[width=\textwidth]{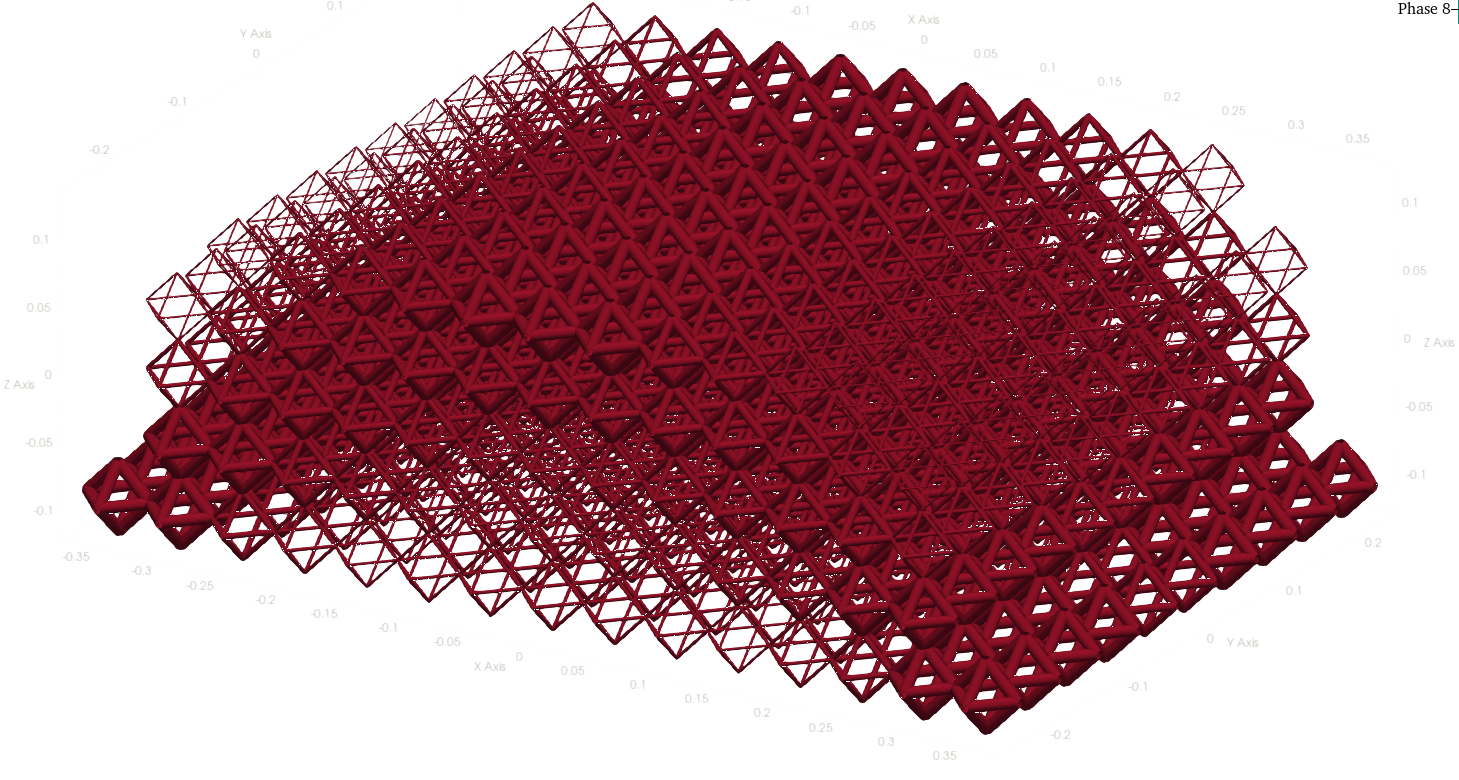}
\caption{Phase 7}
\label{MBBPh7}
\end{subfigure}
\begin{subfigure}[b]{0.23\textwidth}
\centering
\includegraphics[width=\textwidth]{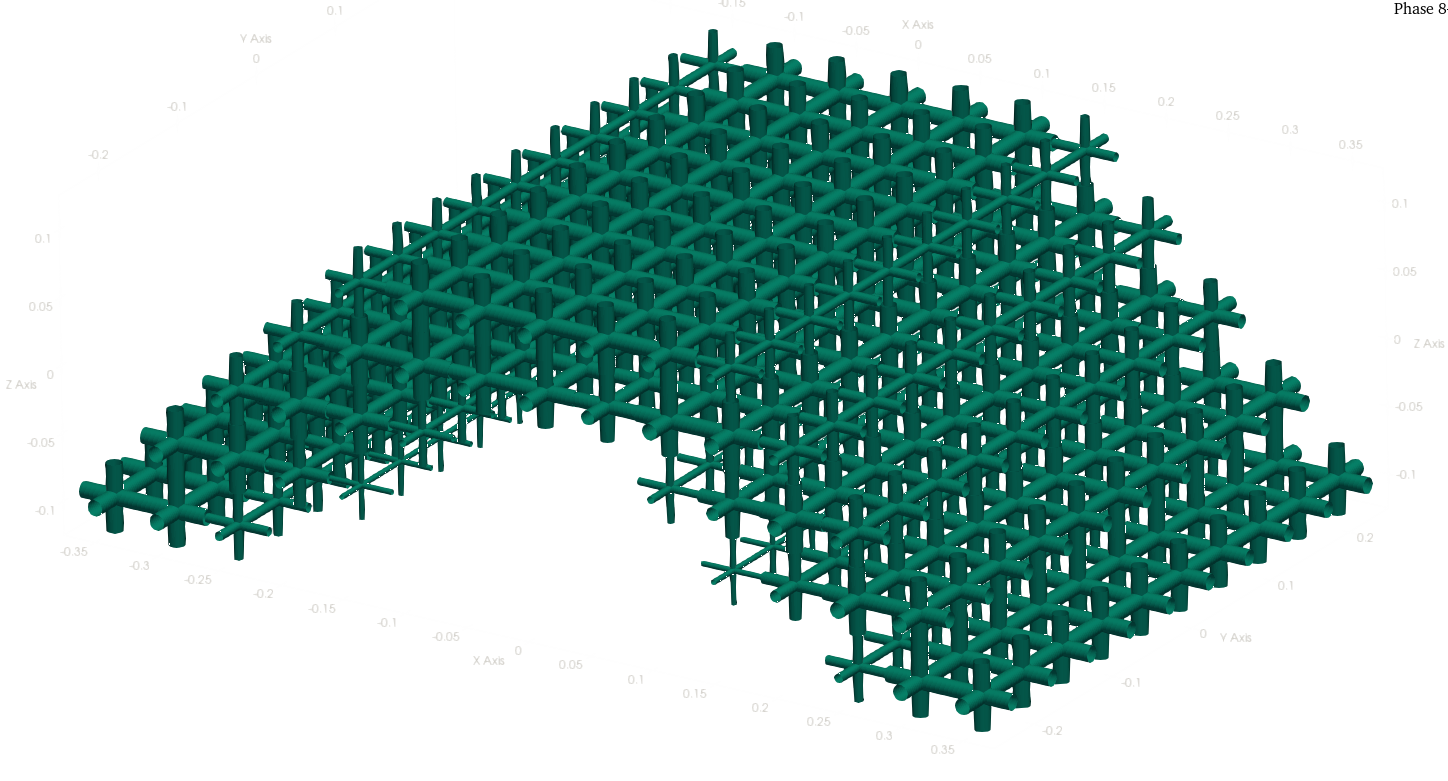}
\caption{Phase 8}
\label{MBBPh8}
\end{subfigure}
    \begin{subfigure}[b]{0.6\textwidth}
        \centering
        \includegraphics[width=\textwidth]{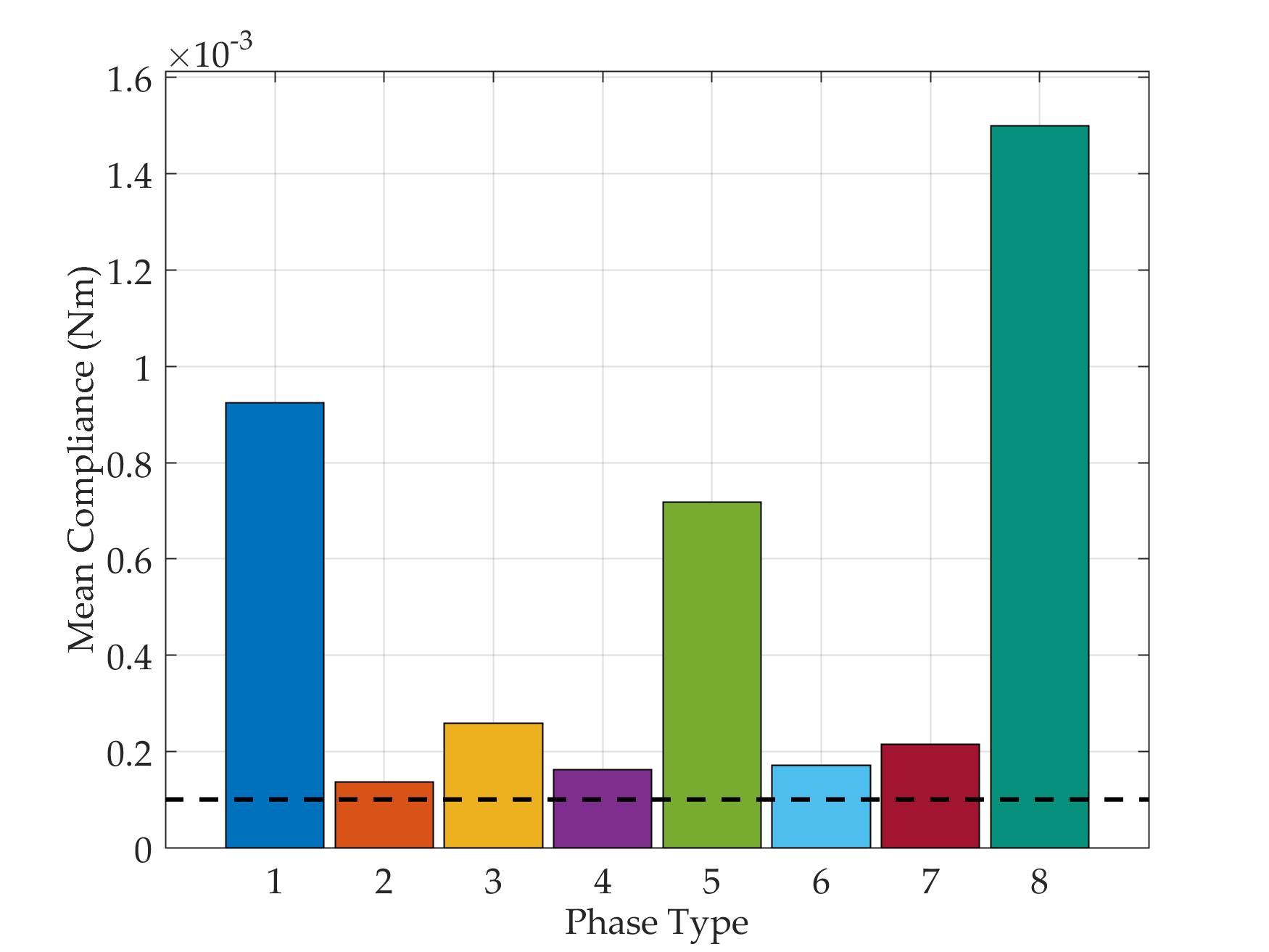}
        \caption{Comparison of stiffness in homogeneous designs (bars in color) and heterogeneous design (black dashed line, Figure \ref{MBBSol1})}
        \label{MBBGraph}
    \end{subfigure}
\caption{Case-Study: Designs for the MBB case }
\label{MBBCS}
\end{figure*}
Here, we consider two realistic problems of larger size than the first two cases. The first is the MBB problem (Figure \ref{MBBBC}) with a rectangular domain, completely fixed at two ends and a central load. The second is the chair problem (Figure \ref{ChairBC}) with an L-shaped 3D domain. The seat and the backrest have forces normal to the surfaces and the four corners on the bottom face are fixed. The optimized mean compliance for MBB case is 1.002E-4 Nm, and for the chair case it is 1.2436E-3 Nm.

The heterogeneity in microstructure topology is visible in the colors. For the MBB case, the middle portions are a combination of phase 3 (yellow) and phase 1 (blue), connecting the two stacks of phase 2 (orange) present at the extreme ends. The place where load is applied has phase 1, 4 (purple) and 7 (maroon). 

In the chair case, the front legs mostly have phase 1 and phase 5 (green), while the back legs only have phase 1. The seat has a lot of purple (phase 4), while the top of the back rest mostly has low density oranges and yellow (phase 2 and 3).   
\end{subsection}
\begin{subsection}{Case Study: Optimal Heterogeneity with all phases vs Optimal Homogeneity with a single phase}
We now consider a case study to highlight the optimality of components with heterogeneous microstructure topology, as compared to components with homogeneous lattice structure. 

We solve the four examples discussed previously. Eight more optimization runs are performed for each case allowing only one type of phase. All of the parameters remain the same and statically equivalent loads are applied. The designs are present in Figures \ref{CantCS}, \ref{TorCS}, \ref{MBBCS}, \ref{ChairCS}. The mean compliance for the optimized designs in heterogeneous and homogeneous cases are also compared.

For a particular boundary condition, the macrogeometry is different for the different phases. In the torsion load case, homogeneous designs for phase 2 and phase 4 have mean compliance very close to the heterogeneous design, a coincidence not observed in other cases. The phases 2 and 4 are the stiffest for the torsion case, as evident by their dominance in the heterogeneous design. We note that phases 3, 6 and 7 are the next best homogeneous designs, having similar optimized homogeneous macrogeometry as phases 2 and 4. They do not appear in the heterogeneous designs due to the volume constraint, and connecting at lesser number of points (8, 12, and 6 respectively). As expected, the mean compliance of the heterogeneous design is better than all of the uniform designs, in all the load cases. The last point has been shown previously for designs where, a finite amount of microstructures are allowed, and the uniform designs do not have size variation. We show this in a more general way.

\end{subsection}

\begin{figure*}
    \centering
    \begin{subfigure}[b]{0.23\textwidth}
        \centering
        \includegraphics[width=\textwidth]{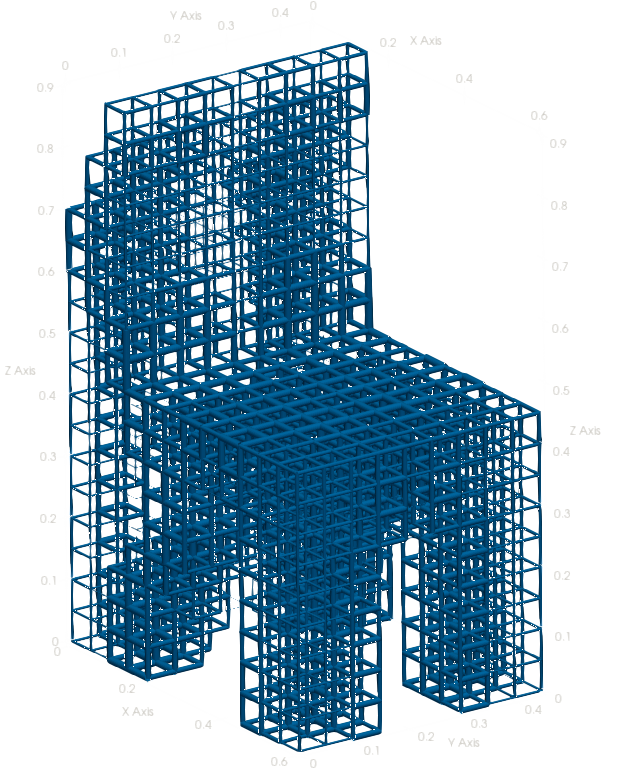}
        \caption{Phase 1}
        \label{ChPh1}
    \end{subfigure}
    \begin{subfigure}[b]{0.23\textwidth}
        \centering
        \includegraphics[width=\textwidth]{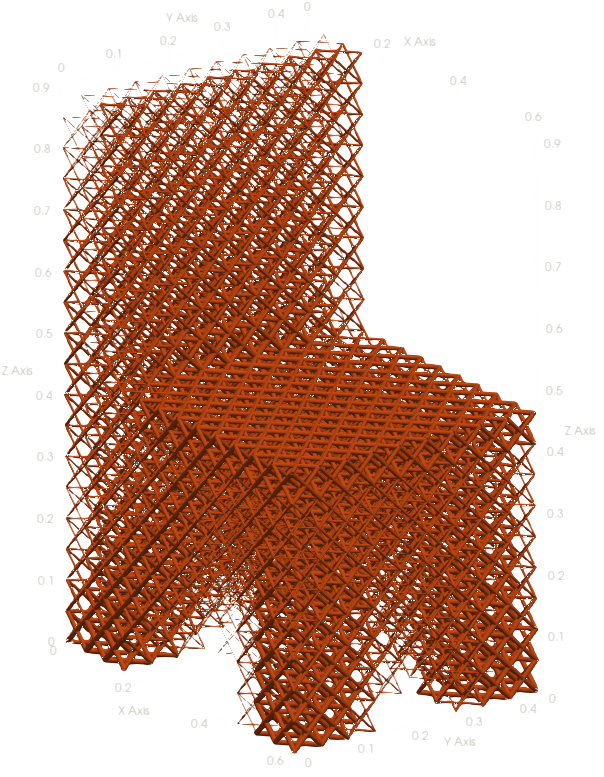}
        \caption{Phase 2}
        \label{ChPh2}
    \end{subfigure}
    \begin{subfigure}[b]{0.23\textwidth}
        \centering
        \includegraphics[width=\textwidth]{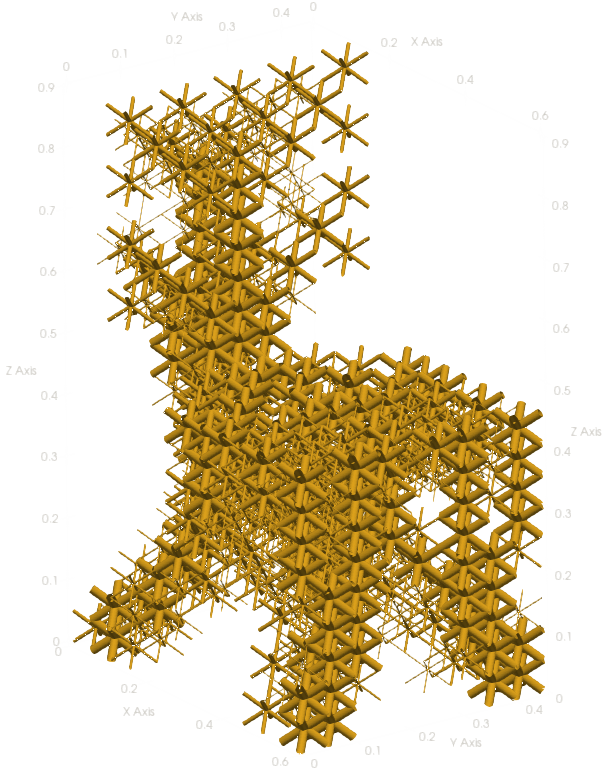}
        \caption{Phase 3}
        \label{ChPh3}
    \end{subfigure}
    \begin{subfigure}[b]{0.23\textwidth}
        \centering
        \includegraphics[width=\textwidth]{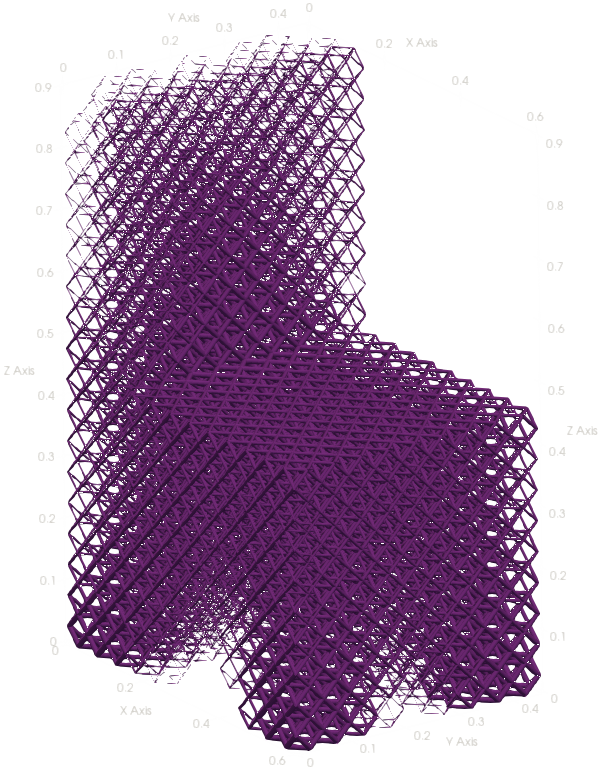}
        \caption{Phase 4}
        \label{ChPh4}
    \end{subfigure}
    \begin{subfigure}[b]{0.23\textwidth}
        \centering
        \includegraphics[width=\textwidth]{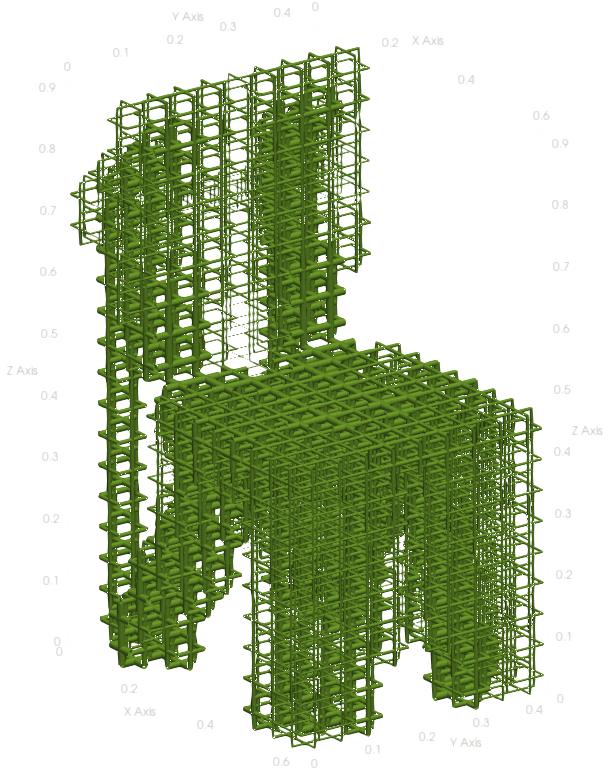}
        \caption{Phase 1}
        \label{ChPh5}
    \end{subfigure}
    \begin{subfigure}[b]{0.23\textwidth}
        \centering
        \includegraphics[width=\textwidth]{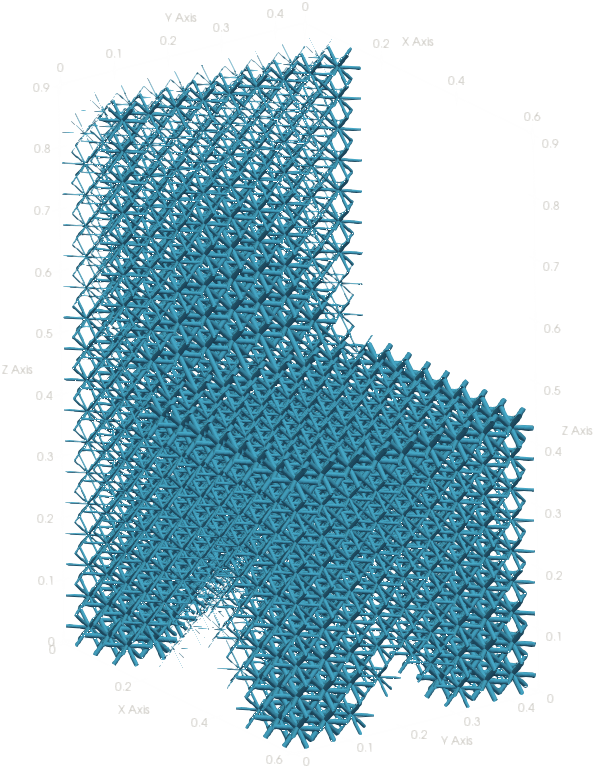}
        \caption{Phase 6}
        \label{ChPh6}
    \end{subfigure}
    \begin{subfigure}[b]{0.23\textwidth}
        \centering
        \includegraphics[width=\textwidth]{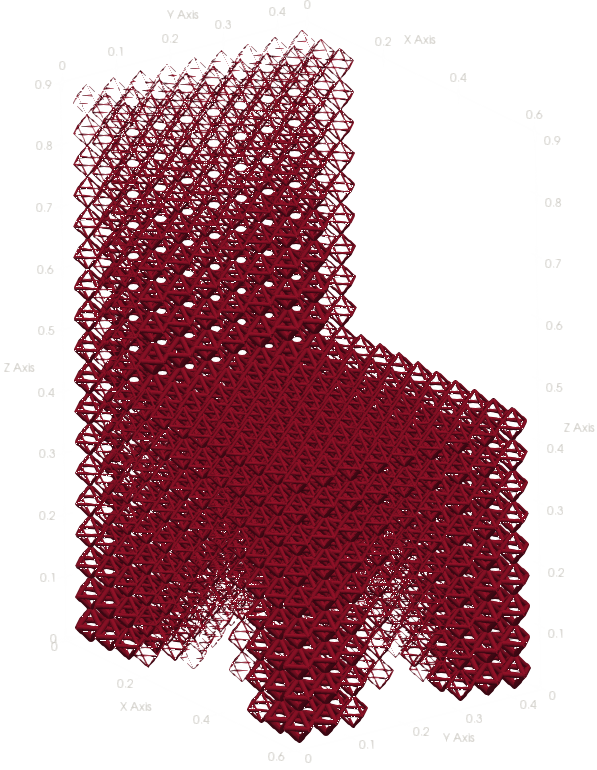}
        \caption{Phase 7}
        \label{ChPh7}
    \end{subfigure}
    \begin{subfigure}[b]{0.23\textwidth}
        \centering
        \includegraphics[width=\textwidth]{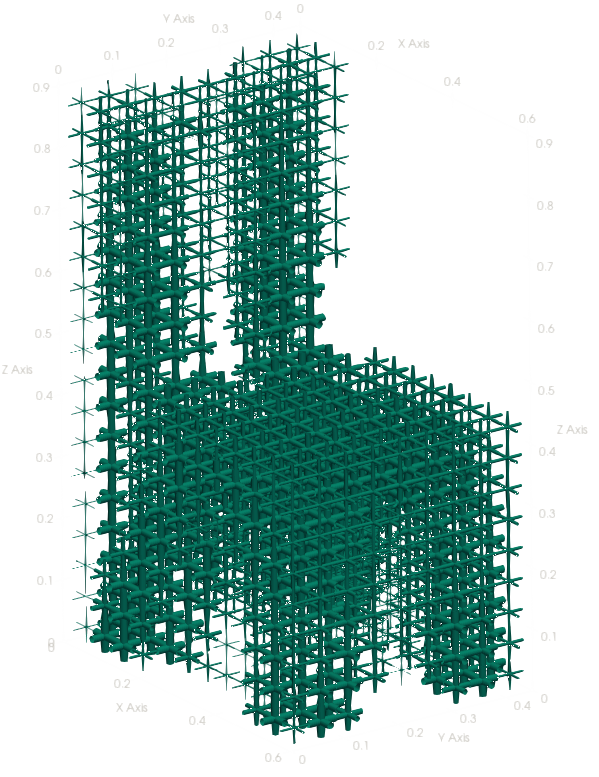}
        \caption{Phase 8}
        \label{ChPh8}
    \end{subfigure}
        \begin{subfigure}[b]{0.6\textwidth}
        \centering
        \includegraphics[width=\textwidth]{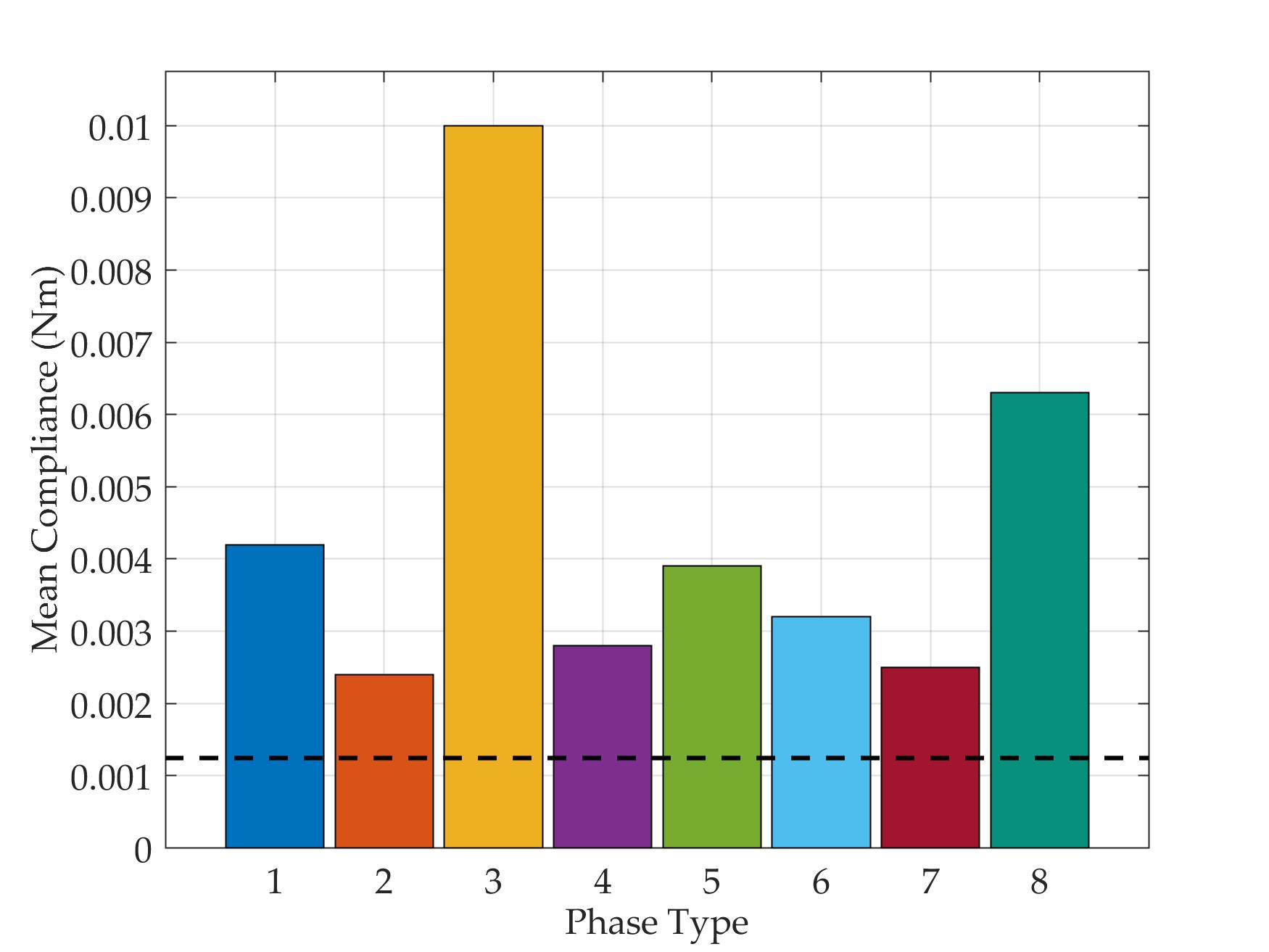}
        \caption{Comparison of stiffness in homogeneous designs (bars in color) and heterogeneous design (black dashed line, Figure \ref{ChairSol1})}
        \label{ChairGraph}
    \end{subfigure}
    \caption{Case-Study: Designs for the Chair case}
    \label{ChairCS}
\end{figure*}
\end{section}
\begin{section}{Interpretation of the Optimal Designs}
As stated earlier, the bulk behaviour of lattice-structure materials is captured by micropolar elasticity \citep{refJasuk,refCossLakes} as opposed to classical elasticity. For cellular materials (such as lattice structures), there is a practical limit on how small the critical volume is for the continuum assumption to hold true \citep{refEringen}. In a subdivision of the domain, the internal moments that arise from the neighbouring subdivisions may not balance out on their own. In this situation, the forces need to contribute to establish moment equilibrium. When the microstructures are homogenized, this aspect is not reflected in the assumptions, leading to discrepancies. We have observed that during the pure bending of a square cross-sectioned beam with a uniform infill of all of the phases, the qualitative behaviour matches the analytical prediction by micropolar elasticity \citep{refLakesSquare}. The cross-section of the beam warps into an S shape (Figure \ref{Sigm}). 

Thus, we need to explore homogenization techniques tailored to extract effective micropolar constants (six, as opposed to two in classical) from a microstructure. Discrete asymptotic homogenization \citep{refTollenaere} has been developed for topologies that consist of beams. The adaptation of that to 3D structures that yield micropolar constants is yet to be done.

In order to interpret the designs, we need to find why a certain combination of phases appear at a location. Our hypothesis is that the internal-force condition on a cube drives the selection of phases. We consider four basic load tests from micropolar elasticity \citep{refStructMat}: tension or compression, shear, torsion and bending. Keeping the volume constant, we apply the loads to all of the phases individually and rank them on the basis of the stiffness. These are represented in the four plots (Figure \ref{PP}). The ranks are strain energies transformed such that 1 represents the stiffest phase and 8 represents the opposite. The ranks in the middle are linearly interpolated between these. The benefit of this is that we get an idea of the closeness of the various phases as well as the transitions. Next, we look at the designs that have pure load cases, axial tension (Figure \ref{Axial}), torsion (Figure \ref{TorSol}) and pure bending (Figure \ref{PB5}). 

\begin{figure*}
    \centering
    \begin{subfigure}[b]{0.4\textwidth}
        \centering
        \includegraphics[width=\textwidth]{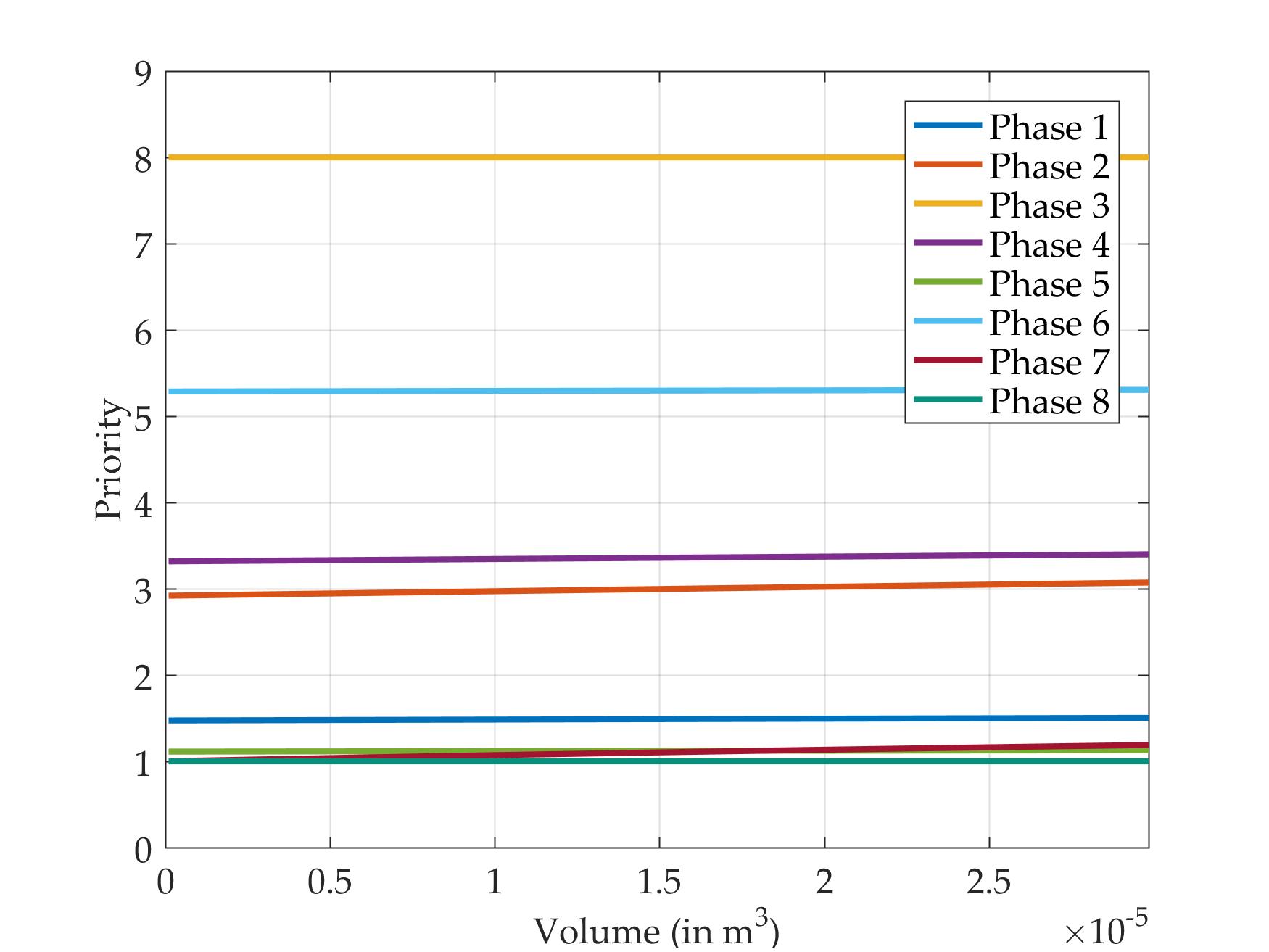}
        \caption{Tension Phase Plot}
        \label{Tens}
    \end{subfigure}
    \begin{subfigure}[b]{0.4\textwidth}
        \centering
        \includegraphics[width=\textwidth]{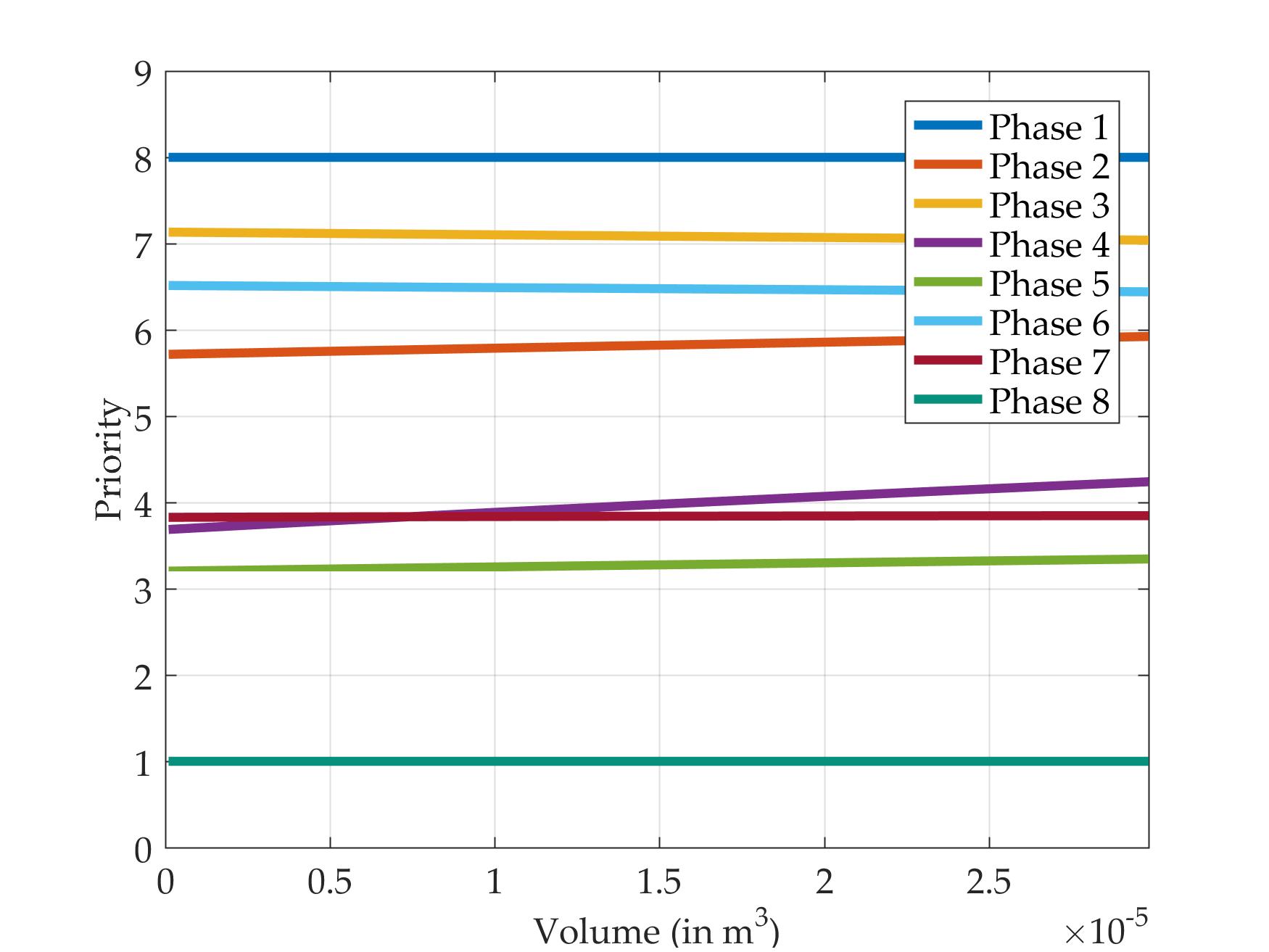}
        \caption{Shear Phase Plot}
        \label{Shear}
    \end{subfigure}
    \begin{subfigure}[b]{0.4\textwidth}
        \centering
        \includegraphics[width=\textwidth]{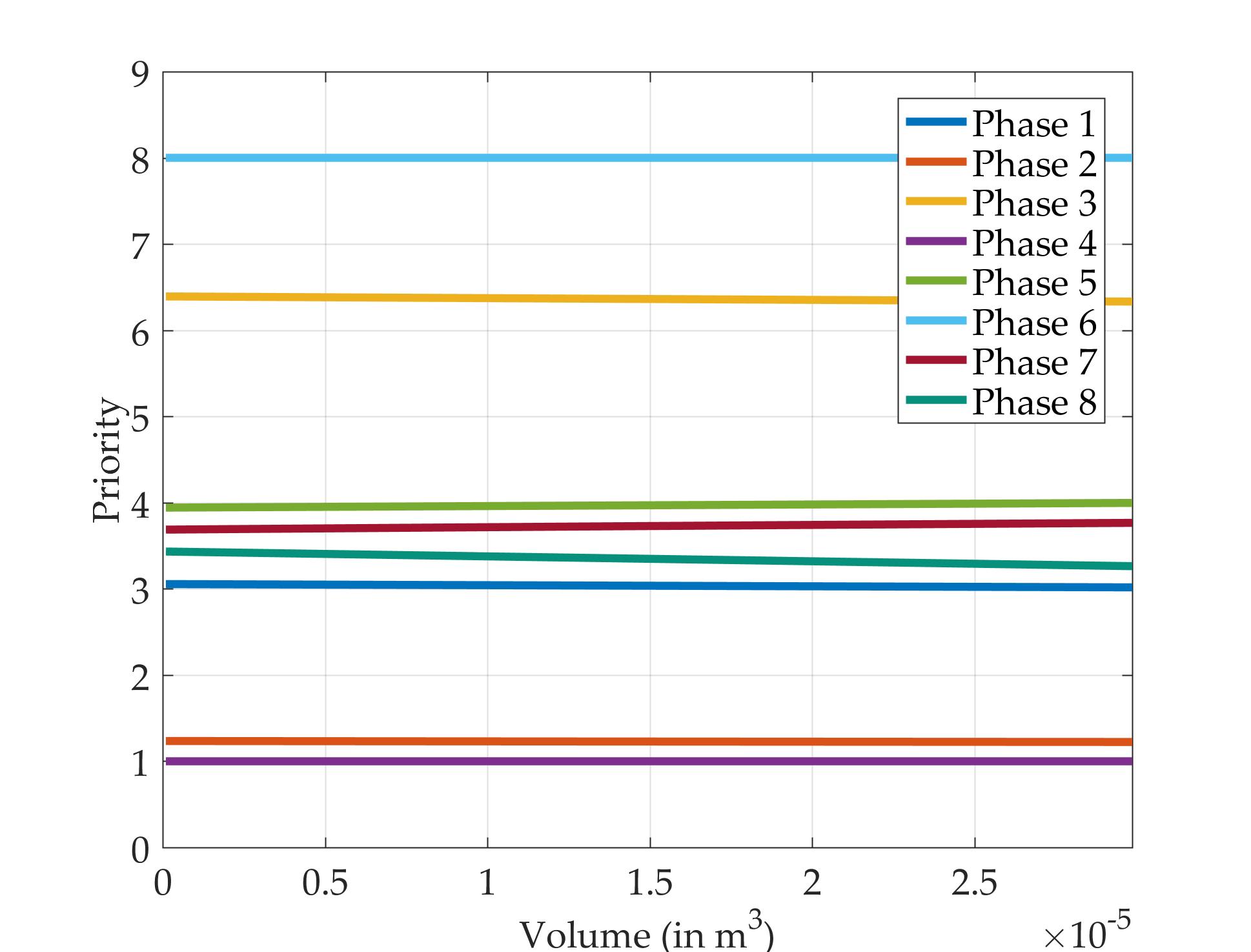}
        \caption{Torsion Phase Plot}
        \label{Tor}
    \end{subfigure}
    \begin{subfigure}[b]{0.4\textwidth}
        \centering
        \includegraphics[width=\textwidth]{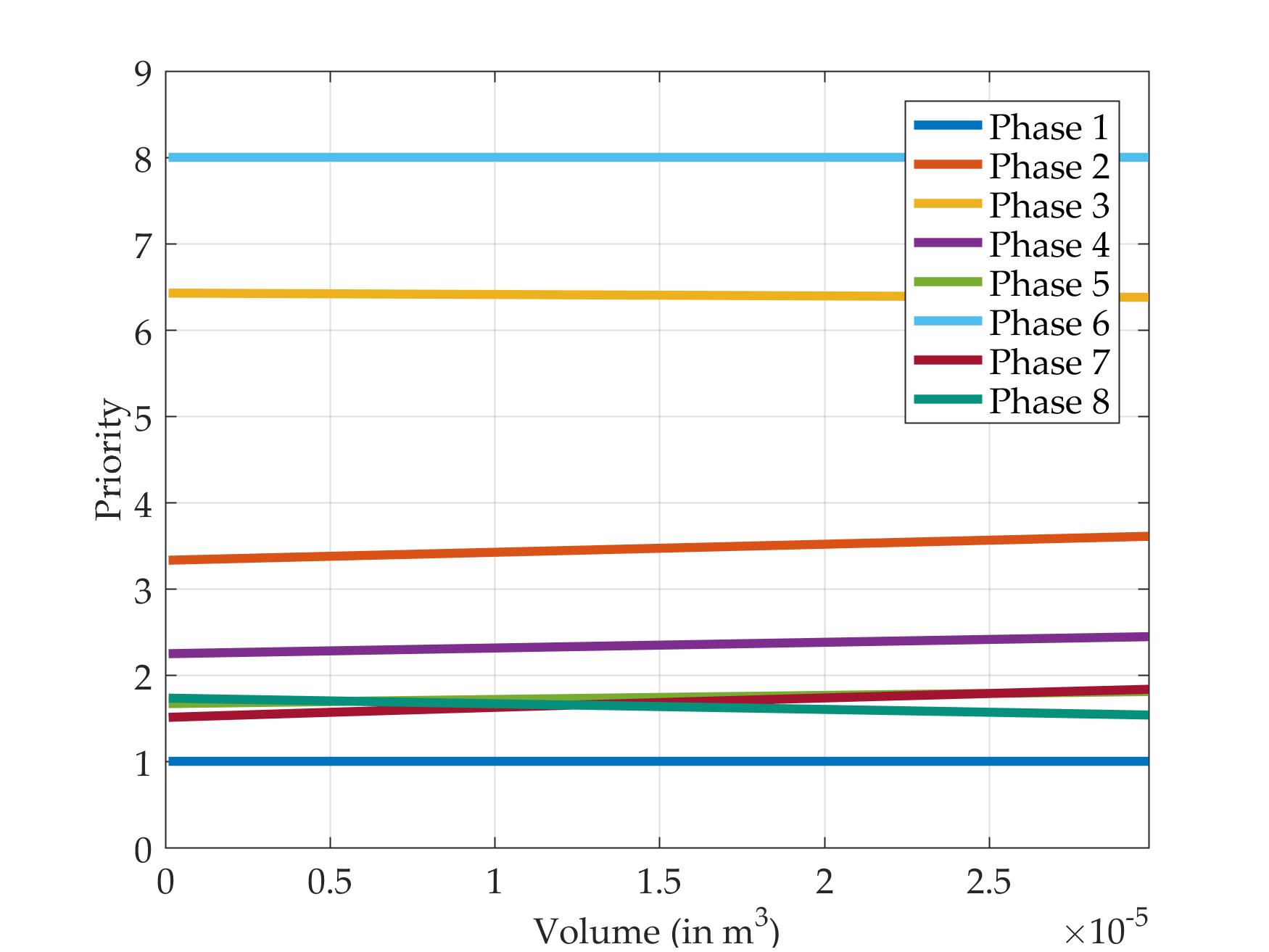}
        \caption{Bending Phase Plot}
        \label{Bend}
    \end{subfigure}
    \caption{Phase Plots for the various load cases}
    \label{PP}
\end{figure*}

\begin{figure*}
    \centering
        \begin{subfigure}[b]{0.4\textwidth}
        \centering
        \includegraphics[width=\textwidth]{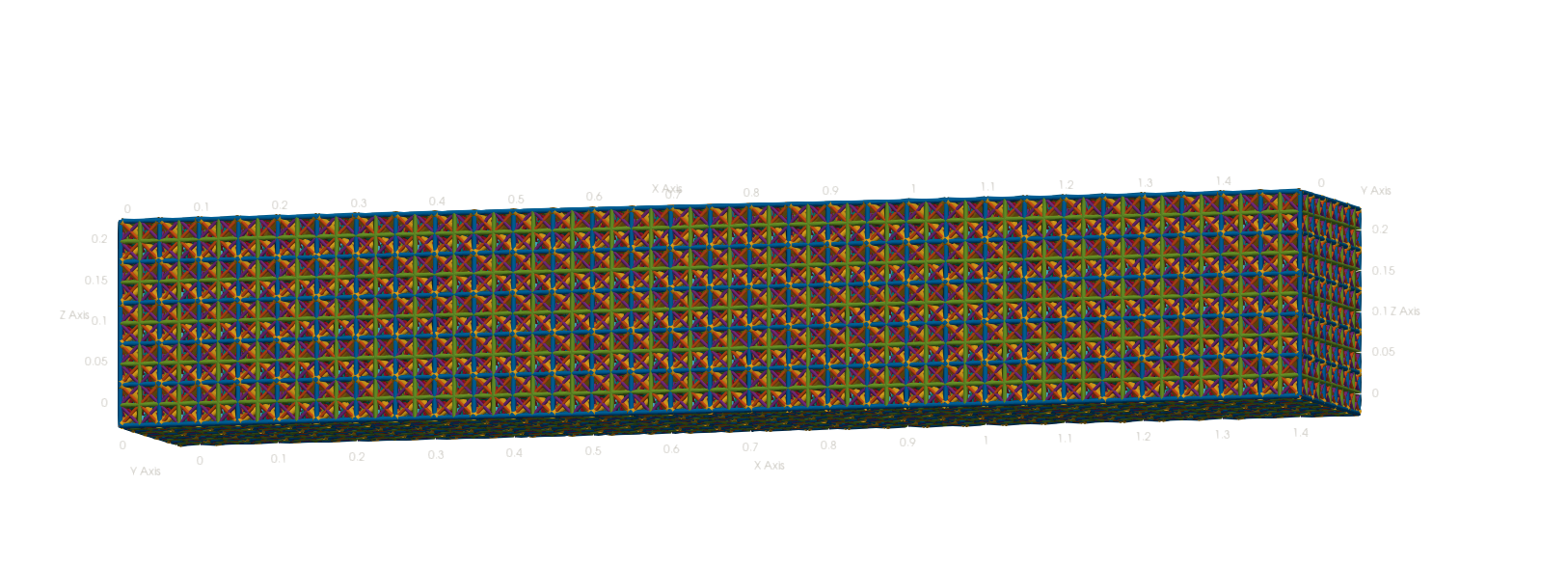}
        \caption{Undeformed square cross section bar}
        \label{Sigmbe4}
    \end{subfigure}
        \begin{subfigure}[b]{0.4\textwidth}
        \centering
        \includegraphics[width=\textwidth]{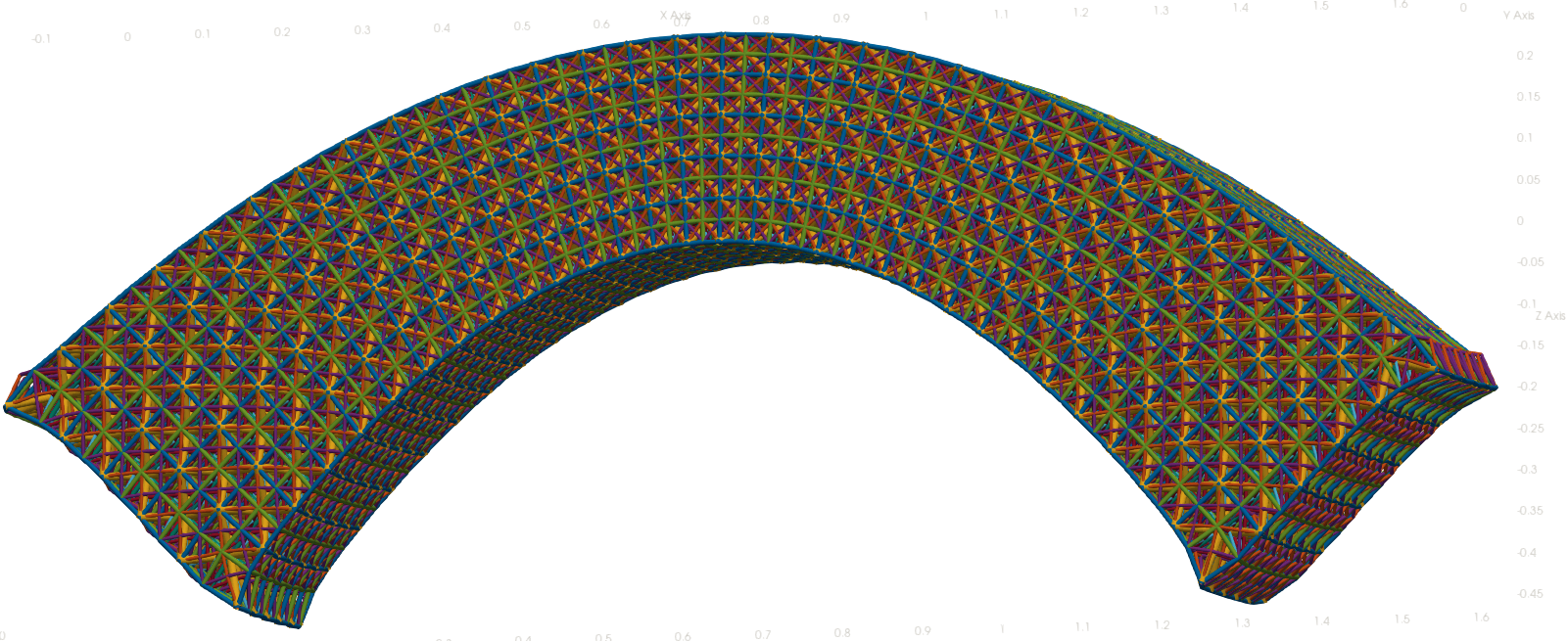}
        \caption{S-type warping of the cross section}
        \label{Sigm}
    \end{subfigure}
    \caption{Sigmoid warpage of the cross section}    
\end{figure*}
\begin{figure*}
    \centering
    \begin{subfigure}[b]{0.23\textwidth}
        \centering
        \includegraphics[width=\textwidth]{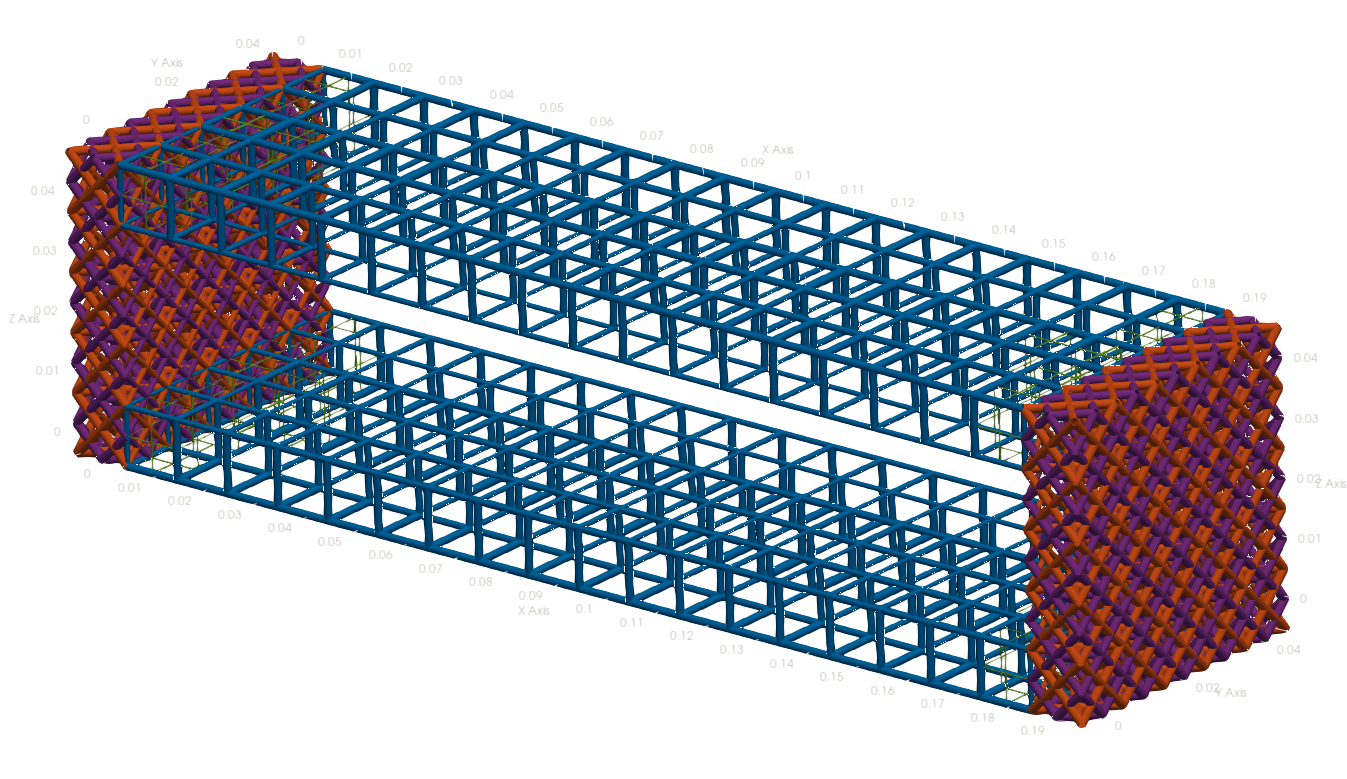}
        \caption{5 \% Volume ratio}
        \label{PB5}
    \end{subfigure}
        \begin{subfigure}[b]{0.23\textwidth}
        \centering
        \includegraphics[width=\textwidth]{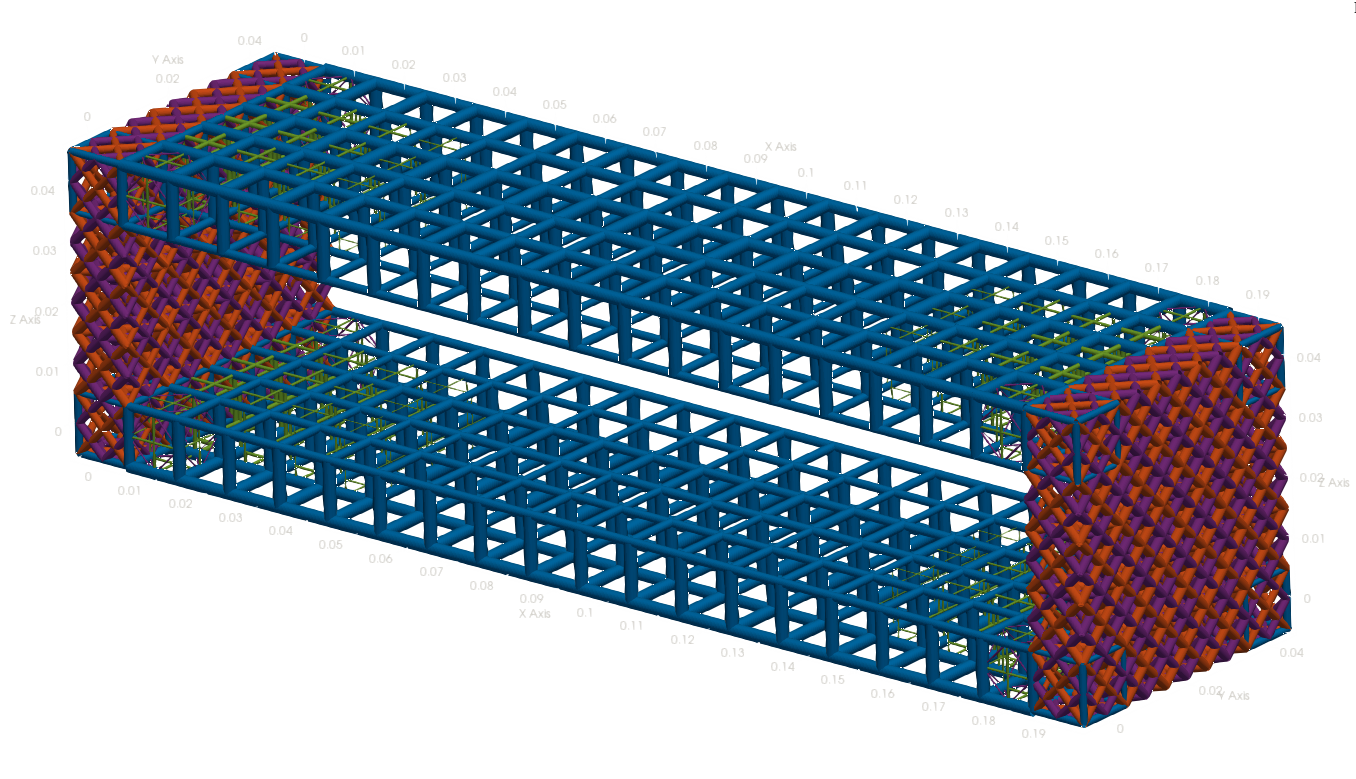}
        \caption{9 \% Volume ratio}
        \label{PB9}
    \end{subfigure}
        \begin{subfigure}[b]{0.23\textwidth}
        \centering
        \includegraphics[width=\textwidth]{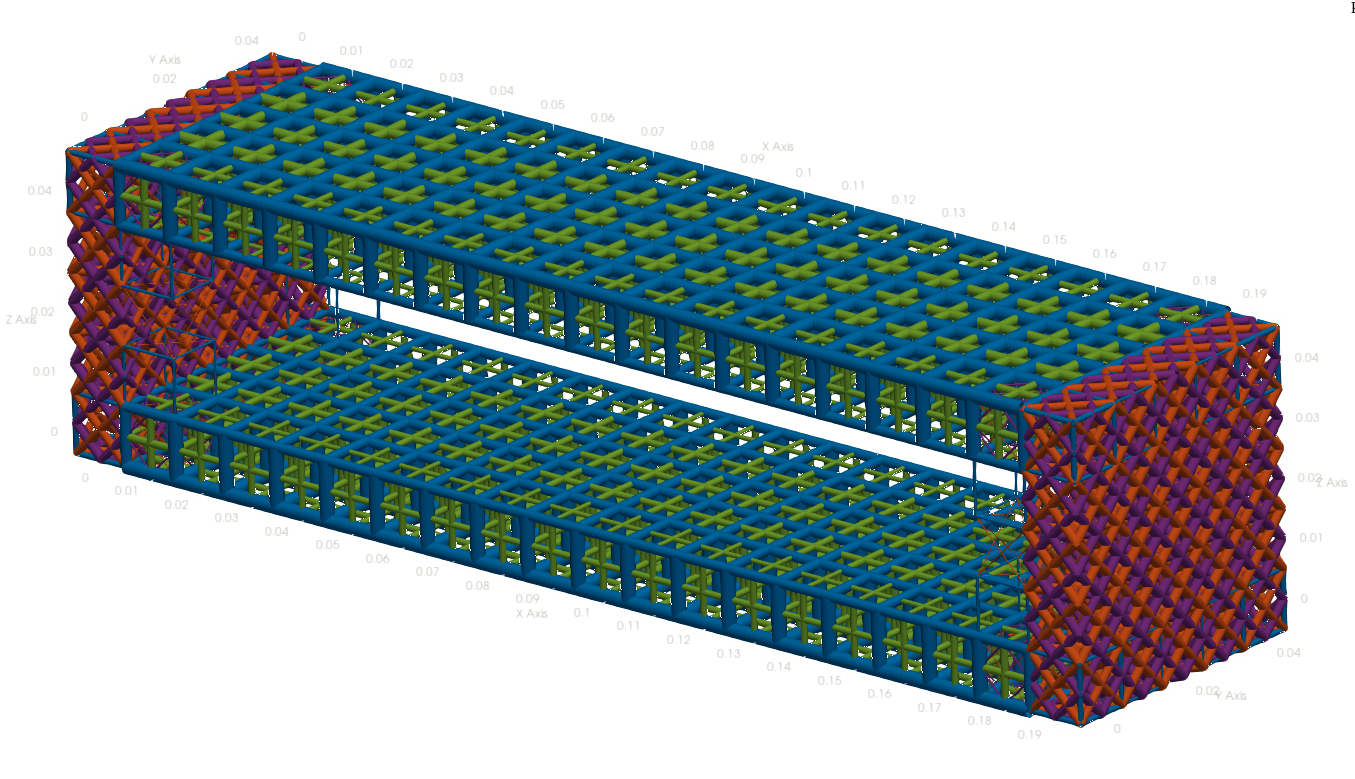}
        \caption{15 \% Volume ratio}
        \label{PB15}
    \end{subfigure}
        \begin{subfigure}[b]{0.23\textwidth}
        \centering
        \includegraphics[width=\textwidth]{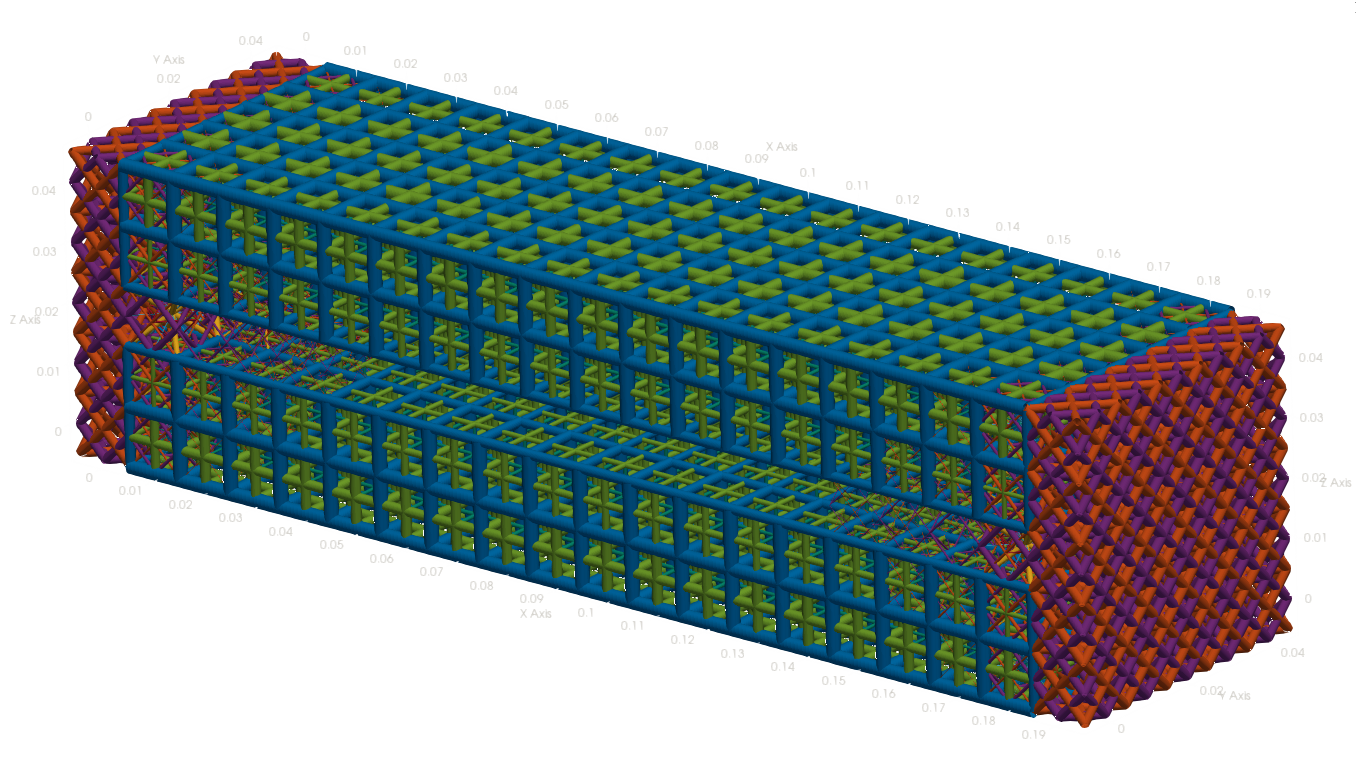}
        \caption{25 \% Volume ratio}
        \label{PB25}
    \end{subfigure}
    \begin{subfigure}[b]{0.23\textwidth}
        \centering
        \includegraphics[width=\textwidth]{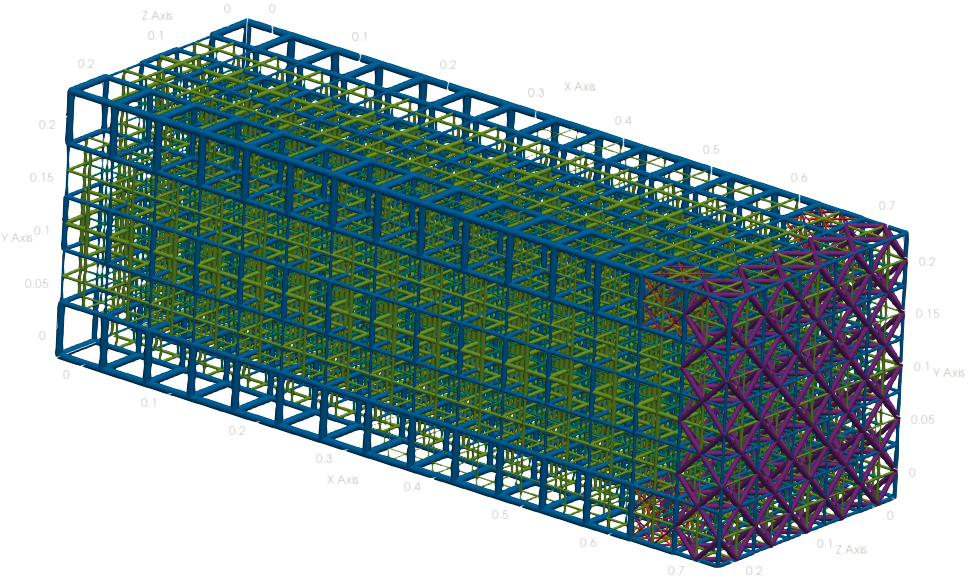}
        \caption{Optimised design for Axial Tension, 5\% Volume ratio}
        \label{Axial}
    \end{subfigure}
    \hfil
    \begin{subfigure}[b]{0.4\textwidth}
        \centering
        \includegraphics[width=\textwidth]{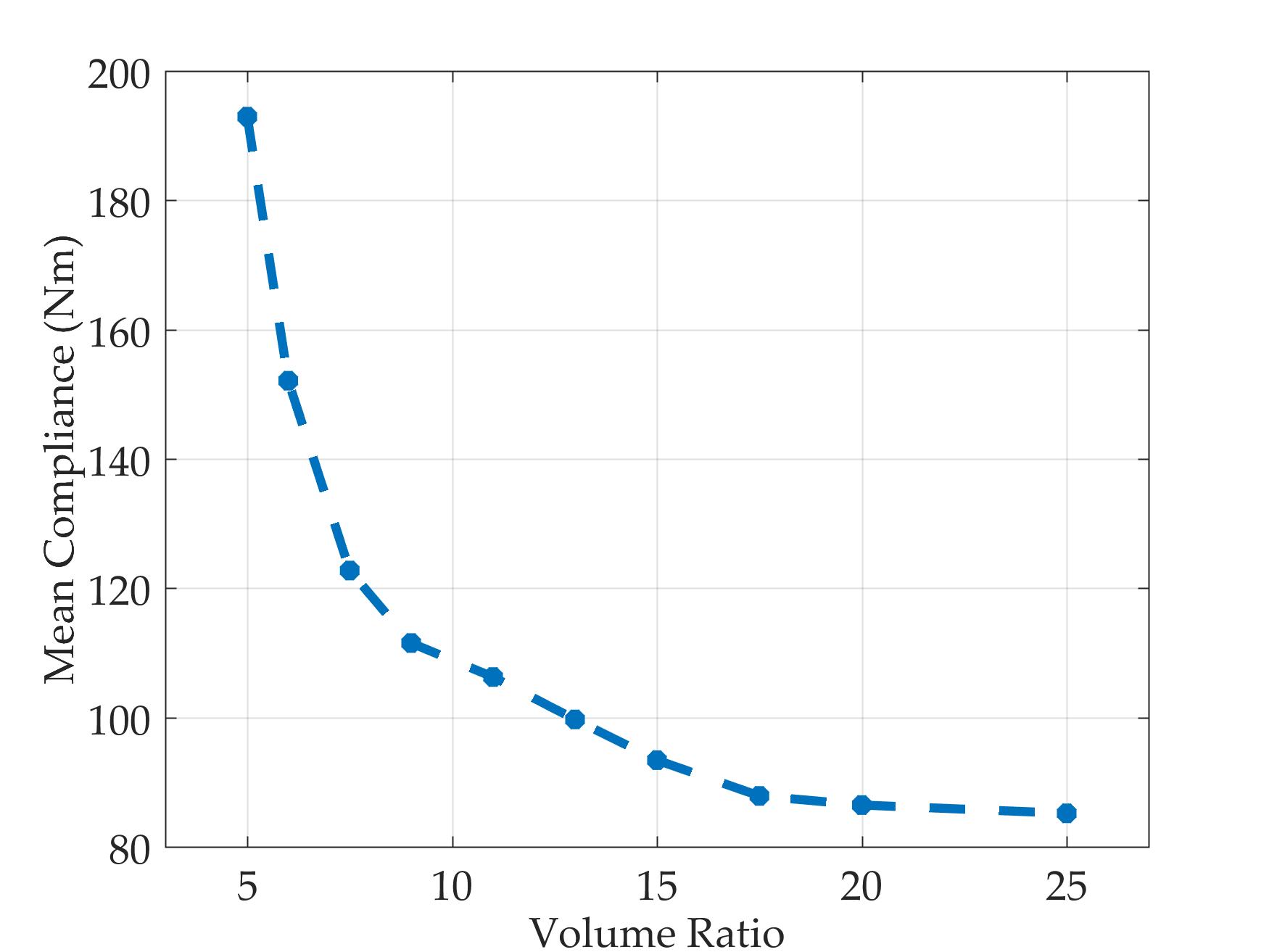}
        \caption{Monotonically decreasing compliance for increasing volume ratio}
        \label{Par}
    \end{subfigure}
     \caption{Optimal designs for Axial tension case and Pure Bending case with various volume ratio}
    \label{PB}
\end{figure*}
\begin{figure*}
    \centering
    \begin{subfigure}[b]{0.3\textwidth}
        \centering
        \includegraphics[width=\textwidth]{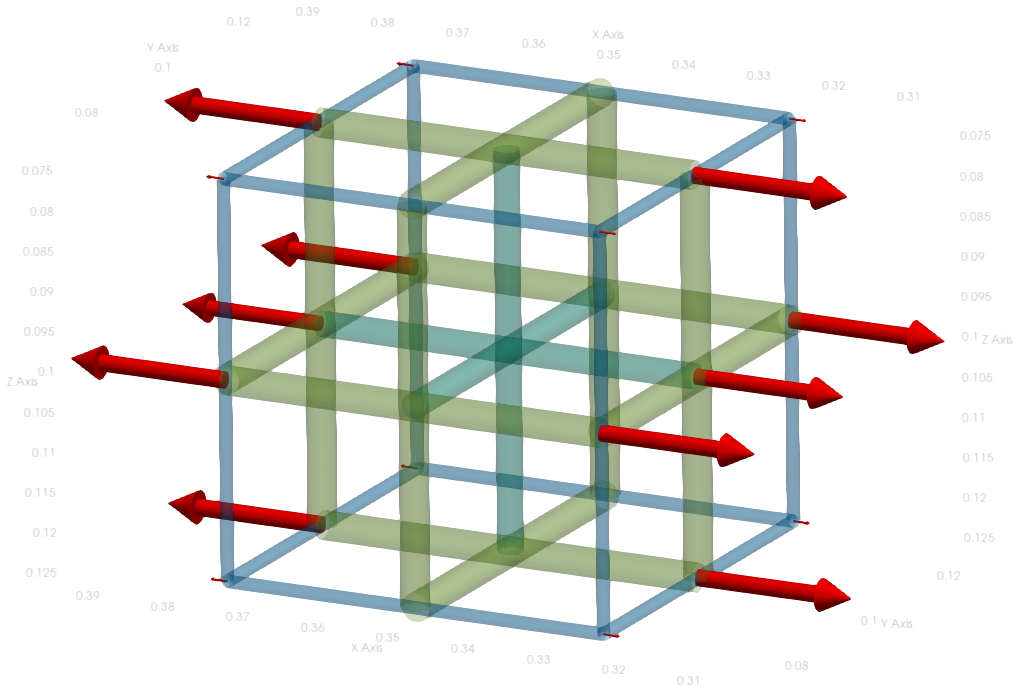}
        \caption{Pure Axial Tension}
        \label{axif}
    \end{subfigure}
        \begin{subfigure}[b]{0.3\textwidth}
        \centering
        \includegraphics[width=\textwidth]{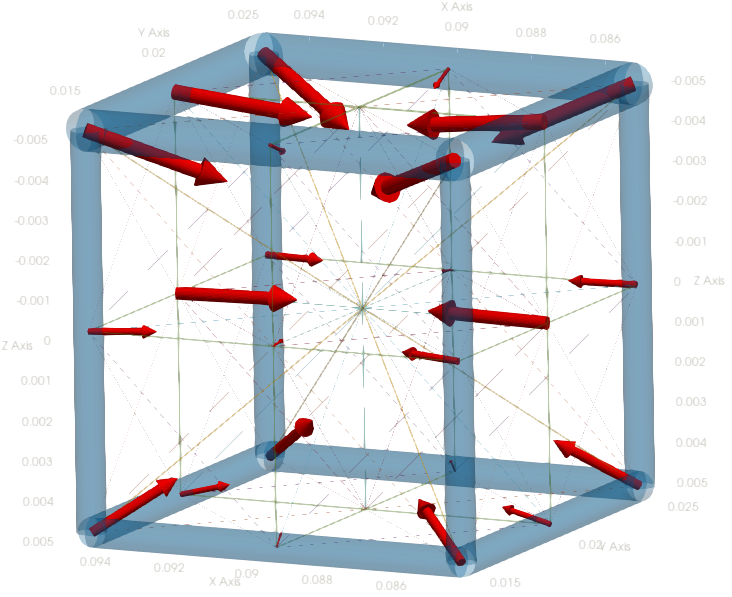}
        \caption{Pure Bending}
        \label{benif}
    \end{subfigure}
    \begin{subfigure}[b]{0.3\textwidth}
        \centering
        \includegraphics[width=\textwidth]{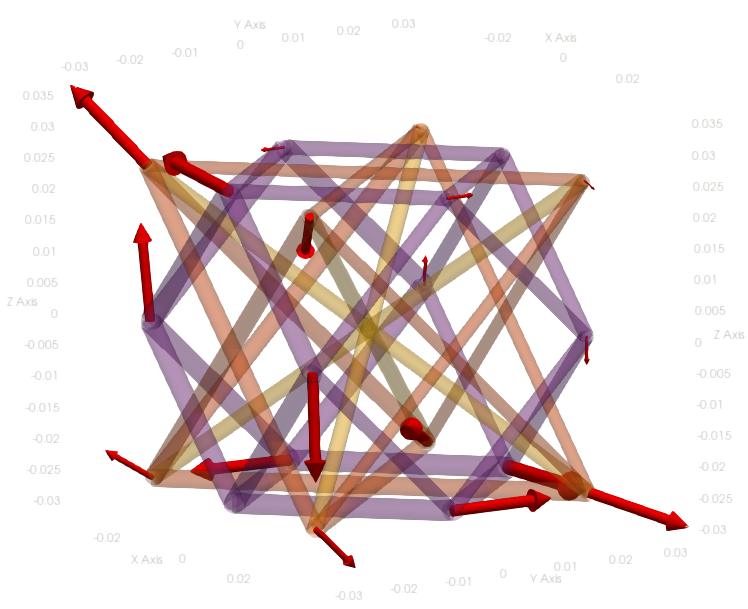}
        \caption{Torsion}
        \label{torif}
    \end{subfigure}
    \caption{Internal Force Condition on a typical cube in optimized designs}
\end{figure*}

To illustrate, we consider the optimized design for the torsion case. For a typical cube, we plot its internal forces (Figure \ref{torif}). The cube is under torsion  and our phase plots will apply here. Phases 2 and 4 dominate and are clearly the best ones for torsion loading in Figure \ref{Tor}.

Similarly, it can be shown that in the optimized design for axial tension case, a typical cube is under tension (Figure \ref{axif}). The design is prismatic and the phases 8, 5, and 1 dominate, the best ones for tension in Figure \ref{Tens}. The pure bending design has material going to the top and bottom stacks Figure \ref{PB5}. Only phase 1 is present at those locations, the best one for bending load in Figure \ref{Bend}. 

We have taken the pure bending case further by performing optimizations for different volume ratios. The designs are present in Figure \ref{PB} alongwith the mean compliance trend, which is monotonically decreasing. It can be seen that in the top and bottom stacks, initially at $\mu=0.05$, only phase 1 is present. The next best phases are phase 2,8 and 5. Among these, only phase 5 connects with phase 1 and its presence increases as the volume ratio increases. Towards the end, we can see that both phase 1 and phase 5 are chosen by the algorithm to populate the locations. This shows that the algorithm chooses the unit cells by accounting both for compatibility and optimality.

The trend observed for axial force, torsion and pure bending design validate our hypothesis. It can be easily seen that the phases chosen at different locations in the optimized design are indeed the ones that perform the best under the internal loads that act at the location, and satisfy the connectivity requirements.  

\end{section}

\begin{section}{Closure}
Lattice structures, a special case of cellular-material structures, are becoming popular in practical applications due to the advent of additive manufacturing. Earlier works on topology optimization consider only one type of lattice unit cell. But then, as is evident in natural materials, heterogeneity is important. Hence, in this paper, we presented design parameterization and a method to obtain optimized heterogeneous lattice structures. We considered, eight types of unit cells and called them phases. Therefore, the obtained result could have \textit{any number of phases} at any location, including none (indicating void). Since we have used Timoshenko beam modeling, we have captured the mechanics that account for shear effects also for not-so-slender beam segments. This makes our results suitable for 3D printing where the unit cell size does not need to be too small compared to the size of the structure. We only considered the mean compliance minimization in our examples but the method presented is equally applicable to other objectives. We show that heterogeneous design is better than any of the eight homogeneous design in all four examples considered here. 

In order to interpret the optimized results, we ranked eight unit cells in resisting axial, bending, and torsion loads. These rankings to indicate why a certain location in the optimized structure has certain type(s) of phase(s). We also discussed why certain unit cell types are not seen in the optimal heterogeneous design due to incompatibility in connectivity, even though they are superior than others when considered alone. While this validation is useful, we suggest that micropolar elasticity approach is a better way to deal with this and to enhance computational efficiency.
\end{section}

\begin{section}{Appendix}

This section is concerned with the finite element formulation of the Timoshenko 3D frame element from the first principles. We have taken the shape functions from \cite{refSpaceAna}. The symbols have their usual meanings unless stated otherwise. The elemental stiffness matrix is derived for the local coordinate system such that $x_1$ direction is along the axis of the beam. As we have taken a circular cross section, the orientation of the other axes is chosen to suit ease of programming, as it does not affect the end result. The elemental matrices are multiplied by suitable rotation matrices during the assembly of the global stiffness matrix.. 

It may be noted that $u_1$, $u_2$    and $u_3$  are the displacements at each point along the three axes in the global coordinate system. The Eulerian description of motion is followed and we assume that the displacements are small. The displacements are assumed to have a form:

\begin{equation}
\begin{split}
&u_1(x_1,x_2,x_3)=u_x(x_1)-\theta_z(x_1) x_2 + \theta_y(x_1) x_3 \\
&u_2(x_1,x_2,x_3)=u_y(x_1)-\theta_x(x_1) x_3 \\
&u_3(x_1,x_2,x_3)=u_z(x_1)+\theta_x(x_1) x_2 
\end{split}
\end{equation}

Here, the six degrees-of-freedom (DOFs), $u_x$, $ u_y $, $u_z$, $\theta_x $, $\theta_y$, $\theta_z $, are all solely functions of $x_1$. Thus, the domain is converted from a 3D bulk solid to a line with the implicit understanding of the equations. The stresses and the strains are:
\begin{equation}
\renewcommand{\arraystretch}{2}
    \begin{array}[width=0.45\textwidth]{l}
        \sigma_{11}=E({u_x}^\prime-{\theta_z}^\prime x_2+{\theta_y}^\prime x_3)\\ \sigma_{12}=\sigma_{21}=K_e G ({u_y}^\prime-\theta_z)-G{\theta_x}^\prime x_3  \\
    \sigma_{13}=\sigma_{31}=K_e G({u_z}^\prime+\theta_y)+G{\theta_x}^\prime x_2  \\
    \epsilon_{11}={u_x}^\prime-{\theta_z}^\prime x_2+{\theta_y}^\prime x_3\\ \epsilon_{12}=\epsilon_{21}=\dfrac{{u_y}^\prime-{\theta_x}^\prime x_3-\theta_z}{2}  \\
    \epsilon_{13}=\epsilon_{31}=\dfrac{{u_z}^\prime+{\theta_x}^\prime x_2+\theta_y}{2}  \\
    \end{array}
\end{equation}
In the shear stresses, we multiply the shear correction factor to the terms arising from bending. We add them here so that it naturally comes in the strain energy functional. 

In order to form the beam element, consider two nodes to be present at the either end of the member. The scalar values of the DOFs at these nodes be denoted by subscripts 1 and 2. In the middle part, the values are interpolated by shape functions.

The axial displacement and the torsion have linear shape functions. Let $\xi = \frac{x_1}{L_e}$ .
\begin{equation}
\begin{split}
&u_x(\xi)=N_1(\xi){u_x}_1 + N_2(\xi){u_x}_2 \\
&\theta_x(\xi)=N_1(\xi){\theta_x}_1 + N_2(\xi){\theta_x}_2
\end{split}
\end{equation}
The displacements related to bending, $ u_y $, $\theta_z $, $u_z$, $\theta_y$, have cubic shape functions. Their form is different from the regular Hermite polynomials as they arise from the analytical solution of the Timoshenko beam  \citep{refSpaceAna}.
\begin{equation}
    \begin{split}
        &u_y=C_0+C_1\xi+C_2\xi^2+C_3\xi^3 \\
        &\theta_z=C_1+C_2\xi+C_3\bigg(3\xi^2+\frac{6EI}{K_eGA}\bigg)
    \end{split} \nonumber
\end{equation}
\begin{equation}
    \begin{split}
        &u_z=C_4+C_5\xi+C_6\xi^2+C_7\xi^3 \\
        &\theta_y=-C_5-C_6\xi-C_7\bigg(3\xi^2+\frac{6EI}{K_eGA}\bigg) \nonumber
    \end{split}
\end{equation}
By condensing this to the shape function form and substituting the DOF values, we get,
\begin{equation}
\begin{split}
&u_y(\xi)=G_1(\xi){u_y}_1+G_2(\xi){\theta_z}_1+G_3(\xi){u_y}_2+G_4(\xi){\theta_z}_2 \\
&\theta_z(\xi)=G_5(\xi){u_y}_1+G_6(\xi){\theta_z}_1+G_7(\xi){u_y}_2+G_8(\xi){\theta_z}_2 \\
&u_z(\xi)=H_1(\xi){u_z}_1+H_2(\xi){\theta_y}_1+H_3(\xi){u_z}_2+H_4(\xi){\theta_y}_2 \\
&\theta_y(\xi)=H_5(\xi){u_z}_1+H_6(\xi){\theta_y}_1+H_7(\xi){u_z}_2+H_8(\xi){\theta_y}_2 \nonumber
\end{split}
\end{equation}
The shape function expressions are listed next.
\begin{equation}
\renewcommand{\arraystretch}{2}
    \begin{array}[width=0.45\textwidth]{l}
    N_1=1-\xi \ \ \ \ \ \ \ \ \ \ \ \ \ \ N_2=\xi  \\
    G_1=1-\dfrac{2\alpha}{2\alpha+1}\xi-\dfrac{3}{2\alpha+1}\xi^2+\dfrac{2}{2\alpha+1}\xi^3      \\
    G_2=\dfrac{\alpha+1}{2\alpha+1}L_e\xi-\dfrac{\alpha+2}{2\alpha+1}L_e\xi^2+\dfrac{1}{2\alpha+1}L_e\xi^3  \\
    G_3=1-G_1  
    \end{array}
    \end{equation}
    \begin{equation}
\renewcommand{\arraystretch}{2}
    \begin{array}[width=0.45\textwidth]{l}
    G_4=-\dfrac{\alpha}{2\alpha+1}L_e\xi-\dfrac{\alpha-1}{2\alpha+1}L_e\xi^2+\dfrac{1}{2\alpha+1}L_e\xi^3  \\
    G_5=\dfrac{6\xi(\xi-1)}{L_e(2\alpha+1)} \\ G_6=-\dfrac{(\xi-1)(2\alpha-3\xi+1)}{2\alpha+1}  \\
    G_8=\dfrac{\xi(2\alpha+3\xi-2)}{2\alpha+1} \ \ \ \ G_7=-G_5\\
    H_1=G_1  \ \ \ \ \ \ \ \ \ \ \ \ \ \ \ \ \ \ \ \ H_2=-G_2\\
    H_3=1-H_1  \ \ \ \ \ \ \ \ \ \ \ \ \ \ \ H_4=-G_4\\
    H_5=G_5  \ \ \ \ \ \ \ \ \ \ \ \ \ \ \ \ \ \ \ \   H_6=G_6\\
    H_7=G_5  \ \ \ \ \ \ \ \ \ \ \ \ \ \ \ \ \ \ \ \ H_8=G_8\\
  \alpha=\dfrac{6EI}{K_eGA{L_e}^2}
    \end{array}
\end{equation}
We now proceed to the variational formulation that will yield the elemental stiffness matrix. In our case, load vectors are not derived as none of the forces will be body forces. We will apply concentrated forces (and moments) on the nodes, that are easy to deal with. The expression for the potential energy is,
\begin{equation}
\Pi=\int_{V}^{}\frac{\sigma_{ij}\epsilon_{ij}}{2}dV+\mathrm{Work \ Potential}
\end{equation}
The first term leads to the stiffness matrix,
\begin{equation}
\begin{split}
&\int_{0}^{L_e}\frac{1}{2}\Bigg[ E(A{{u_x}^\prime}^2+I{{\theta_z}^\prime}^2+I{{\theta_y}^\prime}^2) + GI{{\theta_x}^\prime}^2 \\
&+K_e G (A{{u_y}^\prime}^2+A{\theta_z}^2-2A\theta_z {u_y}^\prime)\\&+K_e G (A{{u_z}^\prime}^2+A{\theta_y}^2+2A\theta_y {u_z}^\prime)   \Bigg]
\end{split}
\end{equation}
In this expression, the first term comes from normal strains that arise due to the axial and bending deformations. The third and fourth come from the shear strains in both of the lateral directions. The second term is due to the torsion. Since we do not consider the distributed forces on a beam, we need not delve into the work potential expression.

The tedious calculations involving the application of the Ritz method, and the subsequent integration steps, are done using the symbolic MATLAB toolbox. For convenience in calculating the sensitivity of the governing equation w.r.t. the area the expressions are converted in the terms of the area of cross-section. The final form of the stiffness matrix is obtained as: 

\begin{equation}
\centering
\left[
\begin{array}{cccccccccccc}
a_2&0&0&0&0&0&-a_2&0&0&0&0&0 \\
&a_4&0&0&0&a_3&0&-a_4&0&0&0&a_3 \\
&&a_4&0&-a_3&0&0&0&-a_4&0&-a_3&0 \\
&&&a_2&0&0&0&0&0&-a_1&0&0 \\
&&&&a_6&0&0&0&a_3&0&a_5&0\\
&&&&&a_6&0&-a_3&0&0&0&a_5\\
&&&&&&a_2&0&0&0&0&0\\
&&&&&&&a_4&0&0&0&-a_3\\
&&&&&&&&a_4&0&a_3&0\\
&&&&&&&&&a_1&0&0\\
&&&&&&&&&&a_6&0\\
&&&&&&&&&&&a_6
\end{array}
\right]
\end{equation}
\begin{equation} \nonumber
\renewcommand{\arraystretch}{2.5}
\begin{array}{lr}
a_1=\dfrac{G}{2\pi {L_e}^2}A^2 & a_2=\dfrac{E}{L_e}A^2 \\
a_3=G K_e c\bigg(\dfrac{A^2}{2cA+1}\bigg) \ \ \ \ \ \ \ &a_4=\dfrac{2a_3}{L_e}\\
a_5=-\dfrac{a_3L_e}{3}(cA-1) & a_6=\dfrac{a_3L_e}{3}(cA+2)\\
G=\dfrac{E}{2(1+\nu)} & c=\dfrac{3(1+\nu)}{\pi K_e {L_e}^2}
\end{array}
\end{equation}
The sensitivity matrix is obtained as:
\begin{equation}
\centering
\left[
\begin{array}{cccccccccccc}
a_1&0&0&0&0&0&-a_1&0&0&0&0&0 \\
&a_4&0&0&0&a_3&0&-a_4&0&0&0&a_3 \\
&&a_4&0&-a_3&0&0&0&-a_4&0&-a_3&0 \\
&&&a_2&0&0&0&0&0&-a_2&0&0 \\
&&&&a_5&0&0&0&a_3&0&a_6&0\\
&&&&&a_5&0&-a_3&0&0&0&a_6\\
&&&&&&a_1&0&0&0&0&0\\
&&&&&&&a_4&0&0&0&-a_3\\
&&&&&&&&a_4&0&a_3&0\\
&&&&&&&&&a_2&0&0\\
&&&&&&&&&&a_5&0\\
&&&&&&&&&&&a_5
\end{array}
\right]
\end{equation}

\begin{equation} \nonumber
\renewcommand{\arraystretch}{2.5}
\begin{array}{l}
a_1=\dfrac{E}{L_e} \ \ \ \ \ \ \ \ a_2=\dfrac{GA}{\pi L_e} \\
a_3=\dfrac{3EA+2\pi K_e G {L_e}^2 c^2 A^2}{\pi {L_e}^2 (2cA+1)^2} \ \ \ \ \ \ \ \ a_4=\dfrac{2}{L_e}a_3\\
a_5=\dfrac{2EA+2Ec^2 A^3+\pi K_e G{L_e}^2 c^2 A^2+2EcA^2 }{\pi {L_e}^2 (2cA+1)^2} \\
a_6=\dfrac{EA-2Ec^2 A^3+\pi K_e G{L_e}^2 c^2 A^2-2EcA^2}{\pi {L_e}^2 (2cA+1)^2}\\
\end{array}
\end{equation}

\end{section}

\begin{acknowledgements}
We thank Prof. Vijay Natarajan from CSA, IISc Bengaluru, for his advice on the visualization of different phases. Valuable discussions with Mr. Jaddivada Siddharth enriched our work in implementing good programming practices. Financial support from the Uchhatar Avishkar Yojana (UAY) of Government of India, Ministry of Heavy Industries with partial support from Eaton Corporation, Pune, India. We also thank Dr. Hari Prasad Konka and his team from Eaton for their regular insights all along the work.  
\end{acknowledgements}
\section*{Conflict of interest}
There is no conflict of interest in this work with any of the authors.
\section*{Replication of Results}
Matlab codes written to produce results are shared at \href{mecheng.iisc.ac.in/suresh/LatticeStructureOptimization}{mecheng.iisc.ac.in/suresh/LatticeStructureOptimization}
\bibliographystyle{spbasic}
\bibliography{LatPapRef}

\end{document}